\documentclass[a4paper,onecolumn,superscriptaddress,11pt,usenames,dvipsnames,svgnames,accepted=2021-04-17]{quantumarticle}
\pdfoutput=1
\usepackage{rotating}
\usepackage{floatpag}
\rotfloatpagestyle{empty}
\renewcommand{\theenumi}{\arabic{enumi}}
\usepackage{hyperref}
\usepackage{dsfont}
\usepackage{amsmath,amsthm,amssymb}
\usepackage{xspace,enumerate,epsfig}%color,
\usepackage{graphicx}
\graphicspath{{.}{./figures/}}
\usepackage{marvosym}

\usepackage{tikzfig}
\usepackage{stmaryrd}
\usepackage{docmute}
\usepackage{keycommand}

\newkeycommand{\pointmap}[style=point][1]{\,\tikz{\node[style=\commandkey{style}] (x) at (0,-0.3) {$#1$};\node [style=none] (3) at (0, 1.0) {};\node [style=none] (3) at (0, -1.0) {};\draw (x)--(0,0.7);}\,}

\newkeycommand{\typedkpoint}[style=kpoint,edgestyle=][2]{%
\,\begin{tikzpicture}
  \begin{pgfonlayer}{nodelayer}
    \node [style=none] (0) at (0, 1) {};
    \node [style=\commandkey{style}] (1) at (0, -0.75) {$#2$};
    \node [style=right label] (2) at (0.25, 0.5) {$#1$};
  \end{pgfonlayer}
  \begin{pgfonlayer}{edgelayer}
    \draw [\commandkey{edgestyle}] (1) to (0.center);
  \end{pgfonlayer}
\end{tikzpicture}\,}

\newkeycommand{\typedkpointdag}[style=kpoint adjoint,edgestyle=][2]{%
\,\begin{tikzpicture}
  \begin{pgfonlayer}{nodelayer}
    \node [style=none] (0) at (0, -1) {};
    \node [style=\commandkey{style}] (1) at (0, 0.75) {$#2$};
    \node [style=right label] (2) at (0.25, -0.5) {$#1$};
  \end{pgfonlayer}
  \begin{pgfonlayer}{edgelayer}
    \draw [\commandkey{edgestyle}] (0.center) to (1);
  \end{pgfonlayer}
\end{tikzpicture}\,}

\newcommand{\bistateadj}[1]{\,\begin{tikzpicture}[yshift=-3mm]
  \begin{pgfonlayer}{nodelayer}
    \node [style=kpointadj, minimum width=1 cm, inner sep=2pt] (0) at (0, 0.5) {$\psi$};
    \node [style=none] (1) at (-0.75, 0.25) {};
    \node [style=none] (2) at (0.75, 0.25) {};
    \node [style=none] (3) at (-0.75, -0.5) {};
    \node [style=none] (4) at (0.75, -0.5) {};
  \end{pgfonlayer}
  \begin{pgfonlayer}{edgelayer}
    \draw (1.center) to (3.center);
    \draw (2.center) to (4.center);
  \end{pgfonlayer}
\end{tikzpicture}\,}

\newkeycommand{\pointketbra}[style1=copoint,style2=point][2]{\,%
\begin{tikzpicture}
  \begin{pgfonlayer}{nodelayer}
    \node [style=none] (0) at (0, -1.5) {};
    \node [style=\commandkey{style1}] (1) at (0, -0.75) {$#1$};
    \node [style=\commandkey{style2}] (2) at (0, 0.75) {$#2$};
    \node [style=none] (3) at (0, 1.5) {};
  \end{pgfonlayer}
  \begin{pgfonlayer}{edgelayer}
    \draw (0.center) to (1);
    \draw (2) to (3.center);
  \end{pgfonlayer}
\end{tikzpicture}\,}

\newkeycommand{\twocopointketbra}[style1=copoint,style2=copoint,style3=point][3]{\,%
\begin{tikzpicture}
  \begin{pgfonlayer}{nodelayer}
    \node [style=none] (0) at (-0.75, -1.5) {};
    \node [style=none] (1) at (0.75, -1.5) {};
    \node [style=\commandkey{style1}] (2) at (-0.75, -0.75) {$#1$};
    \node [style=\commandkey{style2}] (3) at (0.75, -0.75) {$#2$};
    \node [style=\commandkey{style3}] (4) at (0, 0.75) {$#3$};
    \node [style=none] (5) at (0, 1.5) {};
  \end{pgfonlayer}
  \begin{pgfonlayer}{edgelayer}
    \draw (0.center) to (2);
    \draw (1.center) to (3);
    \draw (4) to (5.center);
  \end{pgfonlayer}
\end{tikzpicture}\,}

\newkeycommand{\twopointketbra}[style1=copoint,style2=point,style3=point][3]{\,%
\begin{tikzpicture}
  \begin{pgfonlayer}{nodelayer}
    \node [style=none] (0) at (0, -1.5) {};
    \node [style=none] (1) at (-0.75, 1.5) {};
    \node [style=\commandkey{style1}] (2) at (0, -0.75) {$#1$};
    \node [style=\commandkey{style2}] (3) at (-0.75, 0.75) {$#2$};
    \node [style=\commandkey{style3}] (4) at (0.75, 0.75) {$#3$};
    \node [style=none] (5) at (0.75, 1.5) {};
  \end{pgfonlayer}
  \begin{pgfonlayer}{edgelayer}
    \draw (0.center) to (2);
    \draw (1.center) to (3);
    \draw (4) to (5.center);
  \end{pgfonlayer}
\end{tikzpicture}\,}

\newkeycommand{\pointbraket}[style1=point,style2=copoint][2]{\,%
\begin{tikzpicture}
  \begin{pgfonlayer}{nodelayer}
    \node [style=\commandkey{style1}] (0) at (0, -0.5) {$#1$};
    \node [style=\commandkey{style2}] (1) at (0, 0.5) {$#2$};
  \end{pgfonlayer}
  \begin{pgfonlayer}{edgelayer}
    \draw (0) to (1);
  \end{pgfonlayer}
\end{tikzpicture}\,}

\newkeycommand{\kpointbraket}[style1=kpoint,style2=kpoint adjoint][2]{\,%
\begin{tikzpicture}
  \begin{pgfonlayer}{nodelayer}
    \node [style=\commandkey{style1}] (0) at (0, -0.75) {$#1$};
    \node [style=\commandkey{style2}] (1) at (0, 0.75) {$#2$};
  \end{pgfonlayer}
  \begin{pgfonlayer}{edgelayer}
    \draw (0) to (1);
  \end{pgfonlayer}
\end{tikzpicture}\,}

\newkeycommand{\kpointmap}[style1=kpoint,style2=map][2]{\,%
\begin{tikzpicture}
  \begin{pgfonlayer}{nodelayer}
    \node [style=\commandkey{style1}] (0) at (0, -1.1) {$#1$};
    \node [style=\commandkey{style2}] (1) at (0, 0.2) {$#2$};
    \node [style=none] (2) at (0, 1.5) {};
  \end{pgfonlayer}
  \begin{pgfonlayer}{edgelayer}
    \draw (0) to (1);
    \draw (1) to (2.center);
  \end{pgfonlayer}
\end{tikzpicture}%
\,}

\newkeycommand{\sandwichtwo}[style1=point,style2=map,style3=map,style4=copoint][4]{\,%
\begin{tikzpicture}
  \begin{pgfonlayer}{nodelayer}
    \node [style=\commandkey{style2}] (0) at (0, -0.75) {$#2$};
    \node [style=\commandkey{style3}] (1) at (0, 0.75) {$#3$};
    \node [style=\commandkey{style1}] (2) at (0, -2) {$#1$};
    \node [style=\commandkey{style4}] (3) at (0, 2) {$#4$};
  \end{pgfonlayer}
  \begin{pgfonlayer}{edgelayer}
    \draw (2) to (0);
    \draw (0) to (1);
    \draw (1) to (3);
  \end{pgfonlayer}
\end{tikzpicture}\,}

\newkeycommand{\longbraket}[style1=point,style2=copoint][2]{\,%
\begin{tikzpicture}
  \begin{pgfonlayer}{nodelayer}
    \node [style=\commandkey{style1}] (0) at (0, -2) {$#1$};
    \node [style=\commandkey{style2}] (1) at (0, 2) {$#2$};
  \end{pgfonlayer}
  \begin{pgfonlayer}{edgelayer}
    \draw (0) to (1);
  \end{pgfonlayer}
\end{tikzpicture}\,}

\newkeycommand{\onb}[style=point]{\ensuremath{\left\{\tikz{\node[style=\commandkey{style}] (x) at (0,-0.3) {$j$};\draw (x)--(0,0.7);}\right\}}\xspace}

\newkeycommand{\copointmap}[style=copoint][1]{\,\tikz{\node[style=\commandkey{style}] (x) at (0,0.3) {$#1$};\draw (x)--(0,-0.7);}\,}

\newcommand{\bra}[1]{\ensuremath{\left\langle #1 \right|}}
\newcommand{\ket}[1]{\ensuremath{\left|  #1 \right\rangle}}

\newcommand{\ketbra}[2]{\ensuremath{\ket{#1}\!\bra{#2}}}

\tikzstyle{cdiag}=[matrix of math nodes, row sep=3em, column sep=3em, text height=1.5ex, text depth=0.25ex,inner sep=0.5em]
\tikzstyle{arrow above}=[transform canvas={yshift=0.5ex}]
\tikzstyle{arrow below}=[transform canvas={yshift=-0.5ex}]

\usetikzlibrary{shapes.multipart}

\tikzstyle{bwSpider}=[
       rectangle split,
       rectangle split parts=2,
       rectangle split part fill={black,white},
 minimum size=3.6 mm, inner sep=-2mm, draw=black,scale=0.5,rounded corners=0.8 mm
       ]
 \tikzstyle{wbSpider}=[
       rectangle split,
       rectangle split parts=2,
       rectangle split part fill={white,black},
 minimum size=3.6 mm, inner sep=-2mm, draw=black,scale=0.5,rounded corners=0.8 mm
       ]
\tikzstyle{qWire}=[line width = 1pt, color=quantumviolet]
\tikzstyle{cWire}=[color=quantumgray,thin]%[densely dotted, thick]
\tikzstyle{env}=[copoint,regular polygon rotate=0,minimum width=0.2cm, fill=black]

\tikzstyle{probs}=[shape=semicircle,fill=white,draw=black,shape border rotate=180,minimum width=1.2cm]

\tikzstyle{every picture}=[baseline=-0.25em,scale=0.5]
\tikzstyle{dotpic}=[] % for backwards-compatibility
\tikzstyle{diredges}=[every to/.style={diredge}]
\tikzstyle{math matrix}=[matrix of math nodes,left delimiter=(,right delimiter=),inner sep=2pt,column sep=1em,row sep=0.5em,nodes={inner sep=0pt},text height=1.5ex, text depth=0.25ex]

\tikzstyle{inline text}=[text height=1.5ex, text depth=0.25ex,yshift=0.5mm]
\tikzstyle{label}=[font=\footnotesize,text height=1.5ex, text depth=0.25ex,yshift=0.5mm]
\tikzstyle{left label}=[label,anchor=east,xshift=1.5mm]
\tikzstyle{right label}=[label,anchor=west,xshift=-0.5mm]

\tikzstyle{braceedge}=[decorate,decoration={brace,amplitude=2mm,raise=-1mm}]
\tikzstyle{small braceedge}=[decorate,decoration={brace,amplitude=1mm,raise=-1mm}]

\tikzstyle{doubled}=[line width=1.6pt] % set the line width for all doubled (quantum) maps/wires
\tikzstyle{boldedge}=[doubled,shorten <=-0.17mm,shorten >=-0.17mm]
\tikzstyle{boldedgegray}=[doubled,gray,shorten <=-0.17mm,shorten >=-0.17mm]
\tikzstyle{singleedgegray}=[gray]%,shorten <=-0.1mm,shorten >=-0.1mm]

\tikzstyle{semidoubled}=[line width=1.4pt] % set the line width for all doubled (quantum) maps/wires
\tikzstyle{semiboldedgegray}=[semidoubled,gray,shorten <=-0.17mm,shorten >=-0.17mm]

\tikzstyle{boxedge}=[semiboldedgegray]

\tikzstyle{boldedgedashed}=[very thick,dashed,shorten <=-0.17mm,shorten >=-0.17mm]
\tikzstyle{vboldedgedashed}=[doubled,dashed,shorten <=-0.17mm,shorten >=-0.17mm]
\tikzstyle{left hook arrow}=[left hook-latex]
\tikzstyle{right hook arrow}=[right hook-latex]
\tikzstyle{sembracket}=[line width=0.5pt,shorten <=-0.07mm,shorten >=-0.07mm]

\tikzstyle{causal edge}=[->,thick,gray]
\tikzstyle{causal nondir}=[thick,gray]
\tikzstyle{timeline}=[thick,gray, dashed]

\tikzstyle{cedge}=[<->,thick,gray!70!white]

\tikzstyle{empty diagram}=[draw=gray!40!white,dashed,shape=rectangle,minimum width=1cm,minimum height=1cm]
\tikzstyle{empty diagram small}=[draw=gray!50!white,dashed,shape=rectangle,minimum width=0.6cm,minimum height=0.5cm]

\tikzstyle{dot}=[inner sep=0mm,minimum width=2mm,minimum height=2mm,draw,shape=circle]

\tikzstyle{leak}=[white dot, shape=regular polygon, minimum size=3.3 mm, regular polygon sides=3, outer sep=-0.2mm, regular polygon rotate=270]
\tikzstyle{proj}=[regular polygon,regular polygon sides=4,draw,scale=0.75,inner sep=-0.5pt,minimum width=6mm,fill=white]
\tikzstyle{projOut}=[regular polygon,regular polygon sides=3,draw,scale=0.75,inner sep=-0.5pt,minimum width=7.5mm,fill=white,regular polygon rotate=180]
\tikzstyle{projIn}=[regular polygon,regular polygon sides=3,draw,scale=0.75,inner sep=-0.5pt,minimum width=7.5mm,fill=white]
\tikzstyle{Vleak}=[white dot, shape=regular polygon, minimum size=3.3 mm, regular polygon sides=3, outer sep=-0.2mm, regular polygon rotate=90]
\tikzstyle{dleak}=[white dot, line width=1.6pt, shape=regular polygon, minimum size=3.3 mm, regular polygon sides=3, outer sep=-0.2mm, regular polygon rotate=270]

\tikzstyle{Wsquare}=[white dot, shape=regular polygon, rounded corners=0.8 mm, minimum size=3.3 mm, regular polygon sides=3, outer sep=-0.2mm]
\tikzstyle{Wsquareadj}=[white dot, shape=regular polygon, rounded corners=0.8 mm, minimum size=3.3 mm, regular polygon sides=3, outer sep=-0.2mm, regular polygon rotate=180]
\tikzstyle{ddot}=[inner sep=0mm, doubled, minimum width=2.5mm,minimum height=2.5mm,draw,shape=circle]

\tikzstyle{black dot}=[dot,fill=black]
\tikzstyle{white dot}=[dot,fill=white,,text depth=-0.2mm]
\tikzstyle{white Wsquare}=[Wsquare,fill=gray,,text depth=-0.2mm]
\tikzstyle{white Wsquareadj}=[Wsquareadj,fill=white,,text depth=-0.2mm]
\tikzstyle{green dot}=[white dot] % for backwards-compatibility
\tikzstyle{gray dot}=[dot,fill=gray!40!white,,text depth=-0.2mm]
\tikzstyle{red dot}=[gray dot] % for backwards-compatibility

\tikzstyle{black ddot}=[ddot,fill=black]
\tikzstyle{white ddot}=[ddot,fill=white]
\tikzstyle{gray ddot}=[ddot,fill=gray!40!white]

\tikzstyle{gray edge}=[gray!60!white]

\tikzstyle{small dot}=[inner sep=0.5mm,minimum width=0pt,minimum height=0pt,draw,shape=circle]

\tikzstyle{small black dot}=[small dot,fill=black]
\tikzstyle{small white dot}=[small dot,fill=white]
\tikzstyle{small gray dot}=[small dot,fill=gray!40!white]

\tikzstyle{causal dot}=[inner sep=0.4mm,minimum width=0pt,minimum height=0pt,draw=white,shape=circle,fill=gray!40!white]

\tikzstyle{phase dimensions}=[minimum size=5mm,font=\footnotesize,rectangle,rounded corners=2.5mm,inner sep=0.2mm,outer sep=-2mm]
\tikzstyle{dphase dimensions}=[minimum size=5mm,font=\footnotesize,rectangle,rounded corners=2.5mm,inner sep=0.2mm,outer sep=-2mm]
\tikzstyle{white phase dot}=[dot,fill=white,phase dimensions]
\tikzstyle{white phase ddot}=[ddot,fill=white,dphase dimensions]

\tikzstyle{white rect ddot}=[draw=black,fill=white,doubled,minimum size=5mm,font=\footnotesize,rectangle,rounded corners=2.5mm,inner sep=0.2mm]
\tikzstyle{gray rect ddot}=[draw=black,fill=gray!40!white,doubled,minimum size=6mm,font=\footnotesize,rectangle,rounded corners=3mm]

\tikzstyle{gray phase dot}=[dot,fill=gray!40!white,phase dimensions]
\tikzstyle{gray phase ddot}=[ddot,fill=gray!40!white,dphase dimensions]
\tikzstyle{grey phase dot}=[gray phase dot]
\tikzstyle{grey phase ddot}=[gray phase ddot]

\tikzstyle{small phase dimensions}=[minimum size=4mm,font=\tiny,rectangle,rounded corners=2mm,inner sep=0.2mm,outer sep=-2mm]
\tikzstyle{small dphase dimensions}=[minimum size=4mm,font=\tiny,rectangle,rounded corners=2mm,inner sep=0.2mm,outer sep=-2mm]

\tikzstyle{small gray phase dot}=[dot,fill=gray!40!white,small phase dimensions]
\tikzstyle{small gray phase ddot}=[ddot,fill=gray!40!white,small dphase dimensions]

\tikzstyle{small map}=[draw,shape=rectangle,minimum height=4mm,minimum width=4mm,fill=white]

\tikzstyle{cnot}=[fill=white,shape=circle,inner sep=-1.4pt]

\tikzstyle{asym hadamard}=[fill=white,draw,shape=NEbox,inner sep=0.6mm,font=\footnotesize,minimum height=4mm]
\tikzstyle{asym hadamard conj}=[fill=white,draw,shape=NWbox,inner sep=0.6mm,font=\footnotesize,minimum height=4mm]
\tikzstyle{asym hadamard dag}=[fill=white,draw,shape=SEbox,inner sep=0.6mm,font=\footnotesize,minimum height=4mm]

\tikzstyle{hadamard}=[fill=white,draw,inner sep=0.6mm,font=\footnotesize,minimum height=4mm,minimum width=4mm]
\tikzstyle{small hadamard}=[fill=white,draw,inner sep=0.6mm,minimum height=1.5mm,minimum width=1.5mm]
\tikzstyle{small hadamard rotate}=[small hadamard,rotate=45]
\tikzstyle{dhadamard}=[hadamard,doubled]
\tikzstyle{small dhadamard}=[small hadamard,doubled]
\tikzstyle{small dhadamard rotate}=[small hadamard rotate,doubled]
\tikzstyle{antipode}=[white dot,inner sep=0.3mm,font=\footnotesize]

\tikzstyle{scalar}=[diamond,draw,inner sep=0.5pt,font=\small]
\tikzstyle{dscalar}=[diamond,doubled, draw,inner sep=0.5pt,font=\small]

\tikzstyle{small box}=[rectangle,inline text,fill=white,draw,minimum height=5mm,yshift=-0.5mm,minimum width=5mm,font=\small]
\tikzstyle{small gray box}=[small box,fill=gray!30]
\tikzstyle{medium box}=[rectangle,inline text,fill=white,draw,minimum height=5mm,yshift=-0.5mm,minimum width=10mm,font=\small]
\tikzstyle{square box}=[small box] % for backwards-compatibility
\tikzstyle{medium gray box}=[small box,fill=gray!30]
\tikzstyle{semilarge box}=[rectangle,inline text,fill=white,draw,minimum height=5mm,yshift=-0.5mm,minimum width=12.5mm,font=\small]
\tikzstyle{large box}=[rectangle,inline text,fill=white,draw,minimum height=5mm,yshift=-0.5mm,minimum width=15mm,font=\small]
\tikzstyle{large gray box}=[small box,fill=gray!30]

\tikzstyle{Bayes box}=[rectangle,fill=black,draw, minimum height=3mm, minimum width=3mm]

\tikzstyle{gray square point}=[small box,fill=gray!50]

\tikzstyle{dphase box white}=[dhadamard]
\tikzstyle{dphase box gray}=[dhadamard,fill=gray!50!white]
\tikzstyle{phase box white}=[hadamard]
\tikzstyle{phase box gray}=[hadamard,fill=gray!50!white]

\tikzstyle{point}=[regular polygon,regular polygon sides=3,draw,scale=0.75,inner sep=-0.5pt,minimum width=9mm,fill=white,regular polygon rotate=180]
\tikzstyle{point nosep}=[regular polygon,regular polygon sides=3,draw,scale=0.75,inner sep=-2pt,minimum width=9mm,fill=white,regular polygon rotate=180]
\tikzstyle{copoint}=[regular polygon,regular polygon sides=3,draw,scale=0.75,inner sep=-0.5pt,minimum width=9mm,fill=white]
\tikzstyle{dpoint}=[point,doubled]
\tikzstyle{dcopoint}=[copoint,doubled]

\tikzstyle{pointgrow}=[shape=cornerpoint,kpoint common,scale=0.75,inner sep=3pt]
\tikzstyle{pointgrow dag}=[shape=cornercopoint,kpoint common,scale=0.75,inner sep=3pt]

\tikzstyle{wide copoint}=[fill=white,draw,shape=isosceles triangle,shape border rotate=90,isosceles triangle stretches=true,inner sep=0pt,minimum width=1.5cm,minimum height=6.12mm]
\tikzstyle{wide point}=[fill=white,draw,shape=isosceles triangle,shape border rotate=-90,isosceles triangle stretches=true,inner sep=0pt,minimum width=1.5cm,minimum height=6.12mm,yshift=-0.0mm]
\tikzstyle{wide point plus}=[fill=white,draw,shape=isosceles triangle,shape border rotate=-90,isosceles triangle stretches=true,inner sep=0pt,minimum width=1.74cm,minimum height=7mm,yshift=-0.0mm]

\tikzstyle{wide dpoint}=[fill=white,doubled,draw,shape=isosceles triangle,shape border rotate=-90,isosceles triangle stretches=true,inner sep=0pt,minimum width=1.5cm,minimum height=6.12mm,yshift=-0.0mm]

\tikzstyle{tinypoint}=[regular polygon,regular polygon sides=3,draw,scale=0.55,inner sep=-0.15pt,minimum width=6mm,fill=white,regular polygon rotate=180]

\tikzstyle{white point}=[point]
\tikzstyle{white dpoint}=[dpoint]
\tikzstyle{green point}=[white point] % for backwards-compatibility
\tikzstyle{white copoint}=[copoint]
\tikzstyle{gray point}=[point,fill=gray!40!white]
\tikzstyle{gray dpoint}=[gray point,doubled]
\tikzstyle{red point}=[gray point] % for backwards-compatibility
\tikzstyle{gray copoint}=[copoint,fill=gray!40!white]
\tikzstyle{gray dcopoint}=[gray copoint,doubled]

\tikzstyle{white point guide}=[regular polygon,regular polygon sides=3,font=\scriptsize,draw,scale=0.65,inner sep=-0.5pt,minimum width=9mm,fill=white,regular polygon rotate=180]

\tikzstyle{black point}=[point,fill=black,font=\color{white}]
\tikzstyle{black copoint}=[copoint,fill=black,font=\color{white}]

\tikzstyle{tiny gray point}=[tinypoint,fill=gray!40!white]

\tikzstyle{diredge}=[->]
\tikzstyle{ddiredge}=[<->]
\tikzstyle{rdiredge}=[<-]
\tikzstyle{thickdiredge}=[->, very thick]
\tikzstyle{pointer edge}=[->,very thick,gray]
\tikzstyle{pointer edge part}=[very thick,gray]
\tikzstyle{dashed edge}=[dashed]
\tikzstyle{thick dashed edge}=[very thick,dashed]
\tikzstyle{thick gray dashed edge}=[thick dashed edge,gray!40]
\tikzstyle{thick map edge}=[very thick,|->]

\makeatletter
\newcommand{\boxshape}[3]{%
\pgfdeclareshape{#1}{
\inheritsavedanchors[from=rectangle] % this is nearly a rectangle
\inheritanchorborder[from=rectangle]
\inheritanchor[from=rectangle]{center}
\inheritanchor[from=rectangle]{north}
\inheritanchor[from=rectangle]{south}
\inheritanchor[from=rectangle]{west}
\inheritanchor[from=rectangle]{east}
\backgroundpath{% this is new
\southwest \pgf@xa=\pgf@x \pgf@ya=\pgf@y
\northeast \pgf@xb=\pgf@x \pgf@yb=\pgf@y

\@tempdima=#2
\@tempdimb=#3

\pgfpathmoveto{\pgfpoint{\pgf@xa - 5pt + \@tempdima}{\pgf@ya}}
\pgfpathlineto{\pgfpoint{\pgf@xa - 5pt - \@tempdima}{\pgf@yb}}
\pgfpathlineto{\pgfpoint{\pgf@xb + 5pt + \@tempdimb}{\pgf@yb}}
\pgfpathlineto{\pgfpoint{\pgf@xb + 5pt - \@tempdimb}{\pgf@ya}}
\pgfpathlineto{\pgfpoint{\pgf@xa - 5pt + \@tempdima}{\pgf@ya}}
\pgfpathclose
}
}}

\boxshape{NEbox}{0pt}{3pt}
\boxshape{SEbox}{0pt}{-3pt}
\boxshape{NWbox}{5pt}{0pt}
\boxshape{SWbox}{-5pt}{0pt}
\boxshape{EBox}{-3pt}{3pt}
\boxshape{WBox}{3pt}{-3pt}
\makeatother

\tikzstyle{cloud}=[shape=cloud,draw,minimum width=1.5cm,minimum height=1.5cm]

\tikzstyle{map}=[draw,shape=NEbox,inner sep=1pt,minimum height=4mm,fill=white]
\tikzstyle{dashedmap}=[draw,dashed,shape=NEbox,inner sep=2pt,minimum height=6mm,fill=white]
\tikzstyle{mapdag}=[draw,shape=SEbox,inner sep=1pt,minimum height=4mm,fill=white]
\tikzstyle{mapadj}=[draw,shape=SEbox,inner sep=2pt,minimum height=6mm,fill=white]
\tikzstyle{maptrans}=[draw,shape=SWbox,inner sep=2pt,minimum height=6mm,fill=white]
\tikzstyle{mapconj}=[draw,shape=NWbox,inner sep=2pt,minimum height=6mm,fill=white]

\tikzstyle{medium map}=[draw,shape=NEbox,inner sep=2pt,minimum height=6mm,fill=white,minimum width=7mm]
\tikzstyle{medium map dag}=[draw,shape=SEbox,inner sep=2pt,minimum height=6mm,fill=white,minimum width=7mm]
\tikzstyle{medium map adj}=[draw,shape=SEbox,inner sep=2pt,minimum height=6mm,fill=white,minimum width=7mm]
\tikzstyle{medium map trans}=[draw,shape=SWbox,inner sep=2pt,minimum height=6mm,fill=white,minimum width=7mm]
\tikzstyle{medium map conj}=[draw,shape=NWbox,inner sep=2pt,minimum height=6mm,fill=white,minimum width=7mm]
\tikzstyle{semilarge map}=[draw,shape=NEbox,inner sep=2pt,minimum height=6mm,fill=white,minimum width=9.5mm]
\tikzstyle{semilarge map trans}=[draw,shape=SWbox,inner sep=2pt,minimum height=6mm,fill=white,minimum width=9.5mm]
\tikzstyle{semilarge map adj}=[draw,shape=SEbox,inner sep=2pt,minimum height=6mm,fill=white,minimum width=9.5mm]
\tikzstyle{semilarge map dag}=[draw,shape=SEbox,inner sep=2pt,minimum height=6mm,fill=white,minimum width=9.5mm]
\tikzstyle{semilarge map conj}=[draw,shape=NWbox,inner sep=2pt,minimum height=6mm,fill=white,minimum width=9.5mm]
\tikzstyle{large map}=[draw,shape=NEbox,inner sep=2pt,minimum height=6mm,fill=white,minimum width=12mm]
\tikzstyle{large map conj}=[draw,shape=NWbox,inner sep=2pt,minimum height=6mm,fill=white,minimum width=12mm]
\tikzstyle{very large map}=[draw,shape=NEbox,inner sep=2pt,minimum height=6mm,fill=white,minimum width=17mm]

\tikzstyle{medium dmap}=[draw,doubled,shape=NEbox,inner sep=2pt,minimum height=6mm,fill=white,minimum width=7mm]
\tikzstyle{medium dmap dag}=[draw,doubled,shape=SEbox,inner sep=2pt,minimum height=6mm,fill=white,minimum width=7mm]
\tikzstyle{medium dmap adj}=[draw,doubled,shape=SEbox,inner sep=2pt,minimum height=6mm,fill=white,minimum width=7mm]
\tikzstyle{medium dmap trans}=[draw,doubled,shape=SWbox,inner sep=2pt,minimum height=6mm,fill=white,minimum width=7mm]
\tikzstyle{medium dmap conj}=[draw,doubled,shape=NWbox,inner sep=2pt,minimum height=6mm,fill=white,minimum width=7mm]
\tikzstyle{semilarge dmap}=[draw,doubled,shape=NEbox,inner sep=2pt,minimum height=6mm,fill=white,minimum width=9.5mm]
\tikzstyle{semilarge dmap trans}=[draw,doubled,shape=SWbox,inner sep=2pt,minimum height=6mm,fill=white,minimum width=9.5mm]
\tikzstyle{semilarge dmap adj}=[draw,doubled,shape=SEbox,inner sep=2pt,minimum height=6mm,fill=white,minimum width=9.5mm]
\tikzstyle{semilarge dmap dag}=[draw,doubled,shape=SEbox,inner sep=2pt,minimum height=6mm,fill=white,minimum width=9.5mm]
\tikzstyle{semilarge dmap conj}=[draw,doubled,shape=NWbox,inner sep=2pt,minimum height=6mm,fill=white,minimum width=9.5mm]
\tikzstyle{large dmap}=[draw,doubled,shape=NEbox,inner sep=2pt,minimum height=6mm,fill=white,minimum width=12mm]
\tikzstyle{large dmap conj}=[draw,doubled,shape=NWbox,inner sep=2pt,minimum height=6mm,fill=white,minimum width=12mm]
\tikzstyle{large dmap trans}=[draw,doubled,shape=SWbox,inner sep=2pt,minimum height=6mm,fill=white,minimum width=12mm]
\tikzstyle{large dmap adj}=[draw,doubled,shape=SEbox,inner sep=2pt,minimum height=6mm,fill=white,minimum width=12mm]
\tikzstyle{large dmap dag}=[draw,doubled,shape=SEbox,inner sep=2pt,minimum height=6mm,fill=white,minimum width=12mm]
\tikzstyle{very large dmap}=[draw,doubled,shape=NEbox,inner sep=2pt,minimum height=6mm,fill=white,minimum width=19.5mm]

\tikzstyle{muxbox}=[draw,shape=rectangle,minimum height=3mm,minimum width=3mm,fill=white]
\tikzstyle{dmuxbox}=[muxbox,doubled]

\tikzstyle{box}=[draw,shape=rectangle,inner sep=2pt,minimum height=6mm,minimum width=6mm,fill=white]
\tikzstyle{dbox}=[draw,doubled,shape=rectangle,inner sep=2pt,minimum height=6mm,minimum width=6mm,fill=white]
\tikzstyle{dmap}=[draw,doubled,shape=NEbox,inner sep=2pt,minimum height=6mm,fill=white]
\tikzstyle{dmapdag}=[draw,doubled,shape=SEbox,inner sep=2pt,minimum height=6mm,fill=white]
\tikzstyle{dmapadj}=[draw,doubled,shape=SEbox,inner sep=2pt,minimum height=6mm,fill=white]
\tikzstyle{dmaptrans}=[draw,doubled,shape=SWbox,inner sep=2pt,minimum height=6mm,fill=white]
\tikzstyle{dmapconj}=[draw,doubled,shape=NWbox,inner sep=2pt,minimum height=6mm,fill=white]

\tikzstyle{ddmap}=[draw,doubled,dashed,shape=NEbox,inner sep=2pt,minimum height=6mm,fill=white]
\tikzstyle{ddmapdag}=[draw,doubled,dashed,shape=SEbox,inner sep=2pt,minimum height=6mm,fill=white]
\tikzstyle{ddmapadj}=[draw,doubled,dashed,shape=SEbox,inner sep=2pt,minimum height=6mm,fill=white]
\tikzstyle{ddmaptrans}=[draw,doubled,dashed,shape=SWbox,inner sep=2pt,minimum height=6mm,fill=white]
\tikzstyle{ddmapconj}=[draw,doubled,dashed,shape=NWbox,inner sep=2pt,minimum height=6mm,fill=white]

\boxshape{sNEbox}{0pt}{3pt}
\boxshape{sSEbox}{0pt}{-3pt}
\boxshape{sNWbox}{3pt}{0pt}
\boxshape{sSWbox}{-3pt}{0pt}
\tikzstyle{smap}=[draw,shape=sNEbox,fill=white]
\tikzstyle{smapdag}=[draw,shape=sSEbox,fill=white]
\tikzstyle{smapadj}=[draw,shape=sSEbox,fill=white]
\tikzstyle{smaptrans}=[draw,shape=sSWbox,fill=white]
\tikzstyle{smapconj}=[draw,shape=sNWbox,fill=white]

\tikzstyle{dsmap}=[draw,dashed,shape=sNEbox,fill=white]
\tikzstyle{dsmapdag}=[draw,dashed,shape=sSEbox,fill=white]
\tikzstyle{dsmaptrans}=[draw,dashed,shape=sSWbox,fill=white]
\tikzstyle{dsmapconj}=[draw,dashed,shape=sNWbox,fill=white]

\boxshape{mNEbox}{0pt}{10pt}
\boxshape{mSEbox}{0pt}{-10pt}
\boxshape{mNWbox}{10pt}{0pt}
\boxshape{mSWbox}{-10pt}{0pt}
\tikzstyle{mmap}=[draw,shape=mNEbox]
\tikzstyle{mmapdag}=[draw,shape=mSEbox]
\tikzstyle{mmaptrans}=[draw,shape=mSWbox]
\tikzstyle{mmapconj}=[draw,shape=mNWbox]

\tikzstyle{mmapgray}=[draw,fill=gray!40!white,shape=mNEbox]
\tikzstyle{smapgray}=[draw,fill=gray!40!white,shape=sNEbox]

\makeatletter

\pgfdeclareshape{cornerpoint}{
\inheritsavedanchors[from=rectangle] % this is nearly a rectangle
\inheritanchorborder[from=rectangle]
\inheritanchor[from=rectangle]{center}
\inheritanchor[from=rectangle]{north}
\inheritanchor[from=rectangle]{south}
\inheritanchor[from=rectangle]{west}
\inheritanchor[from=rectangle]{east}
\backgroundpath{% this is new
\southwest \pgf@xa=\pgf@x \pgf@ya=\pgf@y
\northeast \pgf@xb=\pgf@x \pgf@yb=\pgf@y

\pgfmathsetmacro{\pgf@shorten@left}{\pgfkeysvalueof{/tikz/shorten left}}
\pgfmathsetmacro{\pgf@shorten@right}{\pgfkeysvalueof{/tikz/shorten right}}

\pgfpathmoveto{\pgfpoint{0.5 * (\pgf@xa + \pgf@xb)}{\pgf@ya - 5pt}}
\pgfpathlineto{\pgfpoint{\pgf@xa - 8pt + \pgf@shorten@left}{\pgf@yb - 1.5 * \pgf@shorten@left}}
\pgfpathlineto{\pgfpoint{\pgf@xa - 8pt + \pgf@shorten@left}{\pgf@yb}}
\pgfpathlineto{\pgfpoint{\pgf@xb + 8pt - \pgf@shorten@right}{\pgf@yb}}
\pgfpathlineto{\pgfpoint{\pgf@xb + 8pt - \pgf@shorten@right}{\pgf@yb - 1.5 * \pgf@shorten@right}}
\pgfpathclose
}
}

\pgfdeclareshape{cornercopoint}{
\inheritsavedanchors[from=rectangle] % this is nearly a rectangle
\inheritanchorborder[from=rectangle]
\inheritanchor[from=rectangle]{center}
\inheritanchor[from=rectangle]{north}
\inheritanchor[from=rectangle]{south}
\inheritanchor[from=rectangle]{west}
\inheritanchor[from=rectangle]{east}
\backgroundpath{% this is new
\southwest \pgf@xa=\pgf@x \pgf@ya=\pgf@y
\northeast \pgf@xb=\pgf@x \pgf@yb=\pgf@y

\pgfmathsetmacro{\pgf@shorten@left}{\pgfkeysvalueof{/tikz/shorten left}}
\pgfmathsetmacro{\pgf@shorten@right}{\pgfkeysvalueof{/tikz/shorten right}}

\pgfpathmoveto{\pgfpoint{0.5 * (\pgf@xa + \pgf@xb)}{\pgf@yb + 5pt}}
\pgfpathlineto{\pgfpoint{\pgf@xa - 8pt + \pgf@shorten@left}{\pgf@ya + 1.5 * \pgf@shorten@left}}
\pgfpathlineto{\pgfpoint{\pgf@xa - 8pt + \pgf@shorten@left}{\pgf@ya}}
\pgfpathlineto{\pgfpoint{\pgf@xb + 8pt - \pgf@shorten@right}{\pgf@ya}}
\pgfpathlineto{\pgfpoint{\pgf@xb + 8pt - \pgf@shorten@right}{\pgf@ya + 1.5 * \pgf@shorten@right}}
\pgfpathclose
}
}

\makeatother

\pgfkeyssetvalue{/tikz/shorten left}{0pt}
\pgfkeyssetvalue{/tikz/shorten right}{0pt}

\tikzstyle{kpoint common}=[draw,fill=white,inner sep=1pt,minimum height=4mm]
\tikzstyle{kpoint sc}=[shape=cornerpoint,kpoint common]
\tikzstyle{kpoint adjoint sc}=[shape=cornercopoint,kpoint common]
\tikzstyle{kpoint}=[shape=cornerpoint,shorten left=5pt,kpoint common]
\tikzstyle{kpoint adjoint}=[shape=cornercopoint,shorten left=5pt,kpoint common]
\tikzstyle{kpoint conjugate}=[shape=cornerpoint,shorten right=5pt,kpoint common]
\tikzstyle{kpoint transpose}=[shape=cornercopoint,shorten right=5pt,kpoint common]
\tikzstyle{kpoint symm}=[shape=cornerpoint,shorten left=5pt,shorten right=5pt,kpoint common]

\tikzstyle{wide kpoint sc}=[shape=cornerpoint,kpoint common, minimum width=1 cm]
\tikzstyle{wide kpointdag sc}=[shape=cornercopoint,kpoint common, minimum width=1 cm]

\tikzstyle{black kpoint}=[shape=cornerpoint,shorten left=5pt,kpoint common,fill=black,font=\color{white}]

\tikzstyle{black kpoint sm}=[shape=cornerpoint,shorten left=5pt,kpoint common,fill=black,font=\color{white},scale=0.75]

\tikzstyle{black kpoint adjoint}=[shape=cornercopoint,shorten left=5pt,kpoint common,fill=black,font=\color{white}]
\tikzstyle{black kpointadj}=[shape=cornercopoint,shorten left=5pt,kpoint common,fill=black,font=\color{white}]

\tikzstyle{black kpointadj sm}=[shape=cornercopoint,shorten left=5pt,kpoint common,fill=black,font=\color{white},scale=0.75]

\tikzstyle{black dkpoint}=[shape=cornerpoint,shorten left=5pt,kpoint common,fill=black, doubled,font=\color{white}]
\tikzstyle{black dkpoint adjoint}=[shape=cornercopoint,shorten left=5pt,kpoint common,fill=black, doubled,font=\color{white}]
\tikzstyle{black dkpointadj}=[shape=cornercopoint,shorten left=5pt,kpoint common,fill=black, doubled,font=\color{white}]

\tikzstyle{black dkpoint sm}=[shape=cornerpoint,shorten left=5pt,kpoint common,fill=black, doubled,font=\color{white},scale=0.75]
\tikzstyle{black dkpointadj sm}=[shape=cornercopoint,shorten left=5pt,kpoint common,fill=black, doubled,font=\color{white},scale=0.75]

\tikzstyle{kpointdag}=[kpoint adjoint]
\tikzstyle{kpointadj}=[kpoint adjoint]
\tikzstyle{kpointconj}=[kpoint conjugate]
\tikzstyle{kpointtrans}=[kpoint transpose]

\tikzstyle{big kpoint}=[kpoint, minimum width=1.2 cm, minimum height=8mm, inner sep=4pt, text depth=3mm]

\tikzstyle{wide kpoint}=[kpoint, minimum width=1 cm, inner sep=2pt]%, text depth=-0.7 mm]
\tikzstyle{wide kpointdag}=[kpointdag, minimum width=1 cm, inner sep=2pt]%, text depth=0.7 mm]
\tikzstyle{wide kpointconj}=[kpointconj, minimum width=1 cm, inner sep=2pt]%, text depth=-0.7 mm]
\tikzstyle{wide kpointtrans}=[kpointtrans, minimum width=1 cm, inner sep=2pt]%, text depth=0.7 mm]

\tikzstyle{wider kpoint}=[kpoint, minimum width=1.25 cm, inner sep=2pt]%, text depth=-0.7 mm]
\tikzstyle{wider kpointdag}=[kpointdag, minimum width=1.25 cm, inner sep=2pt]%, text depth=0.7 mm]
\tikzstyle{wider kpointconj}=[kpointconj, minimum width=1.25 cm, inner sep=2pt]%, text depth=-0.7 mm]
\tikzstyle{wider kpointtrans}=[kpointtrans, minimum width=1.25 cm, inner sep=2pt]%, text depth=0.7 mm]

\tikzstyle{gray kpoint}=[kpoint,fill=gray!50!white]
\tikzstyle{gray kpointdag}=[kpointdag,fill=gray!50!white]
\tikzstyle{gray kpointadj}=[kpointadj,fill=gray!50!white]
\tikzstyle{gray kpointconj}=[kpointconj,fill=gray!50!white]
\tikzstyle{gray kpointtrans}=[kpointtrans,fill=gray!50!white]

\tikzstyle{gray dkpoint}=[kpoint,fill=gray!50!white,doubled]
\tikzstyle{gray dkpointdag}=[kpointdag,fill=gray!50!white,doubled]
\tikzstyle{gray dkpointadj}=[kpointadj,fill=gray!50!white,doubled]
\tikzstyle{gray dkpointconj}=[kpointconj,fill=gray!50!white,doubled]
\tikzstyle{gray dkpointtrans}=[kpointtrans,fill=gray!50!white,doubled]

\tikzstyle{white label}=[draw,fill=white,rectangle,inner sep=0.7 mm]
\tikzstyle{gray label}=[draw,fill=gray!50!white,rectangle,inner sep=0.7 mm]
\tikzstyle{black label}=[draw,fill=black,rectangle,inner sep=0.7 mm]

\tikzstyle{dkpoint}=[kpoint,doubled]
\tikzstyle{wide dkpoint}=[wide kpoint,doubled]
\tikzstyle{dkpointdag}=[kpoint adjoint,doubled]
\tikzstyle{wide dkpointdag}=[wide kpointdag,doubled]
\tikzstyle{dkcopoint}=[kpoint adjoint,doubled]
\tikzstyle{dkpointadj}=[kpoint adjoint,doubled]
\tikzstyle{dkpointconj}=[kpoint conjugate,doubled]
\tikzstyle{dkpointtrans}=[kpoint transpose,doubled]

\tikzstyle{kscalar}=[kpoint common, shape=EBox, inner xsep=-1pt, inner ysep=3pt,font=\small]
\tikzstyle{kscalarconj}=[kpoint common, shape=WBox, inner xsep=-1pt, inner ysep=3pt,font=\small]

\tikzstyle{spekpoint}=[kpoint sc,minimum height=5mm,inner sep=3pt]
\tikzstyle{spekcopoint}=[kpoint adjoint sc,minimum height=5mm,inner sep=3pt]

\tikzstyle{dspekpoint}=[spekpoint,doubled]
\tikzstyle{dspekcopoint}=[spekcopoint,doubled]

 \tikzstyle{upground}=[circuit ee IEC,thick,ground,rotate=90,scale=2.5]
 \tikzstyle{downground}=[circuit ee IEC,thick,ground,rotate=-90,scale=2.5]
 \tikzstyle{bigground}=[regular polygon,regular polygon sides=3,draw=gray,scale=0.50,inner sep=-0.5pt,minimum width=10mm,fill=gray]
\tikzstyle{arrs}=[-latex,font=\small,auto]
\tikzstyle{arrow plain}=[arrs]
\tikzstyle{arrow dashed}=[dashed,arrs]
\tikzstyle{arrow bold}=[very thick,arrs]
\tikzstyle{arrow hide}=[draw=white!0,-]
\tikzstyle{arrow reverse}=[latex-]
\tikzstyle{cdnode}=[]

\let\olddagger\dagger
\renewcommand{\dagger}{\ensuremath{\olddagger}\xspace}

\newtheorem{innercustomthm}{Postulate}
\newenvironment{postulate}[1]
  {\renewcommand\theinnercustomthm{#1}\innercustomthm}
  {\endinnercustomthm}
\theoremstyle{definition}
\newtheorem{theorem}{Theorem}[section]

\newtheorem{lemma}[theorem]{Lemma}
\newtheorem{proposition}[theorem]{Proposition}

\newtheorem{definition}[theorem]{Definition}

\newtheorem{example}[theorem]{Example}

\newtheorem{example*}[theorem]{Example*}
\newtheorem{examples*}[theorem]{Examples*}
\newtheorem{remark}[theorem]{Remark}
\newtheorem{remark*}[theorem]{Remark*}

\theoremstyle{plain}

\usepackage{color}
\def\bR{\begin{color}{red}}
\def\bB{\begin{color}{blue}}
\def\bM{\begin{color}{magenta}}
\def\bC{\begin{color}{cyan}}
\def\bW{\begin{color}{white}}
\def\bBl{\begin{color}{black}}
\def\bG{\begin{color}{green}}
\def\bY{\begin{color}{yellow}}
\def\e{\end{color}\xspace}
\newcommand{\bit}{\begin{itemize}}
\newcommand{\eit}{\end{itemize}\par\noindent}
\newcommand{\ben}{\begin{enumerate}}
\newcommand{\een}{\end{enumerate}\par\noindent}
\newcommand{\beq}{\begin{equation}}
\newcommand{\eeq}{\end{equation}\par\noindent}
\newcommand{\beqa}{\begin{eqnarray*}}
\newcommand{\eeqa}{\end{eqnarray*}\par\noindent}
\newcommand{\beqn}{\begin{eqnarray}}
\newcommand{\eeqn}{\end{eqnarray}\par\noindent}
\def\jR{\begin{color}{DarkRed}}
\def\jB{\begin{color}{blue}}
\def\jM{\begin{color}{magenta}}
\def\jC{\begin{color}{cyan}}
\def\jW{\begin{color}{white}}
\def\jBl{\begin{color}{black}}
\def\jG{\begin{color}{green}}
\def\jY{\begin{color}{yellow}}

\def\cR{\begin{color}{Crimson}}
\def\cB{\begin{color}{blue}}
\def\cM{\begin{color}{magenta}}
\def\cC{\begin{color}{cyan}}
\def\cW{\begin{color}{white}}
\def\cBl{\begin{color}{black}}
\def\cG{\begin{color}{green}}
\def\cY{\begin{color}{yellow}}
\usepackage{authblk}
\usepackage[numbers,sort&compress]{natbib}
\begin{document}
\title{Reconstructing quantum theory from diagrammatic postulates}
\date{2021-04-17}
\author{John H. Selby}
\affiliation{ICTQT, University of Gda\'nsk, Wita Stwosza 63, 80-308 Gda\'nsk, Poland}
\email{john.h.selby@gmail.com}
\author{Carlo Maria Scandolo}
\email{carlomaria.scandolo@ucalgary.ca}
\affiliation{Department of Mathematics \& Statistics, University of Calgary, Canada}
\affiliation{Institute for Quantum Science and Technology, University of Calgary, Canada}
\author{Bob Coecke}
\email{bob.coecke@cambridgequantum.com}
\affiliation{Cambridge Quantum Computing Ltd}
\begin{abstract}
A reconstruction of quantum theory refers to both a mathematical and a conceptual paradigm that allows one to derive the usual formulation of quantum theory from a set of primitive assumptions. The motivation for doing so is a discomfort with the usual formulation of quantum theory, a discomfort that started with its originator John von Neumann.

We present a reconstruction of finite-dimensional quantum theory where all of the postulates are stated in diagrammatic terms, making them intuitive.  Equivalently, they are stated  in category-theoretic terms, making them mathematically appealing. Again equivalently, they are stated  in process-theoretic terms, establishing  that the conceptual backbone of quantum theory concerns the manner in which systems and processes compose.

Aside from the diagrammatic form, the key novel aspect of this reconstruction is the introduction of a new postulate, symmetric purification. Unlike the ordinary purification postulate, symmetric purification applies equally well to classical theory as well as quantum theory. Therefore we first reconstruct the full process theoretic description of quantum theory, consisting of composite classical-quantum systems and their interactions, before restricting ourselves to just the `fully quantum' systems as the final step.

We propose two novel alternative manners of doing so, `no-leaking' (roughly that information gain causes disturbance) and `purity of cups' (roughly the existence of entangled states). Interestingly, these turn out to be equivalent in any process theory with cups \& caps. Additionally, we show how the standard purification postulate can  be seen as an immediate consequence of the symmetric purification postulate and purity of cups.

Other tangential results concern the specific frameworks of generalised probabilistic theories (GPTs) and process theories (a.k.a.\ CQM). Firstly, we provide a diagrammatic presentation of GPTs, which, henceforth, can be subsumed under process theories.
Secondly, we argue that the `sharp dagger' is indeed the right choice of a dagger structure as this sharpness is vital to the reconstruction.
\end{abstract}

\newpage
\tableofcontents
\newpage

\maketitle

\section{Introduction}
Reconstructions of quantum theory aim to reproduce the standard quantum formalism from assumptions of some desired flavour, which in our case means \emph{diagrammatic}. The idea of reconstructing quantum theory is not at all new; indeed, the first to contribute to the endeavour was John von Neumann.  Merely three years after the publication of his book \cite{vN},  which cemented the mathematical formalism of quantum theory,  he made it clear in a letter to the mathematician Garrett Birkhoff that he was  no longer satisfied with the Hilbert space  formalism \cite{Redei1}.  However, rather than just aiming to reconstruct this formalism, his hope was in fact to find a different formalism that  might also produce new physics.  The actual reconstruction program, building further on von Neumann's work,  was outlined by George Mackey \cite{Mackey}, and mostly completed by Constantin Piron \cite{Piron64} (an academic ancestor of two of the authors) within the arena of so-called property lattices, and by G\"unther Ludwig \cite{Ludwig} within the arena of generalised probabilistic theories (GPTs).  The main motivation for this was a dissatisfaction with accepting the abstract Hilbert space as a given. Instead, the aim was to start off with conceptually justified structures and axioms from which the Hilbert space could be derived.

A more modern perspective is that   reconstructions (by
helping one see  the physical principles  that underlie the Hilbert space structure)
allow one both to derive more results of quantum theory, and to understand how it  can, and should, be modified in order to reconcile it with general relativity.

The flavour of the assumptions that go into a reconstruction has varied substantially, ranging from counterfactual ontology to instrumentalism. However, broadly, there are two kinds of approaches.

The first wave of reconstructions \cite{vN, Mackey, Piron64, Ludwig,alfsen2012geometry,alfsen2003state}, surveyed in \cite{CMW}, took place in the previous century, and the assumptions used therein were taken to be axioms in the mathematical sense.  Nothing was left implicit, and, as a result, the mathematical sophistication of these reconstructions was substantial. For example, Piron's Theorem \cite{Piron64} was actually only finalised in 1995 by Sol\`er \cite{Soler}, while arguably it already took off from von Neumann's book \cite{vN}  which initiated the study of quantum logical principles in 1932. In this first wave, different reconstructions corresponded to different conceptually motivated mathematical structures, e.g.\ lattices \cite{Piron64} or generalised probability spaces \cite{Ludwig}.

In this millennium, a new wave of quantum reconstructions emerged under Lucien Hardy's impetus \cite{HardyAxiom, dakic2009quantum, Masanes, Chiri2, HardyBig}.   In these, there was a broad shift away from the mathematical axiomatics of the first wave.
Instead, they took inspiration from Einstein's derivation   of relativity theory from two `principles',
i.e.\ the constancy of the speed of light and the invariance of laws for different observers.
This shift means that many pieces of structure are often sneaking into the reconstructions ---like in Einstein's derivation of relativity theory--- for example,  dimensions are typically restricted to finite ones, and the structure of real line is put in by hand, usually in the form of probabilities, while in \cite{Piron64} it was constructed. The resulting reconstruction axioms and their presentation do not have the feel of a mathematical domain but are more akin to Einstein's principles of relativity.
Consequently, to distinguish the assumptions used in these approaches from the mathematical axioms of the first wave, the term `postulates' was adopted.
Now,  the conceptual grounding of the postulates is of key importance. These are typically cast
within some  `interpretative school'; for example principles only make reference to measurement \cite{HardyAxiom},  or are information-theoretic \cite{clifton2003characterizing, Chiri2}, or even make reference only to properties of single systems \cite{barnum2014higher}. 
 To simplify the problem, and as a potential stepping stone to a full reconstruction, only finite-dimensional Hilbert spaces  were considered. This stance was not only for mathematical convenience, but also justified by the many new results ---within the context of quantum foundations, information, and computation--- demonstrating that the essential quantum phenomena are all apparent within the finite-dimensional setting.

One  common theme in
this new wave of reconstructions is  a consideration of the role of composite systems. For example, principles such as tomographic locality \cite{Barrett} concerned pairs of systems, rather than  just a single one. This is in sharp contrast with the first wave of reconstructions, where the focus was on single systems under observation. Indeed, for this first wave, the Achilles heel was the description of composite systems at a general axiomatic level: while these approaches were able to recover Hilbert space, they all failed to reproduce the tensor product as part of the formalism. This is probably why ---with the emergence of quantum computation and information, where composition is of course vital--- these approaches have more or less vanished. But, despite this  significant advantage  over the first wave, the postulate-based approaches lacks the mathematical clarity of the earlier attempts, and in many cases the proofs are not so insightful.

Learning from the failures of the earlier axiomatic reconstructions, a rigorous focus on composition of quantum systems and processes has been the subject of \em categorical quantum mechanics \em (CQM)  \cite{AC1, CKbook} for some 15 years now.  Its initial goal was to recast quantum theory in high-level terms, making reasoning and computing more intuitive. While the first reconstructions of the second wave focused on states of physical systems and their geometry, the categorical approach changed the paradigm completely, from states to \emph{processes}  \cite{JTF}. Consequently, one now refers to the theories that CQM is concerned with as \emph{process theories}.
The particular nature of the categorical structures that make up process theories have the great upshot that they admit a full and faithful diagrammatic representation \cite{Kindergarten, SelingerCPM, ContPhys} (a.k.a.\ \em quantum picturalism\em). This has provided an intuitive picture of many aspects of quantum information, computation, and a wide variety of other fields \cite{de2017zx,horsman2011quantum,duncan2012graphical,duncan2010rewriting,
kissinger2019tcount,deBeaudrap2020Paulifusion,SZXCalculus,deBeaudrap2020Techniques,
Backens2020extraction,deBeaudrap2020Tcount,
tull_kleiner_2020,schmid2020structure,schmid2020unscrambling,miguelsignorelli2020compositional,coecke2020foundations,
east2020akltstates,vandewetering2020zxcalculus,coecke2021kindergarden} which lends itself well to computational automation \cite{kissinger2015quantomatic,garvie2017verifying,chancellor2016coherent,duncan2013verifying}.

Meanwhile, borrowing from CQM,  many reconstructionists of the second wave adopted `diagrams of processes'  as their starting point \cite{HardyPicturalism, Chiri1,Chiri2,HardyBig}, hence embracing composition of processes as one of the core ingredients of quantum theory.  More recently, a third wave of reconstruction attempts took off \cite{barnum2013ensemble,barnum2014higher,Royal-road}, which can be seen as resurrecting the mathematical spirit of the first wave, while still embracing the principled underpinning that guides the second wave. Recently, there have been various other reconstructions of quantum theory from a variety of perspectives, for example \cite{goyal2008information,hohn2017toolbox,hohn2017quantum,budiyono2017quantum,tullReconstruction,van2018reconstruction,nakahira2019derivation}.

\subsection{Main result}

This paper provides a reconstruction that is conceptually grounded whilst still being based on crisp mathematical axioms. This is achieved by exploiting the correspondence:
\begin{center}
diagrams $\simeq$ category theory $\simeq$ process theory
\end{center}
Our postulates are now  diagrammatic, i.e.\ category-theoretic, i.e.\ process-theoretic, providing them with an intuitive, elegant, as well as principled underpinning. In short, what we prove is:
\begin{center}
 classical interface + postulates about how processes compose $\Longrightarrow$ quantum processes
\end{center}
More explicitly, the complete list of postulates (introduced formally in Sec.~\ref{sec:Postulates}) that we use to reconstruct quantum theory are as follows:
\ben
    \item the theory is a process theory  (Def.~\ref{post:ProcessTheory}),
    \item with a finite local\footnote{ In the sense of local tomography, a commonly used postulate in reconstructing quantum theory.} classical interface (Defs.~\ref{def:ClassicalInterface} \& \ref{def:LocalCI}),
    \item cups \& caps (Def.~\ref{def:compactStructure}),
    \item a sharp dagger (Def.~\ref{post:SharpDagger}),
    \item and in which all processes admit  essentially unique  symmetric purifications (Def.~\ref{def:SymmetricPurification}).
\een

To understand these postulates and the properties derived from them, one must be familiar with the process theory framework and the associated diagrammatic language. In Sec.~\ref{sec:ProcessTheories} we therefore provide an introduction to this framework and the concepts necessary to read the rest of the paper.  For a more detailed introduction see \cite{CKbook}.

Informally, requiring that the theory be a process theory states that nature is fundamentally grounded on processes, and how they compose. The classical interface describes how we interact with, control, and learn about the theory. Cups and caps represent a fundamental symmetry that ensures that the theory has maximal correlations. The sharp dagger represent another fundamental symmetry, namely time reversal. Informally, the time reverse of a process can be thought of as ``playing the video of that process in reverse''. In particular, for a state this corresponds to a `test' for that state:  we view states as a mapping from the trivial system to a non-trivial system, the time reverse is therefore a map from the non-trivial system to the trivial system (i.e. a measurement outcome), this should single out the state within the state space.    Finally, the existence of symmetric purifications ensures that any lack of purity can always be traced to lack of access to the past or the future of certain systems.

The proof of this result is remarkably simple in contrast to many other reconstructions, indeed, the entire reconstruction can be presented in a simple flowchart (see Fig.~\ref{flowchart}). This  is largely owing to two things, firstly, to the use of diagrammatic proofs, and, secondly, to the use of a standard result, the Koecher-Vinberg theorem \cite{koecher1958geodattischen,vinberg1960homogeneous}.
The latter was borrowed from the works of Barnum, Wilce et al., e.g.\ \cite{barnum2016composites,barnum2014local,Royal-road,barnum2013symmetry,barnum2013ensemble}.

There have been several other reconstructions of quantum theory, from various different perspectives, many results of which have been adapted for this work. It is therefore worth highlighting the key novel features of this particular reconstruction that distinguish it from the others.

\bit
\item Firstly, the postulates we impose are  diagrammatic, moreover, they do not pick out states as being special,  applying instead equally well to all processes. As such, they fit with the spirit of CQM and the process-theoretic understanding of the world, that is, as being about processes and composition. Indeed, this is the first reconstruction where the postulates are entirely diagrammatic. In the past some reconstructions have utilised a diagrammatic formalism, but the basic framework has always involved some non-diagrammatic component -- for example in how the probabilistic structure is introduced. We show that this is unnecessary, and that the postulates and the basic framework can all be stated within a diagrammatic formalism. This use of diagrams is much more than just a stylistic choice, it forms the conceptual underpinning of the work, and it is the natural language in which one should express these postulates. For example, the aforementioned symmetric purification postulate is relatively simple to state and understand diagrammatically, whereas it is,  to the authors at least, much more opaque when translated into algebraic or categorical terms.

\item Secondly, we reconstruct the full process-theoretic description of quantum theory -- that is, the C*-algebraic formalism of quantum theory, including measurements, hybrid quantum-classical systems, superselected quantum systems, and so on. We argue that this is how quantum theory \emph{should} be described, and, moreover, this  perspective allows us to clearly see what it is that separates the fully  quantum systems from the others---specifically, the triviality of leaks, or equivalently, purity of the cups \& caps. Typically, reconstruction efforts have only reconstructed the fully-quantum subtheory, and have neglected measurements and classical processes as an integral part of the reconstructed structure. For example, the majority of the other reconstructions take \emph{transitivity} as a postulate\footnote{Or they have it as a direct corollary of a stronger postulate, such as purification or strong symmetry.}. Transitivity is the property that there is a reversible transformation between any pair of pure states. This does not hold for general C*-algebras, so most other reconstructions immediately rule out the full C*-algebraic theory. This calls into question how natural such postulates were in the first place.

\item Thirdly, the symmetric purification postulate is a new postulate that has not been considered elsewhere. Indeed, as far as the authors are aware, this is a feature of quantum theory which has not been noted elsewhere in the literature. As this applies equally well to both classical and quantum theory, it has the potential to help unify many results that typically rely on a distinct proof for each theory. For instance, many features of quantum cryptography ---such as the impossibility of bit-commitment--- use proofs which ultimately rely on the purification postulate. However, bit-commitment is also impossible in classical theory in which purification fails. Therefore it seems that our symmetric form of purification could be utilized to prove results that immediately hold for both theories.

\item Finally, the reconstruction is relatively simple. In particular, the structure of the reconstruction is clear (see Fig.~\ref{flowchart}), allowing a high-level view of how the different postulates relate to each other and how they are used in each step. This should make it simpler to understand how relaxing or altering any given postulate will lead to new theories. In contrast, in many other reconstructions  (perhaps due to their presentation rather than any intrinsic properties) it is difficult to know precisely which postulates, and which of the assumptions  of the framework, are necessary to obtain each result.
\eit

\subsection{Purification}

The standard purification postulate was first used in \cite{Chiri1} as an operational generalisation of the Stinespring dilation theorem \cite{Stinespring}. It roughly states that any mixed process can be represented as a pure process with an extra output that is discarded. Importantly, this postulate is satisfied by quantum theory, but \emph{not} by classical probability theory, so it can be used to single out quantum theory \cite{Chiri1}.

 Our \emph{symmetric} purification postulate, in contrast, holds for both quantum and classical theory. We therefore reconstruct
 quantum, classical and hybrid systems all together. One can then ask how to single out fully quantum systems. There are many ways to do so, but two of them are of particular interest: the existence of a \emph{pure} `cup' (roughly speaking the existence of entanglement), or, the lack of non-trivial `leaks' (essentially that information gain causes disturbance). Whilst at first glance these appear to be unrelated postulates involving different diagrammatic concepts,  they can actually be shown to be equivalent in any process theory with cups \& caps.

Moreover, it can be shown that the standard purification postulate is implied by the conjunction of the symmetric purification postulate and the existence of a pure cup. Hence our work can be seen as deconstructing the standard purification postulate into two parts ---one which applies to both quantum and classical theory, and the other only to quantum--- therefore refining exactly what is uniquely quantum about purification.

\subsection{Connection to GPTs, OPTs, CQM and CPTs}
While initially GPTs only appealed to single systems, under the impetus of CQM a new hybrid form was proposed (a.k.a.  Operational Probabilistic Theories (OPTs)), where composition supported by a diagrammatic backbone. We demonstrate that, within the context of process theories,  the essential structure of OPTs can be derived from the classical interface postulate.   Hence, OPTs are subsumed under the process-theoretic framework. Independently, a similar result was also obtained in \cite{gogioso2017categorical}, where the framework of Categorical Probabilistic Theories (CPTs) was developed. In that paper, three requirements (Definition 1) were made of a CPT. The first of these is part of the definition of a classical interface, whilst we find that the second two  can be derived from our postulated classical interface. Whilst our postulated classical interface is a mathematically stronger assumption than those made in the definition of a CPT, we find it preferable, as it has a clear physical interpretation.

Cups and caps have been part of the structure of CQM from the start \cite{AC1}, and provide a cup- and a cap-shaped wire for each system: allowing for inputs to be connected to inputs and outputs to outputs. Daggers have also been part of CQM since its very beginning \cite{AC2, SelingerCPM}. Unlike the transpose, which is constructed using cups \& caps, a dagger did not have any other structural requirements besides being compositional.  In order to fully characterise the Hermitian adjoint of quantum theory, a \em sharpness \em constraint was added in \cite{selby2016process}.  We demonstrate that, once we have a classical interface, cups \& caps, and a sharp dagger,  the only additional postulate to be imposed, for a process theory to correspond to the Hilbert space model, is the aforementioned symmetric purification postulate.

\section{Process theoretic concepts}\label{sec:ProcessTheories}

Process theories \cite{CKpaperI, CKbook} are theories that have a particular diagrammatic representation.  A comprehensive introduction can be found in \cite{CKbook}.  In this section we introduce the process-theoretic concepts and tools which will play a role in the reconstruction. Some of these will be postulated in Sec.~\ref{sec:Postulates}, whilst others will be derived from these postulates in Sec.~\ref{sec:Reconstruction}, or simply be used in various proofs throughout.

\begin{definition}[Process theory]\label{def:ProcessTheory}
Process theories consist of a collection of \emph{systems}, denoted by \emph{labelled wires}, and a collection of \emph{processes}, denoted by \emph{labelled boxes}
 with input wires (at the bottom) and output wires  (at the top).
 These processes can be wired together, for example:
\[
\InputIfFileExists{Diagrams/diagram.tikz}{}{\input{./figures/Diagrams/diagram.tikz}}
\]
where the resulting \em diagram \em must also be  a process in the theory; the relevant data for a diagram are:
\bit
\item[i.] the processes that appear in the diagram, and
\item[ii.] how the diagram is \emph{connected}, including the overall ordering of free inputs and outputs;
\eit
the formation of diagrams is constrained by:
\ben
\item connected systems must  have the same type (i.e.\ the same label),
\item outputs are wired up to inputs, and
\item no loops are created.
\een
\end{definition}

\begin{remark} For those who favour an information-based characterisation of quantum theory, one can think of the wires in the diagrams as information flows. For those who favour a more operational interpretation, one can think of the processes as corresponding to a single use of a piece of lab equipment in a single run of an experiment.
\end{remark}

There are three particular types of processes within a theory that are often distinguished: those with no inputs, which are state preparation procedures or \emph{states} for short; those with no outputs, corresponding to the outcome of some destructive measurement, or \emph{effects} for short; and those with neither, known as \emph{scalars}, which typically correspond to probabilities. These are respectively denoted as:
\[%
\begin{tikzpicture}
	\begin{pgfonlayer}{nodelayer}
		\node [style=point] (0) at (0, -0) {$s$};
		\node [style=none] (1) at (0, 0.7500002) {};
	\end{pgfonlayer}
	\begin{pgfonlayer}{edgelayer}
		\draw [qWire](1.center) to (0);
	\end{pgfonlayer}
\end{tikzpicture}
}\qquad,\qquad %
\begin{tikzpicture}
	\begin{pgfonlayer}{nodelayer}
		\node [style=copoint] (0) at (0, 0) {$e$};
		\node [style=none] (1) at (0, -0.75) {};
	\end{pgfonlayer}
	\begin{pgfonlayer}{edgelayer}
		\draw [qWire](1.center) to (0);
	\end{pgfonlayer}
\end{tikzpicture}
}\qquad\text{and}\qquad%
\begin{tikzpicture}
	\begin{pgfonlayer}{nodelayer}
		\node [style=scalar] (0) at (0, -0) {$p$};
	\end{pgfonlayer}
\end{tikzpicture}}\quad.\]
Note that we often drop various labels and the box around scalars when they are clear from context. It is often useful ---particularly in connecting these theories to standard mathematical models--- to introduce two primitive forms of composition: \emph{sequential composition}, symbolically denoted as $g\circ f$, and \emph{parallel composition}, denoted as $f \otimes g$,  which diagrammatically correspond to:
\[%
\InputIfFileExists{Diagrams/sequential.tikz}{}{\input{./figures/Diagrams/sequential.tikz}}\qquad \text{and} \qquad %
\InputIfFileExists{Diagrams/parallel.tikz}{}{\input{./figures/Diagrams/parallel.tikz}} \qquad \text{respectively.}\]
We also
represent the parallel composition of systems as
$A\otimes B$,  hence allowing us to treat parallel wires as a single wire, that is:
\[%
\begin{tikzpicture}
	\begin{pgfonlayer}{nodelayer}
		\node [style=none] (0) at (0, 0.75) {};
		\node [style=none] (1) at (0, -0.75) {};
		\node [style={right label}] (2) at (0, -0.2500001) {$A$};
	\end{pgfonlayer}
	\begin{pgfonlayer}{edgelayer}
		\draw [qWire] (1.center) to (0.center);
	\end{pgfonlayer}
\end{tikzpicture}}%
\begin{tikzpicture}
	\begin{pgfonlayer}{nodelayer}
		\node [style=none] (0) at (0, 0.75) {};
		\node [style=none] (1) at (0, -0.75) {};
		\node [style={right label}] (2) at (0, -0.2500001) {$B$};
	\end{pgfonlayer}
	\begin{pgfonlayer}{edgelayer}
		\draw [qWire] (1.center) to (0.center);
	\end{pgfonlayer}
\end{tikzpicture}}\quad = \quad %
\begin{tikzpicture}
	\begin{pgfonlayer}{nodelayer}
		\node [style=none] (0) at (0, 0.75) {};
		\node [style=none] (1) at (0, -0.75) {};
		\node [style={right label}] (2) at (0, -0.2500001) {$A\otimes B$};
	\end{pgfonlayer}
	\begin{pgfonlayer}{edgelayer}
		\draw [qWire] (1.center) to (0.center);
	\end{pgfonlayer}
\end{tikzpicture}}  \]

 Moreover, to translate into standard symbolic notation,  we denote a process $f$ with input $A$ and output $B$ as $f:A\to B$. To describe processes lacking inputs and/or outputs, we must therefore introduce a fictitious `trivial' system denoted $I$, so that we can, for example, denote the state $s$ of a system $C$ as $s:I\to C$.
To be consistent with our diagrammatic notation, that is, $I$ representing the lack of an input, it must satisfy the equation of $I\otimes A = A = A \otimes I$, i.e.\ appending a trivial system does nothing.

A simple example of a process theory is classical probability theory:
\begin{example}[Classical]\label{Ex:Classical}
Here we restrict ourselves to finite probabilistic models. We can characterise every classical system with a natural number $n$ which we call its dimension. The composite of systems is given by the product of their dimensions, i.e.\ $n\otimes m = nm$, and so the trivial system corresponds to $n=1$. Processes $f:n\to m$ then correspond to $m\times n$ matrices with non-negative matrix elements. Sequential composition is given by matrix multiplication and parallel composition by the standard tensor product.  States therefore correspond to $n\times 1$ matrices, i.e.\ vectors, and effects $1\times m$ matrices, i.e.\ covectors; the sequential composition of a state and effect therefore gives a scalar valued in $\mathds{R}^+$.  Many of the processes in this theory lack a physical interpretation. For instance, the physically meaningful scalars are those in the interval $[0,1]$, that is, those which can be interpreted as probabilities. Nonetheless, the non-physical processes are extremely useful for performing calculations in classical probability theory.  We will return to this issue and characterise the physical processes in the following section.
\end{example}

A more involved example is the process-theoretic description of quantum theory, which includes quantum, classical, composite, and hybrid systems. The not fully quantum systems in this process theory can be interpreted in two ways \cite{coecke2017leaks}: either as the systems that arise from the branching structure of measurements, or as the systems that  are obtained from a generalised decoherence mechanism. In particular, the classical systems are necessary for a process-theoretic description of quantum measurements.

\begin{example}[Quantum]\label{Ex:CStar}
 Here we restrict ourselves to finite-dimensional quantum theory. Systems correspond to finite dimensional C*-algebras, which can be represented as direct sums of complex matrix algebras \cite{bratteli1972inductive}: $\bigoplus_k M\left( \mathds{C}^{n_k}\right)$. The standard tensor product provides composition, hence, the trivial system is given by $M\left( \mathds{C}\right) $. Processes are completely positive  maps of C*-algebras, $f:\bigoplus_k M\left( \mathds{C}^{n_k}\right) \to \bigoplus_l M\left( \mathds{C}^{n_l}\right) $. States therefore correspond to block-diagonal density matrices and effects to block-diagonal POVM elements. Again, scalars correspond to non-negative real numbers.

We can retrieve the fully quantum or fully classical systems by restricting to those of the form $M(\mathds{C}^n)$ or $\bigoplus_{k=1}^m M(\mathds{C})$ respectively, the former system being equivalent to an $n$-level quantum system and the latter to an $m$-dimensional classical system. The processes between systems of these respective types are precisely those that we would expect. For example, states of $M(\mathds{C}^n)$ are just $n$-level density matrices, and states of $\bigoplus_{k=1}^m M(\mathds{C})$ are probability distributions over an $m$-element set.

 Again we view this as the process theory for performing calculations about quantum theory. We will introduce  `physicality' conditions in the following section which tell us which of these processes could be implemented in a lab without post-selection. 
\end{example}

\subsection{Discarding and causal processes}\label{sec:Discard}

A process theory often comes with a discarding effect for each system, which provides a way to `throw away'  or even simply `ignore' systems. We denote this by:
\[%
\begin{tikzpicture}
	\begin{pgfonlayer}{nodelayer}
		\node [style=upground] (0) at (0, 0.25) {};
		\node [style=none] (1) at (0, -0.5) {};
		\node [style=right label] (2) at (0, -0.5) {$A$};
	\end{pgfonlayer}
	\begin{pgfonlayer}{edgelayer}
		\draw [qWire](0) to (1.center);
	\end{pgfonlayer}
\end{tikzpicture}
}\]
where, it should be the case that, discarding two systems independently is the same as discarding the composite system:
\[%
\InputIfFileExists{Diagrams/discardComposition.tikz}{}{\input{./figures/Diagrams/discardComposition.tikz}}.\]
In theories with discarding we can elegantly define causality of processes.

\begin{definition}[Causal processes  and causal subtheory] \label{def:Causal}
A process $f$  is  \emph{causal} \cite{Chiri1,CKbook} if it satisfies:
\beq\label{def:causal}
\InputIfFileExists{Diagrams/causalProcess.tikz}{}{\input{./figures/Diagrams/causalProcess.tikz}}
\eeq
By the \emph{causal subtheory} we then mean the theory of all causal processes.  This is itself a valid process theory as it is easy to show that the causal processes are closed under forming diagrams.
\end{definition}

The connection between Def.~\ref{def:Causal} and standard notions of causality may not be immediately apparent. However,  it can be shown that within the causal subtheory  future measurement choices do not effect current experiments \cite{Chiri1}.  More general, it implies that there is no superluminal signalling in the subtheory \cite{Cnonsig}, and even full compatibility with relativity \cite{kissinger2017equivalence}.

Note that the definition of a causal subtheory automatically implies that the only effects in the subtheory are the discarding effects themselves:
\begin{equation}\label{eq:effectcaus}
\InputIfFileExists{Diagrams/CausalEffect.tikz}{}{\input{./figures/Diagrams/CausalEffect.tikz}}
\end{equation}
It therefore turns out to be beneficial not to work with the causal subtheory, but with the full theory including non-causal processes.  This is standard practice within quantum theory which, besides Dirac  kets, also has \emph{non-causal} Dirac bras.  That is, allowing for non-causal processes gives access to the specific outcomes in a measurement (cf.~``in a measurement we obtain the outcome corresponding to bra $\langle x|$''), or more generally, processes that can only occur probabilistically.
Moreover, just as in the case of kets and bras, the larger theory can admit additional symmetries that allow one to  characterise it more easily. We discuss these symmetries below in Sec.~\ref{sec:cupscaps}.

 We view causal processes as being the  \emph{physically realisable} processes, that is, those that represent the physical evolution of a system independent of what any agent thinks about it. How then should we view measurements, which, as we have just discussed above,  are typically defined by non-causal processes? To view measurements as  causal processes, we must
`bunch' together a collection of non-causal processes  to represent all `branches' of a measurement, or more general non-deterministic process, as a single causal process. In order to do so we  will introduce classical systems ---represented diagrammatically by thin grey lines\footnote{Note that in this paper, whilst a thin  gray wire always indicates a classical system, a thicker solid wire indicates \emph{any} system allowed by the theory, which could be in fact classical.}--- to store the  resultant probability distribution over the possible outcomes: 
\[%
\begin{tikzpicture}
	\begin{pgfonlayer}{nodelayer}
		\node [style=none] (0) at (0, 0.75) {};
		\node [style=none] (1) at (0, -0.75) {};
		\node [style={right label}] (2) at (0, -0.2500001) {$n$};
	\end{pgfonlayer}
	\begin{pgfonlayer}{edgelayer}
		\draw [cWire] (1.center) to (0.center);
	\end{pgfonlayer}
\end{tikzpicture}}\]
where $n\in \mathds{N}$ denotes the number of distinguishable states of the classical system. We can then define measurements within the causal subtheory as follows.
\begin{definition}[Measurements]\label{def:measurement}
Destructive measurements are processes with only a classical output:
\[%
\begin{tikzpicture}
	\begin{pgfonlayer}{nodelayer}
		\node [style={small box}] (0) at (0, -0) {$m$};
		\node [style=none] (1) at (0, 1) {};
		\node [style=none] (2) at (0, -1) {};
	\end{pgfonlayer}
	\begin{pgfonlayer}{edgelayer}
		\draw[style=cWire] (1.center) to (0);
		\draw [qWire](0) to (2.center);
	\end{pgfonlayer}
\end{tikzpicture}
}\]
general measurements are then processes that have both classical and non-classical outputs:
and non-destructive measurements are just processes of the form
\[ %
\begin{tikzpicture}
	\begin{pgfonlayer}{nodelayer}
		\node [style=none] (0) at (0.5, 0.5) {};
		\node [style=none] (1) at (0.5, 1) {};
		\node [style=none] (2) at (-0.25, 1) {};
		\node [style=none] (3) at (-0.25, 0.5) {};
		\node [style=none] (4) at (0, -0.5) {};
		\node [style=none] (5) at (0, -1) {};
		\node [style=none] (6) at (0, -0) {$M$};
		\node [style=none] (7) at (-0.75, 0.5) {};
		\node [style=none] (8) at (-0.75, -0.5) {};
		\node [style=none] (9) at (0.75, -0.5) {};
		\node [style=none] (10) at (1, 0.5) {};
	\end{pgfonlayer}
	\begin{pgfonlayer}{edgelayer}
		\draw [style=cWire] (1.center) to (0.center);
		\draw [qWire](2.center) to (3.center);
		\draw [qWire](4.center) to (5.center);
		\draw (7.center) to (10.center);
		\draw (10.center) to (9.center);
		\draw (9.center) to (8.center);
		\draw (8.center) to (7.center);
	\end{pgfonlayer}
\end{tikzpicture}
}\]
\end{definition}
There is no need to add any more here, once we have classical systems as part of the theory measurements are just a special kind of process rather than a fundamentally different type of object. Moreover, there are other sorts of interesting processes that we can consider once we have such classical systems, for example:
\begin{definition}[Classical control]\label{def:classicalcontrolprocs}
Classical control over state preparation is achieved via processes of the form
\[%
\begin{tikzpicture}
	\begin{pgfonlayer}{nodelayer}
		\node [style={small box}] (0) at (0, 0) {$S$};
		\node [style=none] (1) at (0, -1) {};
		\node [style=none] (2) at (0, 1) {};
	\end{pgfonlayer}
	\begin{pgfonlayer}{edgelayer}
		\draw [style=cWire] (1.center) to (0);
		\draw [qWire](0) to (2.center);
	\end{pgfonlayer}
\end{tikzpicture}
}\]
where the choice of classical state input into $S$ will determine the quantum state that is prepared. More generally we can classically control processes using processes of the form:
\[%
\InputIfFileExists{Diagrams/ccProc.tikz}{}{\input{./figures/Diagrams/ccProc.tikz}}\]
where the choice of classical state input into $P$ will determine which quantum process occurs.
\end{definition}

Theseinteractions between classical systems and the general systems of our theory form the basis of the \emph{classical interface} for a theory that we introduce in section \ref{Sec:ClassicalInterface}.
\begin{example}[Classical]
We define the discarding map to be the covector with every matrix element $1$. The causal subtheory  is the restriction to stochastic matrices, as they will satisfy Eq.~\eqref{def:causal}. In particular, this condition implies that states correspond to probability distributions over an $n$ element set, and that processes are maps from probability distributions over $n$ elements to those over $m$ elements. In particular we denote the states and effects that are everywhere $0$ except for a $1$ at position $i$ as:
\[%
\begin{tikzpicture}
	\begin{pgfonlayer}{nodelayer}
		\node [style=point] (0) at (0, -0) {$i$};
		\node [style=none] (1) at (0, 1) {};
	\end{pgfonlayer}
	\begin{pgfonlayer}{edgelayer}
		\draw [style=cWire] (1.center) to (0);
	\end{pgfonlayer}
\end{tikzpicture}} \quad \text{and} \quad %
\begin{tikzpicture}
	\begin{pgfonlayer}{nodelayer}
		\node [style=copoint] (0) at (0, 0) {$i$};
		\node [style=none] (1) at (0, -1) {};
	\end{pgfonlayer}
	\begin{pgfonlayer}{edgelayer}
		\draw [style=cWire] (1.center) to (0);
	\end{pgfonlayer}
\end{tikzpicture}} \quad \quad\text{respectively.}\]
\end{example}
\begin{example}[Quantum]
We define the discarding maps in this theory to be the (partial) trace, hence causal states correspond to block-diagonal trace-$1$, density matrices with blocks of dimension $\left\{n_k\right\}$,  that is, trace-$1$ elements of $\bigoplus_k M\left( \mathds{C}^{n_k}\right)$. General processes are causal if and only if they are trace-preserving.

This theory contains both quantum and classical theory  as subtheories. Moreover it contains processes mapping between these sectors, such as $f:M\left( \mathds{C}^n\right) \to \bigoplus_{k=1}^m M(\mathds{C})$ mapping from a quantum to a classical system. Such processes can be shown to correspond to POVMs in the standard presentation of quantum theory, where the causality constraint for these measurements is equivalent to the constraint that POVM elements sum to the identity. General processes with both quantum and classical inputs and outputs have similarly clear interpretations: the classical input can be seen as a control system determining which process to implement, the implemented processes can involve some (potentially non-destructive) measurement, the result of which is then encoded in the classical output.
\end{example}

\subsection{Diagrammatic sums}\label{App:Sums}

Let us first, for simplicity, introduce a shorthand notation for certain  `vise'-shaped diagrams that will be useful  for this section and throughout the paper.
\begin{definition}[Vise]
A vise, $\chi$, is defined by a pair of processes and a shared system $(x_\chi, y_\chi, E_\chi)$:
 \[%
\InputIfFileExists{Diagrams/circuitFragment.tikz}{}{\input{./figures/Diagrams/circuitFragment.tikz}}\]
\end{definition}
On the one hand we can view these vises as being simply a process with composite input $B\otimes C$ and composite output $A\otimes D$:
\[%
\InputIfFileExists{Diagrams/vise1.tikz}{}{\input{./figures/Diagrams/vise1.tikz}} \]
and on the other hand we can view them as being a map from the processes with input $A$ and output $B$ to the processes with input $C$ and output $D$:
\[%
\InputIfFileExists{Diagrams/vise2.tikz}{}{\input{./figures/Diagrams/vise2.tikz}}\quad :: \quad %
\InputIfFileExists{Diagrams/vise4.tikz}{}{\input{./figures/Diagrams/vise4.tikz}} \quad \mapsto\quad %
\InputIfFileExists{Diagrams/vise3.tikz}{}{\input{./figures/Diagrams/vise3.tikz}}\]

With this notational short-hand in place, we will now introduce another useful concept: that of a `diagrammatic sum'. Whilst not a strictly diagrammatic notion itself, it can be derived from the classical interface that we will postulate, and it is vital to making connections to more standard linear-algebraic techniques that we will use in various proofs.
\begin{definition}[Diagrammatic summation]\label{def:DiagrammaticSums}
A sum is an associative, commutative, binary operation with a zero element on processes with the same inputs and outputs, i.e.\ given a set of processes $f_i:A\to B$, there is a process  $\sum_i f_i:A\to B$.  Moreover, it distributes over diagrams, that is, for all $\chi$, $C$ and $D$:
\beq \label{eq:SumsDistribute}
\InputIfFileExists{Sums/Dist1.tikz}{}{\input{./figures/Sums/Dist1.tikz}}\quad,
\eeq
 where  $\chi$ is shorthand for a diagram of that shape, i.e.:
 \begin{equation}\label{eq:circuitfragment}%
\InputIfFileExists{Diagrams/circuitFragment.tikz}{}{\input{./figures/Diagrams/circuitFragment.tikz}}\quad.\end{equation}
\end{definition}

\begin{example}[Quantum]\label{Ex:AcausalCStar}
We can now discuss sums of processes within quantum theory.  If we define sums of CP maps in the usual way, then linearity of CP maps and bilinearity of the tensor product ensures that these will distribute over diagrams, hence, are diagrammatic sums in the sense of def.~\ref{def:DiagrammaticSums}. 
 \end{example}
 
 \begin{example}[Classical]
 Similarly, we can discuss sums of processes within classical theory.  That is, if we define sums of substochastic maps in the usual way, then linearity of substochastic maps and bilinearity of the tensor product ensure that these will distribute over diagrams. In this case, by decomposing the identity process, sums allow us to have a convenient `matrix representation' of classical processes:
 \[%
\InputIfFileExists{Diagrams/classicalMatrixRep1.tikz}{}{\input{./figures/Diagrams/classicalMatrixRep1.tikz}} = \sum_{i=1}^n\sum_{j=1}^m %
\InputIfFileExists{Diagrams/classicalMatrixRep2.tikz}{}{\input{./figures/Diagrams/classicalMatrixRep2.tikz}} =: \sum_{ij} \mathsf{f}_{i}^{j}  %
\InputIfFileExists{Diagrams/classicalMatrixRep3.tikz}{}{\input{./figures/Diagrams/classicalMatrixRep3.tikz}}\]

\end{example}

In general such a sum will not exist in the causal subtheory,  at least, in the probabilistic
theories that we are interested in here.
The structure of diagrammatic summation
however gives us some insights into the structure of the causal subtheory.
For example -- in a theory where the scalars are non-negative real numbers  and their sum is given by the usual sum (as will be the case in process theories with a classical interface) -- if $f$ and $g$ are causal processes, then:
\[%
\begin{tikzpicture}
	\begin{pgfonlayer}{nodelayer}
		\node [style=none] (0) at (0.75, -0) {$p$};
		\node [style=small box] (1) at (2, -0) {$f$};
		\node [style=none] (2) at (2, 1) {};
		\node [style=none] (3) at (2, -1) {};
	\end{pgfonlayer}
	\begin{pgfonlayer}{edgelayer}
		\draw [qWire](2.center) to (1);
		\draw [qWire](1) to (3.center);
	\end{pgfonlayer}
\end{tikzpicture}
}\quad\text{and}\quad %
\begin{tikzpicture}
	\begin{pgfonlayer}{nodelayer}
		\node [style=none] (0) at (0, -0) {$(1-p)$};
		\node [style=small box] (1) at (2, -0) {$g$};
		\node [style=none] (2) at (2, 1) {};
		\node [style=none] (3) at (2, -1) {};
	\end{pgfonlayer}
	\begin{pgfonlayer}{edgelayer}
		\draw [qWire](2.center) to (1);
		\draw [qWire](1) to (3.center);
	\end{pgfonlayer}
\end{tikzpicture}
}\]
with $p\in\left(0,1\right)$, are not. However, their sum is a causal process as
\[%
\InputIfFileExists{Sums/convex3.tikz}{}{\input{./figures/Sums/convex3.tikz}}\]
That is, convex combinations of causal processes are themselves causal. Introducing sums in such
 theories  therefore `reveals' the \emph{convex} structure of the subcausal theory. This is a recurring theme in this paper. Many compositional structures we use are defined in the full theory, but, still have an important impact on the causal sub-theory. In particular, this will be the case for postulates~\ref{def:compactStructure} and \ref{post:SharpDagger}, as is discussed in detail in Sec.~\ref{sec:cupscaps} .

\subsection{Classical interface} \label{Sec:ClassicalInterface}

We can now formalise how we access general systems by means of a `classical interface'.
\begin{definition}[Classical interface]\label{def:ClassicalInterface}
A classical interface comprises three parts  (each of which we will formally introduce in this section):
\ben
\item[i.]  a  full classical subtheory,
\item[ii.]  all classically controlled processes, and
\item[iii.]  sufficient  causal-compatible  tomography tests.
\een
\end{definition}

The first of these is  straightforward. To have a classical interface  the process theory must have classical systems and processes as defined in Ex.~\ref{Ex:Classical}.

\begin{definition}[Full classical subtheory]\label{def:FullSubTheory}
To begin, the process theory must have arbitrary classical systems:
\[\left\{%
\begin{tikzpicture}
	\begin{pgfonlayer}{nodelayer}
		\node [style=none] (0) at (0, 0.75) {};
		\node [style=none] (1) at (0, -0.75) {};
		\node [style={right label}] (2) at (0, -0.2500001) {$n$};
	\end{pgfonlayer}
	\begin{pgfonlayer}{edgelayer}
		\draw [cWire] (1.center) to (0.center);
	\end{pgfonlayer}
\end{tikzpicture}}\ \ \middle|\ \  n\in\mathds{N} \right\}\]
and the processes with classical inputs and outputs
\[\left\{%
\InputIfFileExists{Diagrams/cfst.tikz}{}{\input{./figures/Diagrams/cfst.tikz}}\right\}\]
must be precisely the set of classical processes, no more no less. Moreover, they must compose in the same way as they do classically, that is, we must be able to manipulate the diagrams involving only classical systems exactly as in Ex.~\ref{Ex:Classical}.
\end{definition}
Note that an important consequence of this is that the trivial system must belong to this classical subtheory such that, for example, classical states are indeed classical. An important consequence of this is that the scalars of a theory with a classical interface are simply the classical scalars, $\mathds{R}^+$.

 The other two parts of the interface introduce the interactions between this full classical subtheory and the rest of the process theory. Firstly, `classically controlled processes' (c.f. Def.~\ref{def:classicalcontrolprocs}) formalise the idea that we can choose which process out of a family to implement, possibly using randomness (e.g.\ by rolling a die).

\begin{definition}[Classically controlled processes]
\label{def:ClassicalControl}
\[
\text{For a set of processes} \quad \left\{%
\InputIfFileExists{Diagrams/classicalControlDefinition2.tikz}{}{\input{./figures/Diagrams/classicalControlDefinition2.tikz}} \right\}_{i=1}^n \quad \text{a controlled process} \quad%
\InputIfFileExists{Diagrams/classicalControlDefinition3.tikz}{}{\input{./figures/Diagrams/classicalControlDefinition3.tikz}} \quad \text{ is one such that:}
\]
\beq \label{eq:ClassicalControl}
\forall i \quad %
\InputIfFileExists{Diagrams/classicalControlDefinition1.tikz}{}{\input{./figures/Diagrams/classicalControlDefinition1.tikz}}
\eeq
\end{definition}

For example, if we let $\{f_i\}$ be a set of states, this provides the controlled state preparations considered in Sec.~\ref{sec:Discard}.

Secondly, `tomography tests' (c.f. Def.~\ref{def:measurement}) formalise how classical systems can be used to characterise processes by the probabilities obtained in experiments.

\begin{definition}[Tests for finite process tomography]\label{def:Tomography}\label{def:FiniteTomography}\label{def:FiniteCI}
For a pair of systems $\left(A,B\right)$ a controlled test is a  vise of the form
\beq \label{eq:Tomography}
\InputIfFileExists{Diagrams/tomogDef2.tikz}{}{\input{./figures/Diagrams/tomogDef2.tikz}}\quad\text{such that}\qquad
\InputIfFileExists{Diagrams/tomogDef4.tikz}{}{\input{./figures/Diagrams/tomogDef4.tikz}}\quad \iff \quad %
\InputIfFileExists{Diagrams/tomogDef3.tikz}{}{\input{./figures/Diagrams/tomogDef3.tikz}}
\eeq
\end{definition}

For example, if we let  $A$ be the trivial system such that $\tau$ is doing state tomography, then this provides us with the measurements  of Def.~\ref{def:measurement}.

\begin{remark}
Note that the definition of classical theory  we are using automatically implies that $n$ and $m$ are \emph{finite}. This means that if a theory has tomographic tests then it is never necessary to perform an infinite number of distinct experiments to characterise a process. Consequently, the set of probabilities needed to describe a process is finite. Note that in practice it is never possible to perform an infinite number of experiments so we should at least expect this to hold for our best effective theory of nature.
\end{remark}

 We however will demand something stronger than this, namely we will demand that our tomographic tests are compatible with causality. As we do not have a notion of a discarding map for the theory as a whole we instead will introduce a notion of compatibility with the causality of the classical subtheory. 

\begin{definition}[Causal-compatible tomographic tests]\label{def:CausCompTests}
A collection of tomographic tests is said to be causal-compatible if any composite of them such that the resulting process is classical is a causal classical process. For example, one way of composing them to give a classical process would be:
\[%
\InputIfFileExists{Diagrams/CausalTomography.tikz}{}{\input{./figures/Diagrams/CausalTomography.tikz}}\]
then causal-compatibility implies that:
\[%
\InputIfFileExists{Diagrams/CausalTomography1.tikz}{}{\input{./figures/Diagrams/CausalTomography1.tikz}} \quad = \quad %
\InputIfFileExists{Diagrams/CausalTomography2.tikz}{}{\input{./figures/Diagrams/CausalTomography2.tikz}}\]
\end{definition}

It is common to assume that tomography can be performed locally. This expresses the idea that: although we know the world to be non-local (in the sense of \cite{Bell}),  there are still no holistic degrees of freedom, and that the description of two distinct regions of space can be formulated entirely in terms of their individual properties and the correlations between them. One can consider that the ability to be characterised locally is really the defining feature of what we mean by a system: a system is something we can isolate and study in its own right, independent of the rest of the world.

\begin{definition}[Local tomography]\label{def:LocalCI}
The tests $\tau$ in the definition of the classical interface are said to be local if they factorise over parallel composition:
\beq\label{def:LocalTomography}
\InputIfFileExists{Diagrams/localTomog1.tikz}{}{\input{./figures/Diagrams/localTomog1.tikz}}
\eeq
where $n=n_\alpha\cdots n_\beta$ and $m=m_\gamma\cdots m_\delta$. If all of the tests in a classical interface are local then it is said to be a local classical interface, and the theory is said to be locally tomographic.
\end{definition}

 \begin{example}
 Quantum theory, as mentioned earlier, has a classical subtheory and moreover has all possible classically controlled processes and suitable causal-compatible tests for finite, local, tomography.  In particular, local tomography  \cite{araki1980characterization,bergia1980actual} has been used in many reconstructions of quantum theory, e.g.\ \cite{chiribella2016quantum,HardyBig}. This can be extended to the full process-theoretic description of quantum theory by viewing general C*-algebras as restricted quantum systems (in the sense of \cite{coecke2017leaks,SelingerIdempotent,heunen2013completely}.
 \end{example}

 Before moving on to some more basic process theoretic concepts let us first explore some of the consequences of having a classical interface.
Ultimately we will show that a classical interface provides us with most of the structure that is typically assumed in the GPT framework \cite{HardyAxiom,Barrett}. 

Firstly we can show that classically controlled processes are unique.
\begin{proposition}\label{lem:UniqueClassicalControl}
In a theory with a classical interface (Post.~\ref{post:ClassicalInterface}) classically controlled processes (Def.~\ref{def:ClassicalControl}) are unique.  That is, given a set of processes $\{f_i:A\to B\}_{i=1}^n$ there is a unique process $F:A\otimes n \to B$ satisfying Def.~\ref{def:ClassicalControl}. 
\end{proposition}
\proof
 See App.~\ref{proof:unqiueCC}.
\endproof

Next we can see that any theory with a classical interface has diagrammatic sums.  Intuitively, this provides another way to represent the branching structure of probabilistic processes, and so, it is represented by a summation operation which distributes over diagrams.  This is achieved by lifting the sum that exists in the classical subtheory to the entire process theory.

\begin{proposition}\label{lem:ConvexCones}
In a theory with classically controlled processes (Def.~\ref{def:ClassicalControl}) we can define a sum of processes  (for any finite set of processes) as:
\beq \label{eq:DefiningSums}
\InputIfFileExists{Sums/Sums1.tikz}{}{\input{./figures/Sums/Sums1.tikz}}
\eeq
where $F$ is the  unique (Prop.~\ref{lem:UniqueClassicalControl}) classically controlled process satisfying:
\beq\label{eq:controlS}
\InputIfFileExists{Sums/Sums2.tikz}{}{\input{./figures/Sums/Sums2.tikz}}.
\eeq
and
\[%
\begin{tikzpicture}
	\begin{pgfonlayer}{nodelayer}
		\node [style=downground] (0) at (0, -0.5) {};
		\node [style=none] (1) at (0, 0.5) {};
		\node [style={right label}] (2) at (0, 0.25) {};
	\end{pgfonlayer}
	\begin{pgfonlayer}{edgelayer}
		\draw [cWire] (1.center) to (0);
	\end{pgfonlayer}
\end{tikzpicture}}\quad =\quad \sum_i %
\begin{tikzpicture}
	\begin{pgfonlayer}{nodelayer}
		\node [style=point] (0) at (0, -0) {$i$};
		\node [style=none] (1) at (0, 1) {};
	\end{pgfonlayer}
	\begin{pgfonlayer}{edgelayer}
		\draw [style=cWire] (1.center) to (0);
	\end{pgfonlayer}
\end{tikzpicture}}. \]
Note that for classical processes this is the same as the usual sum, and, hence, this extends the classical sum to the rest of the process theory.
\end{proposition}
\proof
 See App.~\ref{proof:diagSums}. 
\endproof

Note that this sum gives us a convenient way to represent the classically controlled processes, specifically it is simple to confirm that we can write:
\beq%
\InputIfFileExists{Diagrams/CProcRep.tikz}{}{\input{./figures/Diagrams/CProcRep.tikz}}\quad =\quad \sum_i %
\InputIfFileExists{Diagrams/CProcRep2.tikz}{}{\input{./figures/Diagrams/CProcRep2.tikz}}\eeq

The classical systems come equipped with a notion of discarding, and so, we can characterise the causal subtheory for the classical part of the process theory. Like we did with the summation, we will now show that we can lift this structure from the classical subtheory to the whole process theory.

\begin{proposition}\label{prop:CausalTheory} If a theory has a 
 local classical interface 
then there is a unique way to characterise the processes in the full theory which are compatible with classical causality. That is, we will see that we can define discarding maps by:
\[%
\InputIfFileExists{Diagrams/LocalCausalTomography5.tikz}{}{\input{./figures/Diagrams/LocalCausalTomography5.tikz}}\]
and then all of the causal-compatible processes can be characterised by:
\[%
\InputIfFileExists{Diagrams/LocalCausalTomography7.tikz}{}{\input{./figures/Diagrams/LocalCausalTomography7.tikz}}.\]
\end{proposition}
\proof
 See App.~\ref{proof:CST}.
\endproof

Given this notion of summation and the characterisation of the causal subtheory, we can prove some basic properties regarding the state spaces of systems and the maps between them.

\begin{proposition}\label{lem:LinearityAndConvexCones}
In a theory with a classical interface (Post.~\ref{post:ClassicalInterface}), the states form a finite-dimensional pointed convex cone. Processes then induce completely positive linear maps between these cones. The causal states are defined by an intersecting hyperplane, and causal processes preserve this hyperplane.
\end{proposition}
\proof
 See App.~\ref{proof:GPTStruct}.
\endproof

\begin{remark}
The systems therefore have much of the structure that is assumed in a GPT \cite{Barrett,HardyAxiom}. What we have not assumed or proved is the necessity of the convex cones to be \emph{closed}. For related work connecting these frameworks see, for example,
 \cite{wilceshortcut,gogioso2017categorical,tull2016operational,barnum2013symmetry,wilce2012symmetry,wilce2011symmetry}.
\end{remark}

\begin{remark}
Note that many of the above results can easily be extended from states to arbitrary processes. For example, the set of processes from $A$ to $B$ will generally form a finite-dimensional proper cone, with a convex subset of causal processes, which are defined by a set of linear equality constraints.
\end{remark}

\subsection{Leaks and purity}

We will need  the process-theoretic definition of purity first presented in \cite{selby2017leaks}, and for this purpose we introduce the notions of  \emph{dilations} and \emph{leaks}.
\begin{definition}[Dilations]
A \emph{dilation} of a process $f:A\to B$ is a process $g:A\to B\otimes C$ such that:
\[\begin{tikzpicture}
	\begin{pgfonlayer}{nodelayer}
		\node [style=box] (0) at (0, 0) {$f$};
		\node [style=none] (1) at (0, 1.25) {};
		\node [style=none] (2) at (0, -1.25) {};
		\node [style=right label] (3) at (0, -1.25) {$A$};
		\node [style=right label] (4) at (0, 1) {$B$};
	\end{pgfonlayer}
	\begin{pgfonlayer}{edgelayer}
		\draw[qWire] (1.center) to (0);
		\draw[qWire] (0) to (2.center);
	\end{pgfonlayer}
\end{tikzpicture}
\ \ =\ \ \begin{tikzpicture}
	\begin{pgfonlayer}{nodelayer}
		\node [style={medium map}] (0) at (0, -0) {$g$};
		\node [style=none] (1) at (0.7500001, 1) {};
		\node [style=none] (2) at (0.7500001, 0.5000001) {};
		\node [style=none] (3) at (-0.5000001, 1.75) {};
		\node [style=none] (4) at (-0.5000001, 0.5000001) {};
		\node [style=none] (5) at (-0.5000001, -0.5000001) {};
		\node [style=none] (6) at (-0.5000001, -1.25) {};
		\node [style=upground] (7) at (0.7500001, 1.25) {};
 		\node [style={right label}] (8) at (-0.5, -1.25) {$A$};
 		\node [style={right label}] (9) at (-0.5, 1.5) {$B$};
 		\node [style=right label] (10) at (1.25, 1) {$C$};
	\end{pgfonlayer}
	\begin{pgfonlayer}{edgelayer}
		\draw [style=qWire] (1.center) to (2.center);
		\draw [style=qWire] (3.center) to (4.center);
		\draw [style=qWire] (5.center) to (6.center);
	\end{pgfonlayer}
\end{tikzpicture}
\]
\end{definition}

\begin{definition}[Leaks]
A \emph{leak}  is a dilation of the identity, that is,  a process \[%
\InputIfFileExists{Diagrams/leakL.tikz}{}{\input{./figures/Diagrams/leakL.tikz}}\] such that
\[%
\InputIfFileExists{Diagrams/leakdisc1.tikz}{}{\input{./figures/Diagrams/leakdisc1.tikz}}\:.\]
\end{definition}

Note that when we have multiple leaks, we will typically  label them with a different colour.

\begin{example}[Trivial leaks]
A leak is said to be trivial if there exists a state $s$ such that
\[%
\InputIfFileExists{Diagrams/TrivialLeak.tikz}{}{\input{./figures/Diagrams/TrivialLeak.tikz}}\]
Note that any $s$ which is causal will define a trivial leak in this way.
\end{example}
\begin{example}[Broadcasting]
A broadcasting map
\[%
\InputIfFileExists{Diagrams/Broadcaster.tikz}{}{\input{./figures/Diagrams/Broadcaster.tikz}}\]
is one that `leaks both ways', that is:
\[%
\InputIfFileExists{Diagrams/BroadcasterDef.tikz}{}{\input{./figures/Diagrams/BroadcasterDef.tikz}}\]
 It is well known that these do not exist for quantum systems (as demonstrated by the `no-broadcasting theorem' \cite{barnum1996noncommuting}) whilst for classical systems they are defined by a stochastic map whose action on the basis states is simply the `copier':
\[%
\InputIfFileExists{Diagrams/BroadcasterClassicalNew.tikz}{}{\input{./figures/Diagrams/BroadcasterClassicalNew.tikz}}\]
\end{example}

We provide a formal classification of the leaks for quantum theory in Prop.~\ref{Prop:LeakClassification}. However, informally, the leaks for a quantum system $\mathcal{A} = \bigoplus_i A_i$ can be seen as leaking the `which branch' information. Therefore, for fully quantum systems all leaks are trivial, whilst for classical systems all leaks are based on the  aforementioned broadcasting map.

Informally, a process is pure if it plays nicely with leaks. More precisely  we demand that two independent conditions are satisfied for a process $f$ to be pure. Firstly, that any dilation, $g$, of a pure process, $f$, can be explained by leaks on the input and output systems:
\[%
\InputIfFileExists{Diagrams/f-box.tikz}{}{\input{./figures/Diagrams/f-box.tikz}}\ \ =\ \ %
\InputIfFileExists{Diagrams/pure1.tikz}{}{\input{./figures/Diagrams/pure1.tikz}}\quad \Longrightarrow\quad  \exists\   \left(%
\InputIfFileExists{Diagrams/plainleak.tikz}{}{\input{./figures/Diagrams/plainleak.tikz}},%
\InputIfFileExists{Diagrams/j2.tikz}{}{\input{./figures/Diagrams/j2.tikz}},%
\InputIfFileExists{Diagrams/dilationFromLeaks2.tikz}{}{\input{./figures/Diagrams/dilationFromLeaks2.tikz}} \right)\ \  : \ \  %
\InputIfFileExists{Diagrams/pure2.tikz}{}{\input{./figures/Diagrams/pure2.tikz}}\ \ = \ \ %
\InputIfFileExists{Diagrams/dilationFromLeaks.tikz}{}{\input{./figures/Diagrams/dilationFromLeaks.tikz}}\]

Second, that pure processes do not interact with `leakable' information, so leaking before or after is equivalent:
\[
 \forall\ \ %
\InputIfFileExists{Diagrams/plainleak.tikz}{}{\input{./figures/Diagrams/plainleak.tikz}}\ \exists \ \ %
\InputIfFileExists{Diagrams/j2.tikz}{}{\input{./figures/Diagrams/j2.tikz}} \ \ \text{and} \ \ \forall \ \ %
\InputIfFileExists{Diagrams/j2.tikz}{}{\input{./figures/Diagrams/j2.tikz}}\ \exists \ \ %
\InputIfFileExists{Diagrams/plainleak.tikz}{}{\input{./figures/Diagrams/plainleak.tikz}}  \ \  : \ \  %
\InputIfFileExists{Diagrams/pure3.tikz}{}{\input{./figures/Diagrams/pure3.tikz}}\ \ = \ \ %
\InputIfFileExists{Diagrams/leakAfter.tikz}{}{\input{./figures/Diagrams/leakAfter.tikz}}
\]

These two conditions can be combined into a single simplified statement which we take to be the definition of purity:

\begin{definition}[Purity of processes]\label{def:pureeq} $f$ is \emph{pure} if and only if
\beq
\InputIfFileExists{Diagrams/f-box.tikz}{}{\input{./figures/Diagrams/f-box.tikz}}\ \ =\ \ %
\InputIfFileExists{Diagrams/pure1.tikz}{}{\input{./figures/Diagrams/pure1.tikz}}\quad \Longrightarrow\quad  \exists\ \  %
\InputIfFileExists{Diagrams/plainleak.tikz}{}{\input{./figures/Diagrams/plainleak.tikz}}\ \& \ \ %
\InputIfFileExists{Diagrams/j2.tikz}{}{\input{./figures/Diagrams/j2.tikz}} \ \ : \ \  %
\InputIfFileExists{Diagrams/pure2.tikz}{}{\input{./figures/Diagrams/pure2.tikz}}\ \ = \ \ %
\InputIfFileExists{Diagrams/pure3.tikz}{}{\input{./figures/Diagrams/pure3.tikz}}\ \ = \ \ %
\InputIfFileExists{Diagrams/leakAfter.tikz}{}{\input{./figures/Diagrams/leakAfter.tikz}}
\eeq
\end{definition}

This definition is motivated by the fact that it gives the right notion of purity of processes both in quantum and classical theories \cite{selby2017leaks}.  Note that this definition is \emph{not} equivalent to commonly proposed definitions of purity based on notions of extremality in the geometry of the space of processes. This is desirable as these more standard definitions either deem the classical identity channel or the discarding map to be mixed.  See \cite{selby2017leaks} for further discussion of this point. We characterize the pure processes for quantum theory in Prop.~\ref{Prop:PureProcesses}; in the case of fully quantum systems we obtain the standard notion of purity, i.e.\ that the process has Kraus rank $1$.

In the special case of states (or similarly for effects) we can obtain a much simpler characterisation of purity by noting that a leak for a trivial system is simply a state preparation for the leaked system:
\[%
\begin{tikzpicture}
	\begin{pgfonlayer}{nodelayer}
		\node [style=leak] (0) at (0, 0) {};
		\node [style=none] (1) at (1, 1) {};
		\node [style={right label}] (2) at (1, 1) {$L$};
	\end{pgfonlayer}
	\begin{pgfonlayer}{edgelayer}
		\draw [style=qWire, in=-90, out=0, looseness=1.00] (0) to (1.center);
	\end{pgfonlayer}
\end{tikzpicture}}\quad = \quad %
\begin{tikzpicture}
	\begin{pgfonlayer}{nodelayer}
		\node [style=point] (0) at (0, 0) {$l$};
		\node [style=none] (1) at (0, 1) {};
		\node [style={right label}] (2) at (0, 1) {$L$};
	\end{pgfonlayer}
	\begin{pgfonlayer}{edgelayer}
		\draw [style=qWire] (0) to (1.center);
	\end{pgfonlayer}
\end{tikzpicture}}\]

\begin{example}\label{Ex:PureStates}
For states this definition of purity reduces to
\[%
\InputIfFileExists{Diagrams/PureStateDef.tikz}{}{\input{./figures/Diagrams/PureStateDef.tikz}}\]
\end{example}
 This notion of purity was put forward by Chiribella in \cite{chiribella2014distinguishability} which intuitively states that a pure state is information that is independent of the surrounding context.

We now formalise the idea of a set of states prepared being  pure and jointly distinguishable. This gives an important information-theoretic property of each system, namely, the amount of classical information they can store reliably. Such states are described by a particular type of causal state preparation $S$.   Recall that causality for a state preparation means:
\[
\InputIfFileExists{Diagrams/sharp1.tikz}{}{\input{./figures/Diagrams/sharp1.tikz}}.
\]
 It can be shown \cite{selby2017leaks} that such a state preparation can only be pure if $n=1$, that is, if it is simply a state. However, we can  capture the notion that each individual state in $S$ is pure by considering purity of the following diagram:
\[%
\InputIfFileExists{Diagrams/sharp2v2.tikz}{}{\input{./figures/Diagrams/sharp2v2.tikz}}\]
That is, the states prepared are pure if and only if the state preparation is pure when we keep a record (the ancillary classical system) of which state was prepared.

\begin{definition}[Testability]\label{def:testability}
A causal state preparation $S:n\to A$ is said to be \emph{testable} if
\[%
\InputIfFileExists{Diagrams/sharp2v2.tikz}{}{\input{./figures/Diagrams/sharp2v2.tikz}}\]
is pure, and if there is a measurement  (Def.~\ref{def:measurement}) $M$ such that
\[%
\InputIfFileExists{Diagrams/sharp3.tikz}{}{\input{./figures/Diagrams/sharp3.tikz}}\]
We say that the state preparation $S$ is \emph{maximal}  testable (for a given system $A$) if $n$ is moreover maximal.
\end{definition}

\begin{example}
In quantum theory the maximal testable state preparations correspond to orthonormal bases of the Hilbert space.  That is, given a quantum system $M(\mathds{C}^n)$, a maximal testable set of states will correspond to a set of orthonormal projectors $\ketbra{i}{i} \in M(\mathds{C}^n)$. The maximal testable state preparation will then be some map $S:\bigoplus_{i=1}^n M(\mathds{C}) \to M(\mathds{C}^n) :: \ket{i}\to\ketbra{i}{i}$.
\end{example}

\subsection{Cups, caps and sharp daggers}\label{sec:cupscaps}
We now formalise what it means for a process theory to be `blind w.r.t.\ inputs and outputs' i.e.\ that inputs can be `bent' into outputs and vice versa, essentially
relaxing the basic constraints on forming diagrams for a process theory (cf.\ point $2.$ of Def.~\ref{def:ProcessTheory}).  In terms of diagrams this is usually referred to as \em string diagrams \em \cite {BaezLNP, CKbook} or \emph{compact structure} \cite{KellyLaplaza, AC1}.
This ability to bend wires can also be represented within a process theory as a particular bipartite state, the \emph{cup}, and effect, the \emph{cap}, for each system:
\[%
\InputIfFileExists{Diagrams2/cup.tikz}{}{\input{./figures/Diagrams2/cup.tikz}} \qquad \text{and} \qquad %
\InputIfFileExists{Diagrams2/cap.tikz}{}{\input{./figures/Diagrams2/cap.tikz}}\]
 A detailed discussion can be found in \cite{CatsII, CKbook}.

\begin{definition}[Cups and caps]\label{def:cupscaps}
A theory has \em cups \em  and \em caps \em if for each system it has processes:
\[%
\InputIfFileExists{Diagrams/cup.tikz}{}{\input{./figures/Diagrams/cup.tikz}} \qquad \text{and} \qquad %
\InputIfFileExists{Diagrams/cap.tikz}{}{\input{./figures/Diagrams/cap.tikz}}\]
which satisfy:
\beq\label{eq:yank}
\InputIfFileExists{Diagrams/yank1.tikz}{}{\input{./figures/Diagrams/yank1.tikz}}\ \quad,\ \quad %
\InputIfFileExists{Diagrams/yank2.tikz}{}{\input{./figures/Diagrams/yank2.tikz}}\qquad\text{and}\qquad %
\InputIfFileExists{Diagrams/yank3.tikz}{}{\input{./figures/Diagrams/yank3.tikz}}
\eeq
Equivalently, this means that in diagrams inputs can be connected to inputs, outputs to outputs, and also that loops are allowed.
\end{definition}

 Cups and caps have a very clear conceptual meaning: they assert that the theory has `maximal' correlations.  Firstly, that the theory must have correlations is easily shown by a simple diagrammatic argument.  Assume, for the sake of contradiction,  that there are no correlations, i.e.\ that all bipartite states \em separate\em:
\[
\InputIfFileExists{Diagrams/ProductState.tikz}{}{\input{./figures/Diagrams/ProductState.tikz}}
\]
then, in particular, the cup separates:
\[
\InputIfFileExists{Diagrams/ProductCup.tikz}{}{\input{./figures/Diagrams/ProductCup.tikz}}
\]
so we have:
\[
\InputIfFileExists{Diagrams/NoCorrelations.tikz}{}{\input{./figures/Diagrams/NoCorrelations.tikz}}
\]
That is, all wires separate, so the theory  does not permit any interactions, correlations or evolution. We say that such a theory is \emph{trivial}.  Any non-trivial theory with cups must therefore have correlations. Intuitively, maximality can  then be deduced from the first of Eqs.~\eqref{eq:yank} in that `via a cup and cap an identity can be realised', so that the cup and cap must allow for a flow of all state-data at the input.  Formally this intuition is substantiated in \cite{coeckeuniqueness}.

Similarly to the above proof, it can now
easily be seen that the cap
cannot be causal, confirming that this simple diagrammatic structure is lost when passing to the causal subtheory.  Indeed, if the cap were causal then the Eq.~\eqref{eq:effectcaus} becomes:
\[
\InputIfFileExists{Diagrams/CausalCap.tikz}{}{\input{./figures/Diagrams/CausalCap.tikz}}
\]
and again all wires would separate and the theory would be trivial:
\[%
\InputIfFileExists{Diagrams2/capNotDiscard.tikz}{}{\input{./figures/Diagrams2/capNotDiscard.tikz}}\]
Consequently, non-trivial causal subtheories will not have cups and caps.

\begin{example}[Quantum] \label{Ex:DaggerCompact}
For the quantum case the cups \& caps realise the Choi-Jamio{\l}kowski isomorphism. That is, the cup and cap are (non-normalised) maximally entangled states and effects respectively. More formally, the cup for a $d$-dimensional system $A$ is given by the super-normalised state written in Dirac notation as:
\[%
\InputIfFileExists{Diagrams/cup.tikz}{}{\input{./figures/Diagrams/cup.tikz}}\ \sim\ \ \sum_{i,j=1}^{d}\ketbra{ii}{jj}\]
similarly the cap is given by the same matrix but interpreted as a POVM element. For general C*-algebras, e.g.\ $\mathcal{A}=\bigoplus_k A_k$ where $A_k$ are $d_k$-dimensional quantum systems, we can write the cup as:
\[ %
\InputIfFileExists{Diagrams2/CStarCup.tikz}{}{\input{./figures/Diagrams2/CStarCup.tikz}}\ \ \sim\ \ \ \bigoplus_k \sum_{i,j=1}^{d_k} \ketbra{ii}{jj}\ =\ \sum_{i,j=1}^{d_1} \ketbra{ii}{jj} + \sum_{i,j=d_1+1}^{d_1+d_2}\ketbra{ii}{jj}+\cdots\]
the cap is again given by the same matrix interpreted as a POVM element.
\end{example}
 \begin{example}[Classical]
In the classical case these cups and caps are given by:
\[%
\InputIfFileExists{Diagrams/classicalCup.tikz}{}{\input{./figures/Diagrams/classicalCup.tikz}}\ =\ \sum_{i=1}^n %
\InputIfFileExists{Diagrams/classicalCup2.tikz}{}{\input{./figures/Diagrams/classicalCup2.tikz}}\qquad \text{and} \qquad %
\InputIfFileExists{Diagrams/classicalCap.tikz}{}{\input{./figures/Diagrams/classicalCap.tikz}}\ =\ \sum_{j=1}^n %
\InputIfFileExists{Diagrams/classicalCap2.tikz}{}{\input{./figures/Diagrams/classicalCap2.tikz}}\]
which corresponds to the super-normalised perfectly correlated state and effect. It is simple to verify that these do indeed satisfy the relevant equations (Eqs.~\ref{eq:yank}), for example:
\[%
\InputIfFileExists{Diagrams/classicalCompact1.tikz}{}{\input{./figures/Diagrams/classicalCompact1.tikz}}\ =\ \sum_{ij} %
\InputIfFileExists{Diagrams/classicalCompact2.tikz}{}{\input{./figures/Diagrams/classicalCompact2.tikz}}\ =\ \sum_{ij}\delta_{ij} %
\begin{tikzpicture}
	\begin{pgfonlayer}{nodelayer}
		\node [style=copoint] (0) at (0, -0.75) {$j$};
		\node [style=none] (1) at (0, -1.75) {};
		\node [style=none] (2) at (0, 1.75) {};
		\node [style=point] (3) at (0, 0.75) {$i$};
	\end{pgfonlayer}
	\begin{pgfonlayer}{edgelayer}
		\draw [cWire] (1.center) to (0);
		\draw [cWire] (2.center) to (3);
	\end{pgfonlayer}
\end{tikzpicture}}\ =\ %
\begin{tikzpicture}
	\begin{pgfonlayer}{nodelayer}
		\node [style=none] (0) at (0, 1.25) {};
		\node [style=none] (1) at (0, -1.25) {};
	\end{pgfonlayer}
	\begin{pgfonlayer}{edgelayer}
		\draw [cWire] (0.center) to (1.center);
	\end{pgfonlayer}
\end{tikzpicture}}\]
\end{example} 

 Cups and caps provide a way to swap inputs and outputs, and hence, an operation of `time-reversing':
\[%
\InputIfFileExists{Diagrams/Transpose.tikz}{}{\input{./figures/Diagrams/Transpose.tikz}}\]
In the case of quantum theory this corresponds to the transpose, but quantum theory has a second manner of doing so, namely the adjoint, a.k.a.\ \em dagger \em \cite{AC2, SelingerCPM}, which turns Dirac kets into Dirac bras and vice versa. The difference between the transpose and the dagger is the conjugate, and hence witnesses the non-trivial involution on the number field of complex numbers.  Hence having a dagger besides cups \& caps is a truly fundamental structure  within quantum theory.

\begin{definition}[Dagger]\label{def:Dagger}
 A dagger is a reflection of processes:
\[%
\InputIfFileExists{Diagrams/sharpDagger1.tikz}{}{\input{./figures/Diagrams/sharpDagger1.tikz}}\]
(where an asymmetry of the box shape has been introduced to make the reflection clear.)
In particular, it reflects the entire diagram structure:
\beq\label{eq:daggerReflection}%
\InputIfFileExists{Diagrams/sharpDagger2.tikz}{}{\input{./figures/Diagrams/sharpDagger2.tikz}}\eeq
\end{definition}

Just as a Dirac bra can be seen as a `test' for the corresponding ket, we would like our general process-theoretic  dagger to do the same, that is, the dagger of a state  can be interpreted as a  \emph{test} for that state \cite{selby2016process}, in that, for pure states, it uniquely identifies the state with certainty:	
\[
\begin{tikzpicture}
	\begin{pgfonlayer}{nodelayer}
		\node [style=copoint] (0) at (0, 0.75) {$\psi$};
		\node [style=point] (1) at (0, -0.75) {$\rho$};
	\end{pgfonlayer}
	\begin{pgfonlayer}{edgelayer}
		\draw [qWire] (1) to (0);
	\end{pgfonlayer}
\end{tikzpicture}} \ \ = \ \ 1 \quad \iff \quad %
\begin{tikzpicture}
	\begin{pgfonlayer}{nodelayer}
		\node [style=none] (0) at (0, 0.5) {};
		\node [style=point] (1) at (0, -0.5) {$\rho$};
	\end{pgfonlayer}
	\begin{pgfonlayer}{edgelayer}
		\draw [qWire] (1) to (0.center);
	\end{pgfonlayer}
\end{tikzpicture}} \ \ = \ \ %
\begin{tikzpicture}
	\begin{pgfonlayer}{nodelayer}
		\node [style=none] (0) at (0, 0.5) {};
		\node [style=point] (1) at (0, -0.5) {$\psi$};
	\end{pgfonlayer}
	\begin{pgfonlayer}{edgelayer}
		\draw [qWire] (1) to (0.center);
	\end{pgfonlayer}
\end{tikzpicture}}
\]

 In  \cite{selby2016process} it was indeed shown that this sharpness assumptions lifts to a dagger structure on all processes.
 This \em sharpness \em assumption then gives a clear conceptual meaning to the dagger\footnote{This is obviously  distinct  to the conceptual meaning of cups \& caps  -- cups \& caps being about correlations whilst the sharp dagger is about tests and measurements. It is therefore surprising that within classical theory the dagger can actually be constructed from the cups \& caps as will be explicitly shown in the following definition.}.
Formally, we will assume this condition for every testable family of states, which in diagrammatic terms means Eq.~\eqref{eq:sharp4} below. Moreover, if $S$ is maximal, then $S^\dagger$ should be causal, as it is a test where no other outcomes are possible.

Also, while in quantum theory testing is a non-trivial structure, due to the invasive nature of measurements, in classical theory it is trivial, so we expect the dagger not to add any new structure on classical processes, and hence to be a symmetry which is already present due to the cups \& caps.

\begin{definition}[Sharp dagger]\label{def:SharpDagger}
A dagger is \em sharp \em if for all testable (cf.~Def.~\ref{def:testability}) causal state preparations $S$, its dagger $S^\dagger$ \em tests \em it, i.e.:
\beq\label{eq:sharp4}
\InputIfFileExists{Diagrams/sharp4.tikz}{}{\input{./figures/Diagrams/sharp4.tikz}}
\eeq
In addition, $S$ is  maximal
if and only if $S^\dagger$ is causal:
\[
\InputIfFileExists{Diagrams/CausalTest.tikz}{}{\input{./figures/Diagrams/CausalTest.tikz}}
\] 
For classical processes this sharp dagger should be trivial, that is, it is given by the compact structure:
\beq \label{eq:IdentityOnScalars}
\InputIfFileExists{Diagrams/classicalDagger.tikz}{}{\input{./figures/Diagrams/classicalDagger.tikz}}
\eeq
\end{definition}

\begin{example}[Quantum]\label{Ex:Dagger}
For quantum theory generally (including the fully quantum and classical cases) the sharp-dagger is provided by the Hermitian adjoint.
\end{example}
\begin{example}[Classical]
As mentioned in the definition, for classical theory a sharp dagger is provided by the compact structure, this can moreover be shown to be unique. It is instructive to consider what this looks like for the matrix representation of a classical process:
\[%
\InputIfFileExists{Diagrams/classicalDag1.tikz}{}{\input{./figures/Diagrams/classicalDag1.tikz}}\]
We therefore find that $(\mathsf{f}^\dagger)_i^j = \mathsf{f}_j^i$, that is, the dagger is just the matrix transpose.
\end{example} 

\begin{remark}
This `sharpening' of the dagger is essentially the same as the definition given in \cite{selby2016process}, the main distinction being that here we consider general state preparation procedures rather than just individual states.
\end{remark}

\section{The postulates}\label{sec:Postulates}

The first postulate provides the basic framework that we will use for the reconstruction. This allows one to describe essentially any conceivable physical theory, in particular, as we show later, encompassing the generalised probabilistic theory framework which has recently served as the basis for other reconstructions such as \cite{HardyAxiom,dakic2009quantum,Chiri2,hardy2011reformulating,Masanes,masanes2013existence,barnum2014higher,chiribella2016quantum}.

\begin{postulate}{1}[The theory is a process theory]\label{post:ProcessTheory} As defined in Def.~\ref{def:ProcessTheory}. \end{postulate}

The second postulate provides a more operational layer to our theory, describing how we interact with the world via classical inputs and outputs.

\begin{postulate}{2}[There is a finite local classical interface]\label{post:ClassicalInterface}  As defined in Def.~\ref{def:ClassicalInterface},  where all classically controlled processes exist, and there are sufficient causally-compatible local tomographic tests,  as defined in Defs.~\ref{def:CausCompTests} and \ref{def:LocalCI}.
\end{postulate}

The next two postulates are based on standard compositional tools from CQM.

\begin{postulate}{3}[The theory has cups \& caps]\label{def:compactStructure}
As defined in Def.~\ref{def:cupscaps}.
\end{postulate}

\begin{postulate}{4}[There is a sharp dagger]\label{post:SharpDagger}
As defined in Def.~\ref{def:SharpDagger}.
\end{postulate}

The final postulate will be defined and discussed in the following subsection. We also include it here for completeness.
\begin{postulate}{5}[Every process has an essentially unique symmetric purification]\label{post:SymmetricPurification}
As will be defined in Def.~\ref{def:SymmetricPurification}.
\end{postulate}

Note that there is some interdependencies amongst these postulates. For example, Postulates 2, 3, 4 \& 5 are all process-theoretic so rely on Postulate 1, moreover, Postulates 4 \& 5 rely on Postulate 2 as they are defined in terms of the discarding maps derived from Postulate 2, finally, Postulate 5 also relies on Postulate 4 as it uses the dagger in it's definition. 

\subsection{Symmetric purification}

The final postulate symmetrises the `standard' purification postulate introduced in \cite{Chiri1}. The standard purification postulate states: for all states $\rho$, there exists a pure bipartite state $\psi$, such that:
\[%
\begin{tikzpicture}
	\begin{pgfonlayer}{nodelayer}
		\node [style=point] (0) at (0, -0.25) {$\rho$};
		\node [style=none] (1) at (0, 1) {};
		\node [style=none] (2) at (2, -0) {$=$};
		\node [style=none] (3) at (3.5, -0) {};
		\node [style=none] (4) at (4.25, -0) {};
		\node [style=none] (5) at (5.25, -0) {};
		\node [style=none] (6) at (6, -0) {};
		\node [style=none] (7) at (4.75, -0.5) {$\psi$};
		\node [style=none] (8) at (4.75, -1) {};
		\node [style=none] (9) at (4.25, 1) {};
		\node [style=none] (10) at (5.25, 0.5) {};
		\node [style=upground] (11) at (5.25, 0.75) {};
	\end{pgfonlayer}
	\begin{pgfonlayer}{edgelayer}
		\draw [qWire](1.center) to (0);
		\draw (3.center) to (6.center);
		\draw (6.center) to (8.center);
		\draw (8.center) to (3.center);
		\draw [qWire](9.center) to (4.center);
		\draw [qWire](10.center) to (5.center);
	\end{pgfonlayer}
\end{tikzpicture}
} \ . \]
Moreover, this purification is `essentially unique' \cite{chiribella2016quantum}, that is, if there are two such purifications $\psi$ and $\phi$ with the same purifying system, then they are related via a reversible\footnote{A transformation $R$ is said to be reversible if an inverse $R^{-1}$ exists and is physically realisable.} transformation $R$:
\[%
\begin{tikzpicture}
	\begin{pgfonlayer}{nodelayer}
		\node [style=point] (0) at (1, -0.25) {$\rho$};
		\node [style=none] (1) at (1, 1) {};
		\node [style=none] (2) at (2.5, -0) {$=$};
		\node [style=none] (3) at (3.5, -0) {};
		\node [style=none] (4) at (4.25, -0) {};
		\node [style=none] (5) at (5.25, -0) {};
		\node [style=none] (6) at (6, -0) {};
		\node [style=none] (7) at (4.75, -0.5) {$\psi$};
		\node [style=none] (8) at (4.75, -1) {};
		\node [style=none] (9) at (4.25, 1) {};
		\node [style=none] (10) at (5.25, 0.5) {};
		\node [style=upground] (11) at (5.25, 0.75) {};
		\node [style=none] (12) at (-3.25, -0) {};
		\node [style=none] (13) at (-2.25, 0.5) {};
		\node [style=none] (14) at (-1.5, -0) {};
		\node [style=none] (15) at (-2.75, -0.5) {$\phi$};
		\node [style=none] (16) at (-2.25, -0) {};
		\node [style=none] (17) at (-0.5, -0) {$=$};
		\node [style=none] (18) at (-4, -0) {};
		\node [style=none] (19) at (-2.75, -1) {};
		\node [style=none] (20) at (-3.25, 1) {};
		\node [style=upground] (21) at (-2.25, 0.75) {};
		\node [style=none] (22) at (8, -0) {$\implies$};
		\node [style=none] (23) at (15.75, -0.5) {$\psi$};
		\node [style=none] (24) at (10.75, 1) {};
		\node [style=none] (25) at (10.75, -0) {};
		\node [style=none] (26) at (10, -0) {};
		\node [style=none] (27) at (15.25, 1.5) {};
		\node [style=none] (28) at (16.25, -0) {};
		\node [style=none] (29) at (11.75, -0) {};
		\node [style=none] (30) at (13.5, -0) {$=$};
		\node [style=none] (31) at (15.25, -0) {};
		\node [style=none] (32) at (17, -0) {};
		\node [style=none] (33) at (11.25, -1) {};
		\node [style=none] (34) at (14.5, -0) {};
		\node [style=none] (35) at (11.75, 1) {};
		\node [style=none] (36) at (11.25, -0.5) {$\phi$};
		\node [style=none] (37) at (12.5, -0) {};
		\node [style=none] (38) at (15.75, -1) {};
		\node [style=small box] (39) at (16.25, 0.75) {$R$};
		\node [style=none] (40) at (16.25, 1.5) {};
	\end{pgfonlayer}
	\begin{pgfonlayer}{edgelayer}
		\draw [qWire](1.center) to (0);
		\draw (3.center) to (6.center);
		\draw (6.center) to (8.center);
		\draw (8.center) to (3.center);
		\draw [qWire](9.center) to (4.center);
		\draw [qWire](10.center) to (5.center);
		\draw (18.center) to (14.center);
		\draw (14.center) to (19.center);
		\draw (19.center) to (18.center);
		\draw [qWire](20.center) to (12.center);
		\draw [qWire](13.center) to (16.center);
		\draw (34.center) to (32.center);
		\draw (32.center) to (38.center);
		\draw (38.center) to (34.center);
		\draw [qWire](27.center) to (31.center);
		\draw (26.center) to (37.center);
		\draw (37.center) to (33.center);
		\draw (33.center) to (26.center);
		\draw [qWire](24.center) to (25.center);
		\draw [qWire](35.center) to (29.center);
		\draw [qWire](40.center) to (39.center);
		\draw [qWire](39.center) to (28.center);
	\end{pgfonlayer}
\end{tikzpicture}
}.\]

This notion of purification is problematic for us for two reasons. Firstly, it is not compatible with the classical interface, as classical theory does \emph{not} satisfy this postulate. Secondly, it is formulated specifically in terms of states: we aim to treat all processes on an equal footing, viewing them as more fundamental entities, states being special instances thereof.  In this section we therefore introduce the postulate of \emph{symmetric purification}, which resolves these issues. Symmetric purification stipulates that every process arises from a pure process by `discarding' a system to both `the future and the past'.

To produce a time-symmetric version of purification we need to have a notion of `discarding' systems in the past. We can think of the standard discarding effect as an operational way to describe a scenario where we have no knowledge about, control over, or interaction with the future of a system. However, we can also imagine a scenario where we have no information about, no control over, and no interaction with the \emph{past} of a system, to represent this we use the time reverse (i.e.\ dagger) of the discarding (to the future) map:
\beq \label{def:MaxMixState}
\begin{tikzpicture}
	\begin{pgfonlayer}{nodelayer}
		\node [style=downground] (0) at (0, -0.5) {};
		\node [style=none] (1) at (0, 0.5) {};
		\node [style={right label}] (2) at (0, .25) {$A$};
	\end{pgfonlayer}
	\begin{pgfonlayer}{edgelayer}
		\draw [qWire](1.center) to (0);
	\end{pgfonlayer}
\end{tikzpicture}
}\ \ := \ \ {\left(%
\begin{tikzpicture}
	\begin{pgfonlayer}{nodelayer}
		\node [style=upground] (0) at (0, .5) {};
		\node [style=none] (1) at (0, -.50) {};
		\node [style={right label}] (2) at (0, -.50) {$A$};
	\end{pgfonlayer}
	\begin{pgfonlayer}{edgelayer}
		\draw[qWire] (1.center) to (0);
	\end{pgfonlayer}
\end{tikzpicture}
}\right)}^\dagger
\eeq
In the cases of quantum and classical theory, the dagger of the discarding map is an unnormalised maximally mixed state (specifically $\mathds{1}$).  More explicitly, using the matrix representation of classical theory and recalling that the classical dagger is nothing but the transpose we find that:
\[%
\begin{tikzpicture}
	\begin{pgfonlayer}{nodelayer}
		\node [style=upground] (0) at (0, 0.5) {};
		\node [style=none] (1) at (0, -0.5) {};
		\node [style={right label}] (2) at (0, -0.5) {};
	\end{pgfonlayer}
	\begin{pgfonlayer}{edgelayer}
		\draw [cWire] (1.center) to (0);
	\end{pgfonlayer}
\end{tikzpicture}} = \sum_i %
\begin{tikzpicture}
	\begin{pgfonlayer}{nodelayer}
		\node [style=copoint] (0) at (0, 0) {$i$};
		\node [style=none] (1) at (0, -1) {};
	\end{pgfonlayer}
	\begin{pgfonlayer}{edgelayer}
		\draw [style=cWire] (1.center) to (0);
	\end{pgfonlayer}
\end{tikzpicture}}\quad\text{and so}\quad %
\begin{tikzpicture}
	\begin{pgfonlayer}{nodelayer}
		\node [style=downground] (0) at (0, -0.5) {};
		\node [style=none] (1) at (0, 0.5) {};
		\node [style={right label}] (2) at (0, 0.25) {};
	\end{pgfonlayer}
	\begin{pgfonlayer}{edgelayer}
		\draw [cWire] (1.center) to (0);
	\end{pgfonlayer}
\end{tikzpicture}}=\sum_i %
\begin{tikzpicture}
	\begin{pgfonlayer}{nodelayer}
		\node [style=point] (0) at (0, -0) {$i$};
		\node [style=none] (1) at (0, 1) {};
	\end{pgfonlayer}
	\begin{pgfonlayer}{edgelayer}
		\draw [style=cWire] (1.center) to (0);
	\end{pgfonlayer}
\end{tikzpicture}} \]

One may worry that this is not actually a valid state (i.e.\ is not normalised/causal), however, discarding to the past is not something that we can `do in the lab', and so it should not correspond to a state we can actually prepare.

\begin{definition}[Essentially unique symmetric purifications]\label{def:SymmetricPurification} If a process $f:A \to B $ can be dilated to a \emph{pure} process $F:A\otimes B \to B \otimes A$ as follows:
\[
\InputIfFileExists{Diagrams/purification1.tikz}{}{\input{./figures/Diagrams/purification1.tikz}}
\]
then we call $F$ a \emph{symmetric purification} of $f$. Moreover, such purifications are said to be \emph{essentially unique}, if any two purifications $F,G:A\otimes C\to B\otimes D$ are connected by a  vise $R$:
\beq\label{eq:ReversiblyConnectedDefinition}
\InputIfFileExists{Diagrams/reversiblyConnectedDefinition.tikz}{}{\input{./figures/Diagrams/reversiblyConnectedDefinition.tikz}}\:,
\eeq
where the `backwards leak' is provided by the time-reverse (i.e.\ dagger) of a leak, and, moreover, $R$ must be `bi-causal':
\beq\label{eq:reversiblyConnectedConstraint} %
\InputIfFileExists{Diagrams/causallyConnected.tikz}{}{\input{./figures/Diagrams/causallyConnected.tikz}}.\eeq
\end{definition}

Symmetric purification expresses the requirement that all processes of the theory are fundamentally pure, and the apparent  lack of purity should arise from lacking information about, and control over, the past and/or future of some environmental systems.  Moreover, the `essential uniqueness' condition states that the purification is unique -- any apparent distinction can always be traced back to a change in the description of the environment.

As it is formulated,  essentially unique  symmetric purification provides a common ground to explain the lack of purity in all systems in the process-theoretic description of quantum theory, whether they be fully quantum, fully classical or more general hybrid and composite systems.
For a proof of this fact see App~\ref{app:symmetricPurification}.

\begin{remark}
Quantum theory actually satisfies a stronger form of symmetric purification where we can demand also that $F=F^\dagger$, i.e.\ the purified process is
invariant under time-reversal.
This strengthening however, whilst an interesting feature of quantum theory, is not necessary for our reconstruction.
\end{remark}

\section{The reconstruction}\label{sec:Reconstruction}

\begin{figure}[!ht]
 \centering
 \includegraphics[width=0.93\linewidth]{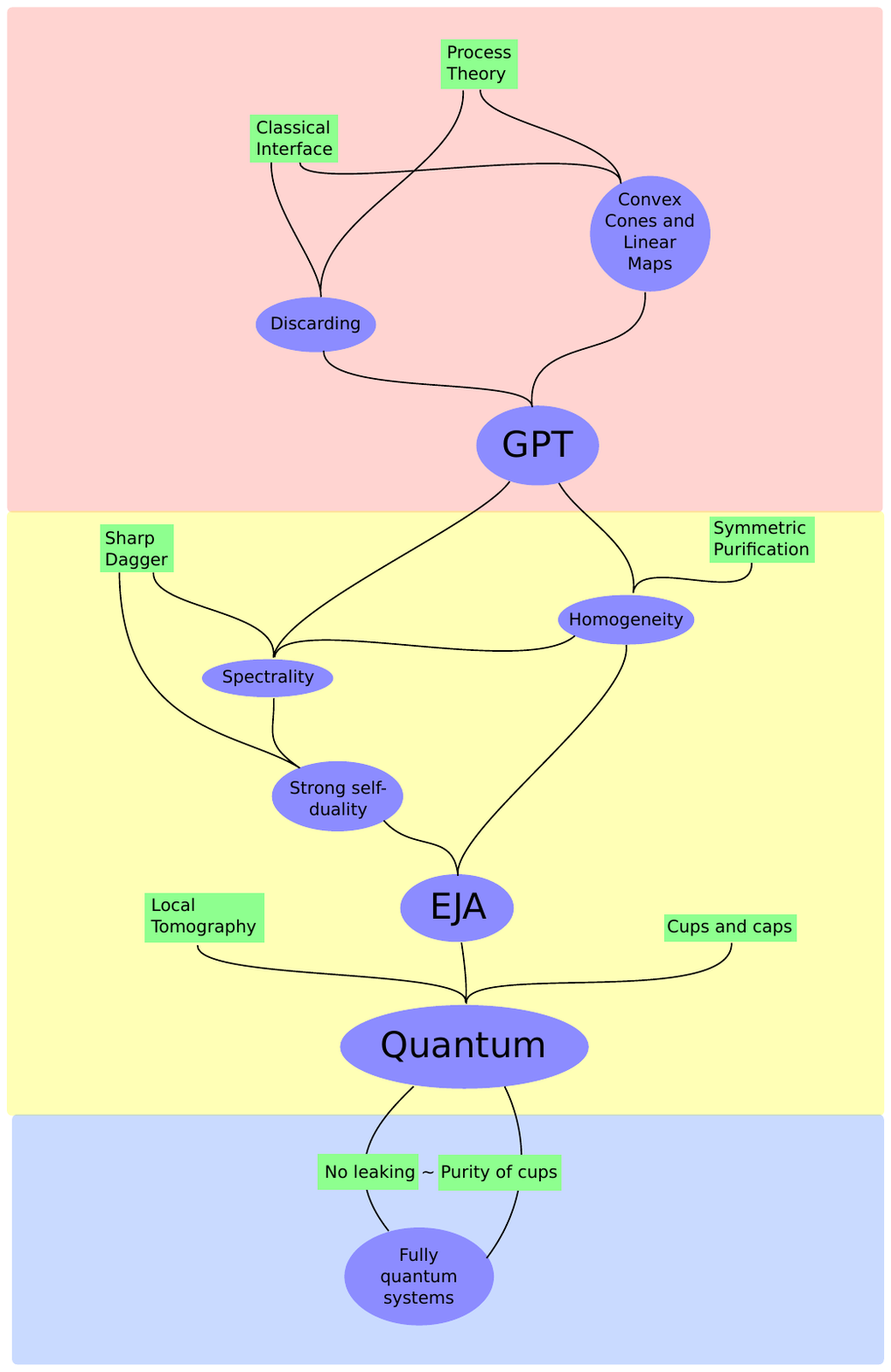}
\caption{\label{flowchart}Flowchart outlining the structure of the reconstruction where the green rectangles correspond to postulates and the blue ellipses to the lemmas and theorems constituting the proof. In the top section we obtain something similar to the framework of generalised probabilistic theory. In the middle section we reconstruct the process-theoretic description of quantum theory consisting of hybrid quantum-classical systems. Finally, in the bottom section we give two equivalent routes to restrict to the fully quantum systems.}
 \end{figure}

A high-level overview of the structure of the reconstruction is provided in Fig.~\ref{flowchart}, showing which postulates are necessary for each lemma, and how these are combined to reconstruct quantum theory.

 To begin let us consider how the sharp dagger interacts with the linear structure coming from the classical interface (see Prop.~\ref{lem:ConvexCones}).

\begin{lemma} In a theory with a classical interface (Post.~\ref{post:ClassicalInterface}) and a sum defined as in Prop.~\ref{lem:ConvexCones}, the dagger is linear.
\end{lemma}
\proof
First recall how we  define the sum:
\[%
\InputIfFileExists{LinearDagger/1.tikz}{}{\input{./figures/LinearDagger/1.tikz}} := %
\InputIfFileExists{LinearDagger/2.tikz}{}{\input{./figures/LinearDagger/2.tikz}} \quad \text{where, for all }i \quad %
\InputIfFileExists{LinearDagger/9.tikz}{}{\input{./figures/LinearDagger/9.tikz}} = %
\begin{tikzpicture}
	\begin{pgfonlayer}{nodelayer}
		\node [style=map] (0) at (0, -0) {$f_i$};
		\node [style=none] (1) at (0, 1) {};
		\node [style=none] (2) at (0, -1) {};
	\end{pgfonlayer}
	\begin{pgfonlayer}{edgelayer}
		\draw [style=qWire] (2.center) to (0);
		\draw [style=qWire] (0) to (1.center);
	\end{pgfonlayer}
\end{tikzpicture}} \]
We can consider how the dagger interacts with this sum:
\begin{align}\nonumber \dagger\left(%
\InputIfFileExists{LinearDagger/1.tikz}{}{\input{./figures/LinearDagger/1.tikz}}\right)\ \ &=\ \ \dagger\left(%
\InputIfFileExists{LinearDagger/2.tikz}{}{\input{./figures/LinearDagger/2.tikz}}\right)\ \ =\ \ %
\InputIfFileExists{LinearDagger/3.tikz}{}{\input{./figures/LinearDagger/3.tikz}}\ \ = \ \ %
\InputIfFileExists{LinearDagger/4.tikz}{}{\input{./figures/LinearDagger/4.tikz}}\\  \nonumber
&=\ \ %
\InputIfFileExists{LinearDagger/5.tikz}{}{\input{./figures/LinearDagger/5.tikz}}\ \ =\ \ %
\InputIfFileExists{LinearDagger/6.tikz}{}{\input{./figures/LinearDagger/6.tikz}}\ \ =\ \ %
\InputIfFileExists{LinearDagger/7.tikz}{}{\input{./figures/LinearDagger/7.tikz}}\\
 &=\ \ \sum_i\dagger\left(%
\begin{tikzpicture}
	\begin{pgfonlayer}{nodelayer}
		\node [style=map] (0) at (0, -0) {$f_i$};
		\node [style=none] (1) at (0, 1) {};
		\node [style=none] (2) at (0, -1) {};
	\end{pgfonlayer}
	\begin{pgfonlayer}{edgelayer}
		\draw [style=qWire] (2.center) to (0);
		\draw [style=qWire] (0) to (1.center);
	\end{pgfonlayer}
\end{tikzpicture}}\right)
\end{align}
This proves that the dagger is \emph{additive}, to extend this to full linearity we can show that:
\begin{align}
\dagger\left(\begin{tikzpicture}
	\begin{pgfonlayer}{nodelayer}
		\node [style=none] (0) at (-1.75, -0) {$\displaystyle\sum_i r_i$};
		\node [style=map] (1) at (0, -0) {$f_i$};
		\node [style=none] (2) at (0, 1) {};
		\node [style=none] (3) at (0, -1) {};
	\end{pgfonlayer}
	\begin{pgfonlayer}{edgelayer}
		\draw [style=qWire] (3.center) to (1);
		\draw [style=qWire] (1) to (2.center);
	\end{pgfonlayer}
\end{tikzpicture}\right) \ \
&= \ \  \sum_i\dagger\left(r_i\ \ \begin{tikzpicture}
	\begin{pgfonlayer}{nodelayer}
		\node [style=map] (0) at (0, -0) {$f_i$};
		\node [style=none] (1) at (0, 1) {};
		\node [style=none] (2) at (0, -1) {};
	\end{pgfonlayer}
	\begin{pgfonlayer}{edgelayer}
		\draw [style=qWire] (2.center) to (0);
		\draw [style=qWire] (0) to (1.center);
	\end{pgfonlayer}
\end{tikzpicture}\right) \\
&= \ \ \sum_i \dagger(r_i) \dagger\left(%
\begin{tikzpicture}
	\begin{pgfonlayer}{nodelayer}
		\node [style=map] (0) at (0, -0) {$f_i$};
		\node [style=none] (1) at (0, 1) {};
		\node [style=none] (2) at (0, -1) {};
	\end{pgfonlayer}
	\begin{pgfonlayer}{edgelayer}
		\draw [style=qWire] (2.center) to (0);
		\draw [style=qWire] (0) to (1.center);
	\end{pgfonlayer}
\end{tikzpicture}}\right) \\
&= \ \ \sum_i r_i \dagger\left(%
\begin{tikzpicture}
	\begin{pgfonlayer}{nodelayer}
		\node [style=map] (0) at (0, -0) {$f_i$};
		\node [style=none] (1) at (0, 1) {};
		\node [style=none] (2) at (0, -1) {};
	\end{pgfonlayer}
	\begin{pgfonlayer}{edgelayer}
		\draw [style=qWire] (2.center) to (0);
		\draw [style=qWire] (0) to (1.center);
	\end{pgfonlayer}
\end{tikzpicture}}\right)
\end{align}
where the first line follows from additivity of the dagger, the second from diagram preservation of the dagger, and the third from the fact that scalars are left invariant by the dagger.
\endproof

We now consider how various key features of the quantum state space arise from our axioms. To begin with, we show that the state cones are homogeneous.

\begin{definition}[Homogeneous cone]\label{def:Homogeneous}
A convex cone $C$ is homogeneous if for every pair of vectors $s_1$, $s_2$ internal to $C$, there exists a cone automorphism $T$ such that $T(s_1)=s_2$.
\end{definition}

\begin{lemma}\label{lem:Homogeneity}
If, in addition to the finite classical interface (Post.~\ref{post:ClassicalInterface}), the theory satisfies \emph{symmetric purification} (Post.~\ref{post:SymmetricPurification}),  and has a \emph{sharp-dagger} (Post.~\ref{post:SharpDagger}), then the state cone is \emph{homogeneous} (Def.~\ref{def:Homogeneous}).
\end{lemma}
\proof
We must show that, for any pair of internal states $s_1$ and $s_2$, there exists a cone automorphism $T$ such that $T(s_1)=s_2$. As $T$ is reversible, this is equivalent to the statement that there is a cone automorphism between a particular chosen internal state and any other. For this proof we take the particular internal state to be the `discarding' state.  That the discarding state is internal follows from the fact that the dagger is linear, provides an isomorphism between the state and effect cones, and that the discarding effect is in the interior of the effect cone. This last point follows from the existence of causally compatible tomographic tests.  We proceed in two steps: firstly, we show that  for any internal state there exists a process that maps the discarding state to it;  secondly, that this map is surjective  on the cone and so ---being linear by Prop.~\ref{lem:ConvexCones}, and as the cones are finite-dimensional---  it is a cone automorphism.

The first part is a simple corollary of symmetric purification. Consider an arbitrary internal state $s$, then its purification $S$ provides a map from the  states of $A$ to itself taking the  discarding state to $s$:
\[%
\InputIfFileExists{Diagrams/statePurification.tikz}{}{\input{./figures/Diagrams/statePurification.tikz}}\]

For the second part, we adapt \cite[Proposition 7]{chiribella2016quantum} to show that $S$ is surjective  on the cone.Note that as $s$ is internal, then any state $a$ is in some decomposition of $s$:
\[%
\InputIfFileExists{Diagrams/internalState.tikz}{}{\input{./figures/Diagrams/internalState.tikz}}\]
where $p\in(0,1)$. Therefore we can construct the following dilation $\sigma$ of $s$  where $%
\begin{tikzpicture}
	\begin{pgfonlayer}{nodelayer}
		\node [style=point] (0) at (0, -0) {$0$};
		\node [style=none] (1) at (0, 0.75) {};
	\end{pgfonlayer}
	\begin{pgfonlayer}{edgelayer}
		\draw [qWire](1.center) to (0);
	\end{pgfonlayer}
\end{tikzpicture}
}$ and $%
\begin{tikzpicture}
	\begin{pgfonlayer}{nodelayer}
		\node [style=point] (0) at (0, -0) {$1$};
		\node [style=none] (1) at (0, 0.75) {};
	\end{pgfonlayer}
	\begin{pgfonlayer}{edgelayer}
		\draw [qWire](1.center) to (0);
	\end{pgfonlayer}
\end{tikzpicture}
}$ are causal, perfectly distinguishable states of some system $B$\footnote{Note that we can take $B$ to be a classical bit to ensure that such states exist.}:
\[%
\InputIfFileExists{Diagrams/internalStateDilation.tikz}{}{\input{./figures/Diagrams/internalStateDilation.tikz}}\]
This has a symmetric purification $\Sigma$, which is moreover a purification of $s$, hence we can construct two purifications of $s$ with the same input and output systems:
\[%
\InputIfFileExists{Diagrams/internalStatePurifications.tikz}{}{\input{./figures/Diagrams/internalStatePurifications.tikz}}\]
where $N_B:=%
\begin{tikzpicture}[scale=0.5]
	\begin{pgfonlayer}{nodelayer}
		\node [style=none] (0) at (0, 0.25) {};
		\node [style=none] (1) at (0, -0.5) {};
		\node [style=upground] (2) at (0, 0.5) {};
	\end{pgfonlayer}
	\begin{pgfonlayer}{edgelayer}
		\draw (0.center) to (1.center);
	\end{pgfonlayer}
\end{tikzpicture}}\circ %
\begin{tikzpicture}[scale=0.5]
	\begin{pgfonlayer}{nodelayer}
		\node [style=none] (0) at (0, -0.25) {};
		\node [style=none] (1) at (0, 0.5) {};
		\node [style=downground] (2) at (0, -0.5) {};
		\node [style=right label] (3) at (0, 0.2500001) {$B$};
	\end{pgfonlayer}
	\begin{pgfonlayer}{edgelayer}
		\draw (0.center) to (1.center);
	\end{pgfonlayer}
\end{tikzpicture}}$  is non-zero as the discarding map and its dagger are both internal to their associated convex cones (see also the end of App.~\ref{proof:GPTStruct} for an explicit proof that $N_B$ is non-zero).  Then, by the definition of symmetric purification, these are related by some $R$ as:
\[%
\InputIfFileExists{Diagrams/internalStatePurificationsConnected.tikz}{}{\input{./figures/Diagrams/internalStatePurificationsConnected.tikz}}\]
In the second step we use that $S$ is pure and so, by the definition of pure processes, we can replace the leak afterwards with a leak before. It is then clear that there is a state, $\alpha$, that is mapped to $p a$ by $S$:
\[%
\InputIfFileExists{Diagrams/surjective.tikz}{}{\input{./figures/Diagrams/surjective.tikz}}\]
Hence, $\frac{1}{p}\alpha$ is mapped to $a$. As $a$ is an arbitrary state, $S$ is surjective  on the cone,and hence a cone automorphism. We therefore have homogeneity of the state cone.
\endproof

Secondly, we show that state spaces satisfy spectrality.
\begin{definition}[Spectrality]\label{def:Spectral}
A theory is said to be spectral if any (causal) state can be written as
\[%
\InputIfFileExists{Diagrams/spectrality.tikz}{}{\input{./figures/Diagrams/spectrality.tikz}}\]
where $p$ is a (causal) classical state and $S$ is a maximal testable state preparation (Def.~\ref{def:testability}).
\end{definition}
\begin{lemma}\label{Lem:Spectral}
If, in addition to the finite classical interface (Post.~\ref{post:ClassicalInterface}) and the Homogeneity from Lem.~\ref{lem:Homogeneity}  (from Posts.~\ref{post:ClassicalInterface} and \ref{post:SymmetricPurification}),  the theory has a \emph{sharp-dagger} (Post.~\ref{post:SharpDagger}) then the state cone is \emph{spectral} (Def.~\ref{def:Spectral}).
\end{lemma}
\proof
Firstly note that, for any reversible transformation $T$ that:
\[ %
\InputIfFileExists{Diagrams/sharp2v3.tikz}{}{\input{./figures/Diagrams/sharp2v3.tikz}} \ \ \text{pure} \quad\implies\quad  %
\InputIfFileExists{Diagrams/reversiblePreservesPurity.tikz}{}{\input{./figures/Diagrams/reversiblePreservesPurity.tikz}}\ \ \ \text{pure.}\]
To see this consider an arbitrary dilation, $D$, of this process:
\[%
\InputIfFileExists{Diagrams/reversibleProof1.tikz}{}{\input{./figures/Diagrams/reversibleProof1.tikz}}\]
Any such dilation will also provide  a dilation of the original process by composing it with $T^{-1}$, that is:
\[%
\InputIfFileExists{Diagrams/newSpectralityProofDiagram.tikz}{}{\input{./figures/Diagrams/newSpectralityProofDiagram.tikz}}\]
As this original process is, by assumption, pure we therefore have:
\[%
\InputIfFileExists{Diagrams/reversibleProof2.tikz}{}{\input{./figures/Diagrams/reversibleProof2.tikz}}\]
 where the first equality  follows from the definition of purity (Def. \ref{def:pureeq}) and the fact that all classical leaks can be written in terms of the broadcasting map (see \cite{selby2017leaks} for details). The second equality follows from associativity of broadcasting.
Hence, by composing this with $T$ we find that this arbitrary dilation $D$, can be rewritten as a leak before or a leak after the dilated process, i.e.:
\[%
\InputIfFileExists{Diagrams/newSpectralityProofDiagram2.tikz}{}{\input{./figures/Diagrams/newSpectralityProofDiagram2.tikz}}\]
Therefore, the conditions necessary such that the process is pure are satisfied.

Now to obtain spectrality, note that every system $A$ must have a (generally non-unique) maximal testable state preparation $S:n\to A$ (Def.~\ref{def:testability}) although it could be trivial (i.e.\ $n=0$). For such a maximal testable state preparation, by the definition of the sharp dagger (Def.~\ref{def:SharpDagger}), we have:
\[%
\InputIfFileExists{Diagrams/spectralityProof0.tikz}{}{\input{./figures/Diagrams/spectralityProof0.tikz}}\]
and so taking the dagger of this equation we find that:
\[%
\InputIfFileExists{Diagrams/spectralityProof1.tikz}{}{\input{./figures/Diagrams/spectralityProof1.tikz}}\]
Using this together with homogeneity, means that for any internal state $t$, there exists a reversible map $T$ such that:
\[%
\InputIfFileExists{Diagrams/SpectralNew1.tikz}{}{\input{./figures/Diagrams/SpectralNew1.tikz}}\]
We then define a classical state and effect, $p$ and $\bar{p}$, by:
\[%
\InputIfFileExists{Diagrams/spectralityProof3.tikz}{}{\input{./figures/Diagrams/spectralityProof3.tikz}} \qquad \text{and}\qquad %
\InputIfFileExists{Diagrams/spectralityProof4.tikz}{}{\input{./figures/Diagrams/spectralityProof4.tikz}}\qquad \text{respectively.}\]
 Note that, $\bar{p}$ necessarily exists as $p$ is an internal state. To see that $p$ is indeed internal note that this is equivalent to the statement that:
\[\forall i \quad %
\begin{tikzpicture}
	\begin{pgfonlayer}{nodelayer}
		\node [style=copoint] (0) at (-1, 0.75) {$i$};
		\node [style=point] (1) at (-1, -0.25) {$p$};
	\end{pgfonlayer}
	\begin{pgfonlayer}{edgelayer}
		\draw [style=cWire] (0) to (1);
	\end{pgfonlayer}
\end{tikzpicture}} \ \neq \ 0\] 
Now, we can compute these scalars to be:
\[%
\InputIfFileExists{Diagrams/SpecNew2.tikz}{}{\input{./figures/Diagrams/SpecNew2.tikz}}\]
where the first equality follows from the definition of $p$, the second from the definition of the classical cup, and the third from the fact that $S$ is a controlled state preparation. Note that this means that the $s_i$ are valid states and so live inside the state cone. Moreover, note that $T$ is a cone automorphism (see Def.~\ref{def:Homogeneous}) so must map these states to valid non-zero states, and as the hyperplane defined by $%
\begin{tikzpicture}[scale=0.5]
	\begin{pgfonlayer}{nodelayer}
		\node [style=none] (0) at (0, 0.25) {};
		\node [style=none] (1) at (0, -0.5) {};
		\node [style=upground] (2) at (0, 0.5) {};
	\end{pgfonlayer}
	\begin{pgfonlayer}{edgelayer}
		\draw (0.center) to (1.center);
	\end{pgfonlayer}
\end{tikzpicture}}\circ s=1$ intersects the cone (see Prop.~\ref{lem:ConvexCones}) we find, for all $i$, that:
\[ %
\begin{tikzpicture}
	\begin{pgfonlayer}{nodelayer}
		\node [style=map] (0) at (0, -0) {$T$};
		\node [style=none] (1) at (0, -1.5) {};
		\node [style=point] (2) at (0, -1.5) {$s_i$};
		\node [style=none] (3) at (0, 1) {};
	\end{pgfonlayer}
	\begin{pgfonlayer}{edgelayer}
		\draw [qWire] (3.center) to (0);
		\draw [qWire] (0) to (2);
	\end{pgfonlayer}
\end{tikzpicture}} \ \neq \ 0 \quad \implies\quad%
\InputIfFileExists{Diagrams/SpecNew3.tikz}{}{\input{./figures/Diagrams/SpecNew3.tikz}} \ \neq \ 0 \quad \implies\quad  %
\begin{tikzpicture}
	\begin{pgfonlayer}{nodelayer}
		\node [style=copoint] (0) at (-1, 0.75) {$i$};
		\node [style=point] (1) at (-1, -0.25) {$p$};
	\end{pgfonlayer}
	\begin{pgfonlayer}{edgelayer}
		\draw [style=cWire] (0) to (1);
	\end{pgfonlayer}
\end{tikzpicture}} \ \neq \ 0\]
Returning to our state of interest $t$ we can now write it in the following form:
\[%
\InputIfFileExists{Diagrams/spectralityProof2.tikz}{}{\input{./figures/Diagrams/spectralityProof2.tikz}}\]
We can then define a perfectly distinguishing measurement by:
\[%
\InputIfFileExists{Diagrams/spectralityProof5.tikz}{}{\input{./figures/Diagrams/spectralityProof5.tikz}}\]
One may worry that $T^{-1}$ is not guaranteed to be a physical transformation, however, regardless, this measurement is well defined as each of the effects that make it up must be physical (as $T$ induces an automorphism on the effect cone so does $T^{-1}$). Hence, this measurement can then be defined as a classically controlled process. It is then simple to check that the pair $(\tau, M)$ satisfy the conditions for the sharp dagger. Firstly that $\tau$ is causal:
\[%
\InputIfFileExists{Diagrams/SpectralNew2.tikz}{}{\input{./figures/Diagrams/SpectralNew2.tikz}}\]
and secondly that $M$ perfectly distinguishes the states of $\tau$:
\[%
\InputIfFileExists{Diagrams/SpectralNew3.tikz}{}{\input{./figures/Diagrams/SpectralNew3.tikz}}\]
Therefore the definition of the sharp dagger implies that $\tau$ is perfectly distinguished by $\tau^\dagger$:
\[%
\InputIfFileExists{Diagrams/spectralityProof6.tikz}{}{\input{./figures/Diagrams/spectralityProof6.tikz}}\]
Moreover $\tau$ is maximal as $S$ was maximal and so:
\[%
\InputIfFileExists{Diagrams/spectralityProof7.tikz}{}{\input{./figures/Diagrams/spectralityProof7.tikz}}\]
Therefore,
\[%
\InputIfFileExists{Diagrams/spectralityProof8.tikz}{}{\input{./figures/Diagrams/spectralityProof8.tikz}}\]
satisfies the conditions of spectrality.

So far we have proved spectrality of the interior of the cone, we will now show that this can be extended to the full cone, and moreover, the entire vector space in which the cone lives. To do so we can adapt Corollary $21$ from \cite{Chiri2}. Note that any vector can be written as the difference of two internal states which can each be spectrally decomposed:
 \[%
\InputIfFileExists{Diagrams/extendingSpectrality1.tikz}{}{\input{./figures/Diagrams/extendingSpectrality1.tikz}}\]
Then define,
\[R:=\max_i\left\{%
\begin{tikzpicture}
	\begin{pgfonlayer}{nodelayer}
		\node [style=copoint] (0) at (0, 0.5) {$i$};
		\node [style=point] (1) at (0, -0.5) {$r_2$};
	\end{pgfonlayer}
	\begin{pgfonlayer}{edgelayer}
		\draw [style=cWire] (0) to (1);
	\end{pgfonlayer}
\end{tikzpicture}
}\right\}+\epsilon,\]
where $\epsilon>0$. Then
\[%
\InputIfFileExists{Diagrams/extendingSpectrality3.tikz}{}{\input{./figures/Diagrams/extendingSpectrality3.tikz}}\]
is a state as all of the elements of the classical vectors are strictly positive thanks to the definition of $R$, and moreover, is internal as there is a decomposition of the state as $s+\epsilon%
\begin{tikzpicture}[scale=0.5]
	\begin{pgfonlayer}{nodelayer}
		\node [style=none] (0) at (0, -0.25) {};
		\node [style=none] (1) at (0, 0.5) {};
		\node [style=downground] (2) at (0, -0.5) {};
	\end{pgfonlayer}
	\begin{pgfonlayer}{edgelayer}
		\draw (0.center) to (1.center);
	\end{pgfonlayer}
\end{tikzpicture}
}$ where $s$ is a vector in the cone. Therefore, as this is an internal state it has a spectral decomposition:
\[%
\InputIfFileExists{Diagrams/extendingSpectrality4.tikz}{}{\input{./figures/Diagrams/extendingSpectrality4.tikz}}\]
and so we can write:
\[%
\InputIfFileExists{Diagrams/extendingSpectrality5.tikz}{}{\input{./figures/Diagrams/extendingSpectrality5.tikz}}\]
We therefore have spectral decompositions for arbitrary vectors.
\endproof

Finally, we show that the state cones are strongly self dual.
\begin{definition}[Strongly self dual cone]\label{def:SelfDual}
A convex cone $C$ is strongly self dual if there exists an inner product $\left<\ ,\ \right>$ on the vector space spanned by $C$,  such that
\[x\in C \quad \iff \quad \left<x,c\right>\geq 0\ \ \forall\ c \in C.\]
\end{definition}
\begin{lemma}\label{lem:selfDual}
If, in addition to the finite classical interface (Post.~\ref{post:ClassicalInterface}) and Spectrality from Lem.~\ref{Lem:Spectral}  (from Posts.~\ref{post:ClassicalInterface},\ref{post:SymmetricPurification} and \ref{post:SharpDagger}),  the theory has a \emph{sharp-dagger} (Post.~\ref{post:SharpDagger}), then the state cone is \emph{strongly self dual} (Def.~\ref{def:SelfDual}).
\end{lemma}
\proof
First we will show that the sharp-dagger provides an inner product defined as:
\beq \label{eq:DaggerInnerProduct}
\langle s_1,s_2 \rangle \ \ := \ \ %
\begin{tikzpicture}
	\begin{pgfonlayer}{nodelayer}
		\node [style=copoint] (0) at (0, 0.5) {$s_1$};
		\node [style=point] (1) at (0, -0.5) {$s_2$};
	\end{pgfonlayer}
	\begin{pgfonlayer}{edgelayer}
		\draw [qWire](0) to (1);
	\end{pgfonlayer}
\end{tikzpicture}
};
\eeq
and secondly we show that the state cone is strongly self dual with respect to this inner product. To show that this is a valid inner product, firstly, we check that this is symmetric:
\[\langle s_1,s_2 \rangle \ \ = \ \ %
\begin{tikzpicture}
	\begin{pgfonlayer}{nodelayer}
		\node [style=copoint] (0) at (0, 0.5) {$s_1$};
		\node [style=point] (1) at (0, -0.5) {$s_2$};
	\end{pgfonlayer}
	\begin{pgfonlayer}{edgelayer}
		\draw [qWire](0) to (1);
	\end{pgfonlayer}
\end{tikzpicture}
}  \ \ \stackrel{(\ref{eq:IdentityOnScalars})}{=} \ \  \dagger\left(%
\begin{tikzpicture}
	\begin{pgfonlayer}{nodelayer}
		\node [style=copoint] (0) at (0, 0.5) {$s_1$};
		\node [style=point] (1) at (0, -0.5) {$s_2$};
	\end{pgfonlayer}
	\begin{pgfonlayer}{edgelayer}
		\draw [qWire](0) to (1);
	\end{pgfonlayer}
\end{tikzpicture}
}\right)\ \ \stackrel{(\ref{eq:daggerReflection})}{=} \ \ %
\begin{tikzpicture}
	\begin{pgfonlayer}{nodelayer}
		\node [style=point] (0) at (0, -0.5) {$s_1$};
		\node [style=copoint] (1) at (0, 0.5) {$s_2$};
	\end{pgfonlayer}
	\begin{pgfonlayer}{edgelayer}
		\draw [qWire](0) to (1);
	\end{pgfonlayer}
\end{tikzpicture}
} \ \ =\ \ \langle s_2,s_1 \rangle  \]
Secondly that it is  bilinear follows immediately from linearity of effects given by Lem.~\ref{lem:LinearityAndConvexCones}:
\beq\label{eq:DaggerLinearity}\langle s_1, \alpha s_2 + \beta s_3 \rangle \ \ = \ \ %
\InputIfFileExists{Diagrams/innerProductLinear.tikz}{}{\input{./figures/Diagrams/innerProductLinear.tikz}} \ \ =\ \ \alpha \langle s_1, s_2 \rangle + \beta \langle s_1, s_3 \rangle\eeq
Finally, positivity follows easily from spectrality of arbitrary vectors:
\[\langle v,v\rangle\ \ = \ \ %
\InputIfFileExists{Diagrams/positivityInnerProduct.tikz}{}{\input{./figures/Diagrams/positivityInnerProduct.tikz}}\ \ \geq \ \ 0\]
where equality implies that $r=0$ and hence $v=0$.
Eq.~\ref{eq:DaggerInnerProduct} therefore defines a valid inner product.

Note that, if all elements of $r$ are strictly positive, we have an internal state; if they are non-negative, then we have a state; and if some are negative, then the vector cannot be a state as it would give a negative probability for some effect. It is then simple to check strong self-duality. Firstly, if $s$ is an element of the state cone $C$, then $\langle s,c\rangle \geq 0$ for all $c\in C$ as $\langle s,\cdot\rangle = s^\dagger$ is an effect, and so it evaluates to a positive real number on the cone of states. Conversely, if $v\not\in C$, there is a negative coefficient in the spectral decomposition; without loss of generality we label this element $i$. There then exists some $c\in C$ such that $\langle c,v\rangle <0$, that is:
\[\langle c, v\rangle\ \ =\ \ %
\InputIfFileExists{Diagrams/strongselfduality.tikz}{}{\input{./figures/Diagrams/strongselfduality.tikz}}\ \ <\ \ 0\]
The state cone is therefore strongly self dual with respect to the inner product defined by the sharp dagger.
\endproof

These properties, in particular homogeneity and strong self duality, are well known to get us close to quantum theory, specifically, by using the Koecher-Vinberg theorem \cite{koecher1958geodattischen,vinberg1960homogeneous,farautanalysis},  we get that the state cones correspond to Euclidean Jordan Algebras (EJAs).

\begin{theorem}[Koecher-Vinberg theorem] There is a one-to-one correspondence between Euclidean Jordan Algebras and symmetric cones (convex cones that are closed, pointed, homogeneous, and self-dual)  with  an appropriate choice of order unit \cite{farautanalysis}.
\end{theorem}

\begin{lemma}\label{lem:EJA}
The systems in our theory  (specifically, a process theory (Post.~\ref{post:ProcessTheory}) satisfying Posts.~\ref{post:ClassicalInterface}, \ref{post:SharpDagger} and \ref{post:SymmetricPurification}) correspond to finite-dimensional Euclidean Jordan Algebras.
\end{lemma}
\proof
By using the Koecher-Vinberg theorem above we need simply to demonstrate that our state cones are indeed symmetric cones, and hence also correspond to EJAs.

First note that given a  cone $C$ in a vector space $V$ we define the dual cone by \[C':=\{v | v\in V\ \text{s.t.}\ \langle v,u\rangle \geq 0\ \ \forall\  u\in C\}\]
This implies that the dual cone is closed as it is the intersection of closed half spaces, one for each $u\in C$ defined by $\langle v,u\rangle \geq 0$. Hence, strong self-duality implies that the state cone must be closed too. This therefore follows for our systems from Lem.~\ref{lem:selfDual}. Prop.~\ref{lem:ConvexCones} then implies that the cones are pointed, finite-dimensional and Lem.~\ref{lem:Homogeneity} that they are homogeneous. Hence the state spaces are finite-dimensional symmetric cones. It therefore immediately follows that each system in our theory corresponds to a finite-dimensional EJA.
\endproof

There is a well known classification result for finite dimensional EJAs \cite{jordan1993algebraic}: they correspond to direct sums of five types of simple EJAs. There are two important properties of each of these, their \emph{rank}, which corresponds to the number of states in a maximal testable state preparation, and their \emph{dimension}, the minimal number of effects necessary for state tomography.

\bit
\item[i.] $\mathds{C}_n$, the algebra of self-adjoint $n\times n$ complex matrices. These have rank $n$ and dimension $n^2$.
\item[ii.] $\mathds{R}_n$, the algebra of self-adjoint $n\times n$ real matrices. These have rank $n$ and dimension $\frac{n(n+1)}{2}$.
\item [iii.] $\mathds{H}_n$, the algebra of self-adjoint $n\times n$ quaternionic matrices. These have rank $n$ and dimension $n(2n-1)$.
\item[iv.] $\mathds{O}_3$, the algebra of self-adjoint $3\times 3$ octonionic matrices. This has rank $3$ and dimension $3^3$.
\item[v.] $\mathbf{Spin}_K$, the spin factors. These have rank $2$ and dimension $K$.
\eit
Note that for the spin factors in the case of $K=3$ coincides with $\mathds{R}_2$, $K=4$ with $\mathds{C}_2$ and $K=6$ with $\mathds{H}_2$. Given this classification, we are in a position to ask which of these EJAs is compatible with our compositional structure.  Before that, we will prove the following useful lemma which will be useful for the proof:
\begin{lemma}[Generalised no-restriction hypothesis]\label{lem:NRH}
 Cups \& caps  (Post.~\ref{def:compactStructure}) imply that our theories (satisfying Posts.~\ref{post:ProcessTheory}, \ref{post:ClassicalInterface}, \ref{post:SharpDagger} and \ref{post:SymmetricPurification}) satisfy a generalised version of the no-restriction hypothesis \cite{Chiri1}. If a process is logically possible, then it is physically possible.
\end{lemma}
\proof
A process $f$ is said to be logically possible, if for all states $\psi$, effects $\phi$, and auxillary systems $C$ it results in positive scalars:
\beq
\begin{tikzpicture}
	\begin{pgfonlayer}{nodelayer}
		\node [style=small box] (0) at (0, 0) {$f$};
		\node [style=none] (1) at (-0.75, 1.25) {};
		\node [style=none] (2) at (2.25, 1.25) {};
		\node [style=none] (3) at (0.75, 2.5) {};
		\node [style=none] (4) at (0, 1.25) {};
		\node [style=none] (5) at (1.5, 1.25) {};
		\node [style=none] (6) at (0.75, 1.75) {$\phi$};
		\node [style=none] (7) at (-0.75, -1.25) {};
		\node [style=none] (8) at (2.25, -1.25) {};
		\node [style=none] (9) at (0.75, -2.5) {};
		\node [style=none] (10) at (0, -1.25) {};
		\node [style=none] (11) at (1.5, -1.25) {};
		\node [style=none] (12) at (0.75, -1.75) {$\psi$};
		\node [style=right label] (13) at (0, -1) {$A$};
		\node [style=right label] (14) at (1.5, -0.75) {$C$};
		\node [style=right label] (15) at (0, 0.75) {$B$};
	\end{pgfonlayer}
	\begin{pgfonlayer}{edgelayer}
		\draw (1.center) to (3.center);
		\draw (3.center) to (2.center);
		\draw (2.center) to (1.center);
		\draw (7.center) to (9.center);
		\draw (9.center) to (8.center);
		\draw (8.center) to (7.center);
		\draw [qWire] (4.center) to (0);
		\draw [qWire] (0) to (10.center);
		\draw [qWire] (5.center) to (11.center);
	\end{pgfonlayer}
\end{tikzpicture}
\quad \geq 0
\eeq
Now, as this is true for all $\psi$, then strong-self duality implies that:
\beq\begin{tikzpicture}
	\begin{pgfonlayer}{nodelayer}
		\node [style=small box] (0) at (0, 0) {$f$};
		\node [style=none] (1) at (-0.75, 1.25) {};
		\node [style=none] (2) at (2.25, 1.25) {};
		\node [style=none] (3) at (0.75, 2.5) {};
		\node [style=none] (4) at (0, 1.25) {};
		\node [style=none] (5) at (1.5, 1.25) {};
		\node [style=none] (6) at (0.75, 1.75) {$\phi$};
		\node [style=none] (10) at (0, -1.25) {};
		\node [style=none] (11) at (1.5, -1.25) {};
		\node [style=right label] (13) at (0, -1) {$A$};
		\node [style=right label] (14) at (1.5, -0.75) {$C$};
		\node [style=right label] (15) at (0, 0.75) {$B$};
	\end{pgfonlayer}
	\begin{pgfonlayer}{edgelayer}
		\draw (1.center) to (3.center);
		\draw (3.center) to (2.center);
		\draw (2.center) to (1.center);
		\draw [qWire] (4.center) to (0);
		\draw [qWire] (0) to (10.center);
		\draw [qWire] (5.center) to (11.center);
	\end{pgfonlayer}
\end{tikzpicture}
\eeq
is a valid, physical, effect for any $\phi$ and $C$. In particular, this means that:
\beq
\begin{tikzpicture}
	\begin{pgfonlayer}{nodelayer}
		\node [style=small box] (0) at (0, 0) {$f$};
		\node [style=none] (4) at (0, 1.25) {};
		\node [style=none] (5) at (1.5, 1.25) {};
		\node [style=none] (10) at (0, -1.25) {};
		\node [style=none] (11) at (1.5, -1.25) {};
		\node [style=right label] (13) at (0, -1) {$A$};
		\node [style=right label] (14) at (1.5, -0.75) {$B$};
		\node [style=right label] (15) at (0, 0.75) {$B$};
	\end{pgfonlayer}
	\begin{pgfonlayer}{edgelayer}
		\draw [qWire] (4.center) to (0);
		\draw [qWire] (0) to (10.center);
		\draw [qWire] (5.center) to (11.center);
		\draw [qWire, bend left=90, looseness=2.00] (4.center) to (5.center);
	\end{pgfonlayer}
\end{tikzpicture}
\eeq
is a physical effect, and hence we can use cups \& caps to show that $f$ must also be physical:
\beq
\begin{tikzpicture}
	\begin{pgfonlayer}{nodelayer}
		\node [style=small box] (0) at (0, 0) {$f$};
		\node [style=none] (4) at (0, 1.25) {};
		\node [style=none] (5) at (1.5, 1.25) {};
		\node [style=none] (10) at (0, -2.5) {};
		\node [style=none] (11) at (1.5, -1.25) {};
		\node [style=right label] (13) at (0, -1.5) {$A$};
		\node [style=right label] (14) at (1.5, -0.75) {$B$};
		\node [style=right label] (15) at (0, 0.75) {$B$};
		\node [style=none] (18) at (1.5, -1.25) {};
		\node [style=none] (19) at (3, -1.25) {};
		\node [style=none] (21) at (3, 2.5) {};
		\node [style=right label] (23) at (3, 1.25) {$B$};
	\end{pgfonlayer}
	\begin{pgfonlayer}{edgelayer}
		\draw [qWire] (4.center) to (0);
		\draw [qWire] (0) to (10.center);
		\draw [qWire] (5.center) to (11.center);
		\draw [qWire, bend left=90, looseness=2.00] (4.center) to (5.center);
		\draw [qWire] (19.center) to (21.center);
		\draw [qWire, bend right=90, looseness=2.00] (18.center) to (19.center);
	\end{pgfonlayer}
\end{tikzpicture}
\quad = \quad
\begin{tikzpicture}
	\begin{pgfonlayer}{nodelayer}
		\node [style=small box] (0) at (0, 0) {$f$};
		\node [style=none] (4) at (0, 1.25) {};
		\node [style=none] (10) at (0, -1.25) {};
		\node [style=right label] (13) at (0, -1) {$A$};
		\node [style=right label] (15) at (0, 0.75) {$B$};
	\end{pgfonlayer}
	\begin{pgfonlayer}{edgelayer}
		\draw [qWire] (4.center) to (0);
		\draw [qWire] (0) to (10.center);
	\end{pgfonlayer}
\end{tikzpicture}
\eeq
\endproof

\begin{theorem}
Given that state cones correspond to EJAs  (Lem.~\ref{lem:EJA} from Posts.~ \ref{post:ProcessTheory}, ~\ref{post:ClassicalInterface}, \ref{post:SharpDagger} and \ref{post:SymmetricPurification}) and that the theory satisfies the generalised no restriction hypothesis (Lem.~\ref{lem:NRH} from Posts.~\ref{post:ProcessTheory}, \ref{post:ClassicalInterface},
\ref{def:compactStructure}, \ref{post:SharpDagger} and \ref{post:SymmetricPurification})), then local tomography (Post.~\ref{post:ClassicalInterface}, Def.~\ref{def:LocalCI}) and the existence of cups \& caps (Post.~\ref{def:compactStructure}) restrict us to quantum theory (Ex.~\ref{Ex:CStar}).
\end{theorem}
\proof

First note that $\otimes$ is bilinear, and so distributes over $\oplus$.  If we have some systems $A$ and $B$ which decompose into simple components as $A=\bigoplus_i A_i$  and $B=\bigoplus_j B_j$ then $A\otimes B = \bigoplus_{ij}A_i\otimes B_j$. This structure is most suitably captured by noting that for such systems the identity decomposes as the sum of orthogonal projectors, one for each of the simple components:
\beq
\begin{tikzpicture}
	\begin{pgfonlayer}{nodelayer}
		\node [style=none] (0) at (2.25, 0.5) {};
		\node [style=none] (1) at (2.25, -0.5) {};
		\node [style=none] (2) at (3.25, -0.5) {};
		\node [style=none] (3) at (3.25, 0.5) {};
		\node [style=none] (5) at (2.75, 0.5) {};
		\node [style=none] (6) at (2.75, -0.5) {};
		\node [style=none] (9) at (2.75, -1) {};
		\node [style=none] (10) at (2.75, 1) {};
		\node [style=none] (12) at (2.75, 0) {$\Pi_i$};
		\node [style=none] (27) at (-0.5, 1) {};
		\node [style=none] (28) at (-0.5, -1) {};
		\node [style=none] (31) at (1.25, 0) {$= \sum_i$};
		\node [style=none] (32) at (-1, 0.5) {};
		\node [style=none] (33) at (-1, -0.5) {};
		\node [style=none] (34) at (0, -0.5) {};
		\node [style=none] (35) at (0, 0.5) {};
		\node [style=right label] (36) at (-0.5, -1) {$A$};
	\end{pgfonlayer}
	\begin{pgfonlayer}{edgelayer}
		\draw (0.center) to (3.center);
		\draw (3.center) to (2.center);
		\draw (2.center) to (1.center);
		\draw (1.center) to (0.center);
		\draw [qWire] (10.center) to (5.center);
		\draw [qWire] (6.center) to (9.center);
		\draw [qWire] (27.center) to (28.center);
		\draw [style=thick gray dashed edge] (32.center) to (35.center);
		\draw [style=thick gray dashed edge] (35.center) to (34.center);
		\draw [style=thick gray dashed edge] (34.center) to (33.center);
		\draw [style=thick gray dashed edge] (33.center) to (32.center);
	\end{pgfonlayer}
\end{tikzpicture}
\eeq
note that these projectors are necessarily physical due to lemma \ref{lem:NRH}. Consider the composite of a system $A$ with itself, then we obtain $A\otimes A = \bigoplus_{ij}A_i\otimes A_j$ or diagrammatically:
\beq
\begin{tikzpicture}
	\begin{pgfonlayer}{nodelayer}
		\node [style=none] (0) at (2.25, 0.5) {};
		\node [style=none] (1) at (2.25, -0.5) {};
		\node [style=none] (2) at (3.25, -0.5) {};
		\node [style=none] (3) at (3.25, 0.5) {};
		\node [style=none] (5) at (2.75, 0.5) {};
		\node [style=none] (6) at (2.75, -0.5) {};
		\node [style=none] (9) at (2.75, -1) {};
		\node [style=none] (10) at (2.75, 1) {};
		\node [style=none] (12) at (2.75, 0) {$\Pi_i$};
		\node [style=none] (27) at (-2, 1) {};
		\node [style=none] (28) at (-2, -1) {};
		\node [style=none] (31) at (1.25, 0) {$= \sum_{ij}$};
		\node [style=none] (32) at (-2.5, 0.5) {};
		\node [style=none] (33) at (-2.5, -0.5) {};
		\node [style=none] (34) at (0, -0.5) {};
		\node [style=none] (35) at (0, 0.5) {};
		\node [style=right label] (36) at (-2, -1) {$A$};
		\node [style=none] (37) at (-0.5, 1) {};
		\node [style=none] (38) at (-0.5, -1) {};
		\node [style=right label] (39) at (-0.5, -1) {$A$};
		\node [style=none] (40) at (3.75, 0.5) {};
		\node [style=none] (41) at (3.75, -0.5) {};
		\node [style=none] (42) at (4.75, -0.5) {};
		\node [style=none] (43) at (4.75, 0.5) {};
		\node [style=none] (44) at (4.25, 0.5) {};
		\node [style=none] (45) at (4.25, -0.5) {};
		\node [style=none] (46) at (4.25, -1) {};
		\node [style=none] (47) at (4.25, 1) {};
		\node [style=none] (48) at (4.25, 0) {$\Pi_j$};
	\end{pgfonlayer}
	\begin{pgfonlayer}{edgelayer}
		\draw (0.center) to (3.center);
		\draw (3.center) to (2.center);
		\draw (2.center) to (1.center);
		\draw (1.center) to (0.center);
		\draw [qWire] (10.center) to (5.center);
		\draw [qWire] (6.center) to (9.center);
		\draw [qWire] (27.center) to (28.center);
		\draw [style=thick gray dashed edge] (32.center) to (35.center);
		\draw [style=thick gray dashed edge] (35.center) to (34.center);
		\draw [style=thick gray dashed edge] (34.center) to (33.center);
		\draw [style=thick gray dashed edge] (33.center) to (32.center);
		\draw [qWire] (37.center) to (38.center);
		\draw (40.center) to (43.center);
		\draw (43.center) to (42.center);
		\draw (42.center) to (41.center);
		\draw (41.center) to (40.center);
		\draw [qWire] (47.center) to (44.center);
		\draw [qWire] (45.center) to (46.center);
	\end{pgfonlayer}
\end{tikzpicture}
\eeq
What we want to show is that this is the most refined decomposition of the composite system, in other words, that $A_i\otimes A_j$ are simple. Without loss of generality let us consider the component $A_1\otimes A_1$ and show that this must be simple. For the sake of contradiction, let us assume that $A_1\otimes A_1 = B\oplus C$, diagrammatically:
\beq
\begin{tikzpicture}
	\begin{pgfonlayer}{nodelayer}
		\node [style=none] (0) at (-2.75, 0.5) {};
		\node [style=none] (1) at (-2.75, -0.5) {};
		\node [style=none] (2) at (-0.75, -0.5) {};
		\node [style=none] (3) at (-0.75, 0.5) {};
		\node [style=none] (4) at (-1.25, 0.5) {};
		\node [style=none] (5) at (-2.25, 0.5) {};
		\node [style=none] (6) at (-2.25, -0.5) {};
		\node [style=none] (7) at (-1.25, -0.5) {};
		\node [style=none] (8) at (-1.25, -1) {};
		\node [style=none] (9) at (-2.25, -1) {};
		\node [style=none] (10) at (-2.25, 1) {};
		\node [style=none] (11) at (-1.25, 1) {};
		\node [style=none] (12) at (-1.75, 0) {$\Pi_B$};
		\node [style=none] (13) at (2.75, 1) {};
		\node [style=none] (14) at (1.25, 0.5) {};
		\node [style=none] (15) at (3.25, -0.5) {};
		\node [style=none] (16) at (1.25, -0.5) {};
		\node [style=none] (17) at (2.75, -1) {};
		\node [style=none] (18) at (1.75, 1) {};
		\node [style=none] (19) at (1.75, 0.5) {};
		\node [style=none] (20) at (1.75, -1) {};
		\node [style=none] (21) at (3.25, 0.5) {};
		\node [style=none] (22) at (2.25, 0) {$\Pi_C$};
		\node [style=none] (23) at (1.75, -0.5) {};
		\node [style=none] (24) at (2.75, -0.5) {};
		\node [style=none] (25) at (2.75, 0.5) {};
		\node [style=none] (26) at (0.25, 0) {$+$};
		\node [style=none] (31) at (-4.25, 0) {$=$};
		\node [style=none] (39) at (-8, 0.5) {};
		\node [style=none] (40) at (-7, -0.5) {};
		\node [style=none] (41) at (-8, -0.5) {};
		\node [style=none] (43) at (-7.5, 1) {};
		\node [style=none] (44) at (-7.5, 0.5) {};
		\node [style=none] (45) at (-7.5, -1) {};
		\node [style=none] (46) at (-7, 0.5) {};
		\node [style=none] (47) at (-7.5, 0) {$\Pi_1$};
		\node [style=none] (48) at (-7.5, -0.5) {};
		\node [style=none] (52) at (-6.5, 0.5) {};
		\node [style=none] (53) at (-5.5, -0.5) {};
		\node [style=none] (54) at (-6.5, -0.5) {};
		\node [style=none] (55) at (-6, 1) {};
		\node [style=none] (56) at (-6, 0.5) {};
		\node [style=none] (57) at (-6, -1) {};
		\node [style=none] (58) at (-5.5, 0.5) {};
		\node [style=none] (59) at (-6, 0) {$\Pi_1$};
		\node [style=none] (60) at (-6, -0.5) {};
	\end{pgfonlayer}
	\begin{pgfonlayer}{edgelayer}
		\draw (0.center) to (3.center);
		\draw (3.center) to (2.center);
		\draw (2.center) to (1.center);
		\draw (1.center) to (0.center);
		\draw [qWire] (10.center) to (5.center);
		\draw [qWire] (6.center) to (9.center);
		\draw [qWire] (11.center) to (4.center);
		\draw [qWire] (7.center) to (8.center);
		\draw (14.center) to (21.center);
		\draw (21.center) to (15.center);
		\draw (15.center) to (16.center);
		\draw (16.center) to (14.center);
		\draw [qWire] (18.center) to (19.center);
		\draw [qWire] (23.center) to (20.center);
		\draw [qWire] (13.center) to (25.center);
		\draw [qWire] (24.center) to (17.center);
		\draw (39.center) to (46.center);
		\draw (46.center) to (40.center);
		\draw (40.center) to (41.center);
		\draw (41.center) to (39.center);
		\draw [qWire] (43.center) to (44.center);
		\draw [qWire] (48.center) to (45.center);
		\draw (52.center) to (58.center);
		\draw (58.center) to (53.center);
		\draw (53.center) to (54.center);
		\draw (54.center) to (52.center);
		\draw [qWire] (55.center) to (56.center);
		\draw [qWire] (60.center) to (57.center);
	\end{pgfonlayer}
\end{tikzpicture}
\eeq
which means that we can decompose the identity for the composite system as:
\beq
\InputIfFileExists{Diagrams/decompAA.tikz}{}{\input{./figures/Diagrams/decompAA.tikz}}\quad =:\quad \begin{tikzpicture}
	\begin{pgfonlayer}{nodelayer}
		\node [style=none] (0) at (-2.75, 0.5) {};
		\node [style=none] (1) at (-2.75, -0.5) {};
		\node [style=none] (2) at (-0.75, -0.5) {};
		\node [style=none] (3) at (-0.75, 0.5) {};
		\node [style=none] (4) at (-1.25, 0.5) {};
		\node [style=none] (5) at (-2.25, 0.5) {};
		\node [style=none] (6) at (-2.25, -0.5) {};
		\node [style=none] (7) at (-1.25, -0.5) {};
		\node [style=none] (8) at (-1.25, -1) {};
		\node [style=none] (9) at (-2.25, -1) {};
		\node [style=none] (10) at (-2.25, 1) {};
		\node [style=none] (11) at (-1.25, 1) {};
		\node [style=none] (12) at (-1.75, 0) {$\Pi_B$};
		\node [style=none] (13) at (2.75, 1) {};
		\node [style=none] (14) at (1.25, 0.5) {};
		\node [style=none] (15) at (3.25, -0.5) {};
		\node [style=none] (16) at (1.25, -0.5) {};
		\node [style=none] (17) at (2.75, -1) {};
		\node [style=none] (18) at (1.75, 1) {};
		\node [style=none] (19) at (1.75, 0.5) {};
		\node [style=none] (20) at (1.75, -1) {};
		\node [style=none] (21) at (3.25, 0.5) {};
		\node [style=none] (22) at (2.25, 0) {$\Pi_C$};
		\node [style=none] (23) at (1.75, -0.5) {};
		\node [style=none] (24) at (2.75, -0.5) {};
		\node [style=none] (25) at (2.75, 0.5) {};
		\node [style=none] (26) at (0.25, 0) {$+$};
		\node [style=none] (38) at (6.75, 1) {};
		\node [style=none] (39) at (5.25, 0.5) {};
		\node [style=none] (40) at (7.25, -0.5) {};
		\node [style=none] (41) at (5.25, -0.5) {};
		\node [style=none] (42) at (6.75, -1) {};
		\node [style=none] (43) at (5.75, 1) {};
		\node [style=none] (44) at (5.75, 0.5) {};
		\node [style=none] (45) at (5.75, -1) {};
		\node [style=none] (46) at (7.25, 0.5) {};
		\node [style=none] (47) at (6.25, 0) {$\Pi_X$};
		\node [style=none] (48) at (5.75, -0.5) {};
		\node [style=none] (49) at (6.75, -0.5) {};
		\node [style=none] (50) at (6.75, 0.5) {};
		\node [style=none] (51) at (4.25, 0) {$+$};
	\end{pgfonlayer}
	\begin{pgfonlayer}{edgelayer}
		\draw (0.center) to (3.center);
		\draw (3.center) to (2.center);
		\draw (2.center) to (1.center);
		\draw (1.center) to (0.center);
		\draw [qWire] (10.center) to (5.center);
		\draw [qWire] (6.center) to (9.center);
		\draw [qWire] (11.center) to (4.center);
		\draw [qWire] (7.center) to (8.center);
		\draw (14.center) to (21.center);
		\draw (21.center) to (15.center);
		\draw (15.center) to (16.center);
		\draw (16.center) to (14.center);
		\draw [qWire] (18.center) to (19.center);
		\draw [qWire] (23.center) to (20.center);
		\draw [qWire] (13.center) to (25.center);
		\draw [qWire] (24.center) to (17.center);
		\draw (39.center) to (46.center);
		\draw (46.center) to (40.center);
		\draw (40.center) to (41.center);
		\draw (41.center) to (39.center);
		\draw [qWire] (43.center) to (44.center);
		\draw [qWire] (48.center) to (45.center);
		\draw [qWire] (38.center) to (50.center);
		\draw [qWire] (49.center) to (42.center);
	\end{pgfonlayer}
\end{tikzpicture}
\eeq
where the sum is understood as running over all pairs $(i,j)$ except $i=j=1$.

We can use this to define a leak $\Delta$ for system $A$ by:
\[%
\InputIfFileExists{Diagrams/decompAALeaks.tikz}{}{\input{./figures/Diagrams/decompAALeaks.tikz}}\]
It is straightforward to check that this is indeed a leak, and moreover, using the orthogonality of the projectors $\Pi$  and recalling that sums distribute over diagrams, it is not hard to show that:
\beq\label{eq:AALeak}%
\InputIfFileExists{Diagrams/decompAALeaksEquation.tikz}{}{\input{./figures/Diagrams/decompAALeaksEquation.tikz}}\eeq
Now consider the effect of the leak on pure causal states $\chi$. The definition of purity immediately implies that:
\beq\label{eq:leakonstate}%
\InputIfFileExists{Diagrams/AALeakOnPure.tikz}{}{\input{./figures/Diagrams/AALeakOnPure.tikz}}
\eeq
 Now we can define orthogonal projectors   $\pi^\chi_b$, $\pi^\chi_c$ and $\pi^\chi_x$ by:
\beq\label{eq:deforthogproj}
\begin{tikzpicture}
	\begin{pgfonlayer}{nodelayer}
		\node [style=none] (20) at (-0.25, -0.75) {$\rho_\chi$};
		\node [style=none] (21) at (-1.5, -0.25) {};
		\node [style=none] (22) at (-0.25, -1.5) {};
		\node [style=none] (23) at (1, -0.25) {};
		\node [style=none] (24) at (-1, -0.25) {};
		\node [style=none] (25) at (-0.25, -0.25) {};
		\node [style=none] (26) at (0.5, -0.25) {};
		\node [style=none] (27) at (-1, 2.25) {};
		\node [style=none] (28) at (-0.25, 0.25) {};
		\node [style=copoint] (29) at (0.5, 0.25) {$0$};
		\node [style=none] (30) at (1.25, 0.25) {};
		\node [style=none] (31) at (1.25, -2.5) {};
		\node [style=none] (32) at (-1.75, 1.5) {};
		\node [style=none] (33) at (1.75, 1.5) {};
		\node [style=none] (34) at (1.75, -1.75) {};
		\node [style=none] (35) at (-1.75, -1.75) {};
	\end{pgfonlayer}
	\begin{pgfonlayer}{edgelayer}
		\draw (21.center) to (22.center);
		\draw (22.center) to (23.center);
		\draw (21.center) to (23.center);
		\draw [qWire] (27.center) to (24.center);
		\draw [qWire] (28.center) to (25.center);
		\draw [style=cWire] (29) to (26.center);
		\draw [style=qWire, bend left=90, looseness=2.00] (28.center) to (30.center);
		\draw [style=qWire] (30.center) to (31.center);
		\draw [thick gray dashed edge] (32.center) to (35.center);
		\draw [thick gray dashed edge] (35.center) to (34.center);
		\draw [thick gray dashed edge] (34.center) to (33.center);
		\draw [thick gray dashed edge] (33.center) to (32.center);
	\end{pgfonlayer}
\end{tikzpicture}
\quad = \quad
\InputIfFileExists{Diagrams/decompAProjb.tikz}{}{\input{./figures/Diagrams/decompAProjb.tikz}} \qquad \text{,} \qquad \begin{tikzpicture}
	\begin{pgfonlayer}{nodelayer}
		\node [style=none] (20) at (-0.25, -0.75) {$\rho_\chi$};
		\node [style=none] (21) at (-1.5, -0.25) {};
		\node [style=none] (22) at (-0.25, -1.5) {};
		\node [style=none] (23) at (1, -0.25) {};
		\node [style=none] (24) at (-1, -0.25) {};
		\node [style=none] (25) at (-0.25, -0.25) {};
		\node [style=none] (26) at (0.5, -0.25) {};
		\node [style=none] (27) at (-1, 2.25) {};
		\node [style=none] (28) at (-0.25, 0.25) {};
		\node [style=copoint] (29) at (0.5, 0.25) {$1$};
		\node [style=none] (30) at (1.25, 0.25) {};
		\node [style=none] (31) at (1.25, -2.5) {};
		\node [style=none] (32) at (-1.75, 1.5) {};
		\node [style=none] (33) at (1.75, 1.5) {};
		\node [style=none] (34) at (1.75, -1.75) {};
		\node [style=none] (35) at (-1.75, -1.75) {};
	\end{pgfonlayer}
	\begin{pgfonlayer}{edgelayer}
		\draw (21.center) to (22.center);
		\draw (22.center) to (23.center);
		\draw (21.center) to (23.center);
		\draw [qWire] (27.center) to (24.center);
		\draw [qWire] (28.center) to (25.center);
		\draw [style=cWire] (29) to (26.center);
		\draw [style=qWire, bend left=90, looseness=2.00] (28.center) to (30.center);
		\draw [style=qWire] (30.center) to (31.center);
		\draw [thick gray dashed edge] (32.center) to (35.center);
		\draw [thick gray dashed edge] (35.center) to (34.center);
		\draw [thick gray dashed edge] (34.center) to (33.center);
		\draw [thick gray dashed edge] (33.center) to (32.center);
	\end{pgfonlayer}
\end{tikzpicture}
\quad=\quad%
\InputIfFileExists{Diagrams/decompAProjc.tikz}{}{\input{./figures/Diagrams/decompAProjc.tikz}} \eeq 
and
\beq
\begin{tikzpicture}
	\begin{pgfonlayer}{nodelayer}
		\node [style=none] (20) at (-0.25, -0.75) {$\rho_\chi$};
		\node [style=none] (21) at (-1.5, -0.25) {};
		\node [style=none] (22) at (-0.25, -1.5) {};
		\node [style=none] (23) at (1, -0.25) {};
		\node [style=none] (24) at (-1, -0.25) {};
		\node [style=none] (25) at (-0.25, -0.25) {};
		\node [style=none] (26) at (0.5, -0.25) {};
		\node [style=none] (27) at (-1, 2.25) {};
		\node [style=none] (28) at (-0.25, 0.25) {};
		\node [style=copoint] (29) at (0.5, 0.25) {$2$};
		\node [style=none] (30) at (1.25, 0.25) {};
		\node [style=none] (31) at (1.25, -2.5) {};
		\node [style=none] (32) at (-1.75, 1.5) {};
		\node [style=none] (33) at (1.75, 1.5) {};
		\node [style=none] (34) at (1.75, -1.75) {};
		\node [style=none] (35) at (-1.75, -1.75) {};
	\end{pgfonlayer}
	\begin{pgfonlayer}{edgelayer}
		\draw (21.center) to (22.center);
		\draw (22.center) to (23.center);
		\draw (21.center) to (23.center);
		\draw [qWire] (27.center) to (24.center);
		\draw [qWire] (28.center) to (25.center);
		\draw [style=cWire] (29) to (26.center);
		\draw [style=qWire, bend left=90, looseness=2.00] (28.center) to (30.center);
		\draw [style=qWire] (30.center) to (31.center);
		\draw [thick gray dashed edge] (32.center) to (35.center);
		\draw [thick gray dashed edge] (35.center) to (34.center);
		\draw [thick gray dashed edge] (34.center) to (33.center);
		\draw [thick gray dashed edge] (33.center) to (32.center);
	\end{pgfonlayer}
\end{tikzpicture}
\quad=\quad\begin{tikzpicture}
	\begin{pgfonlayer}{nodelayer}
		\node [style=none] (15) at (4.5, 0) {$=:$};
		\node [style=small box] (16) at (6, 0) {$\pi^\chi_x$};
		\node [style=none] (17) at (6, 1.25) {};
		\node [style=none] (18) at (6, -1.25) {};
		\node [style=none] (20) at (1.5, 0.5) {};
		\node [style=none] (21) at (1.5, -0.5) {};
		\node [style=none] (22) at (2, -0.5) {};
		\node [style=none] (23) at (2, 0.5) {};
		\node [style=none] (24) at (3, 0.5) {};
		\node [style=none] (25) at (3.5, 0.5) {};
		\node [style=none] (26) at (3.5, -0.5) {};
		\node [style=none] (27) at (3, -0.5) {};
		\node [style=none] (28) at (2, 1) {};
		\node [style=none] (29) at (3, 1.75) {};
		\node [style=none] (30) at (3, -1.75) {};
		\node [style=point] (31) at (2, -1.25) {$\chi$};
		\node [style=upground] (32) at (2, 1.25) {};
		\node [style=none] (33) at (2.5, 0) {$\Pi_X$};
	\end{pgfonlayer}
	\begin{pgfonlayer}{edgelayer}
		\draw [qWire] (17.center) to (16);
		\draw [qWire] (16) to (18.center);
		\draw (20.center) to (25.center);
		\draw (25.center) to (26.center);
		\draw (26.center) to (21.center);
		\draw (21.center) to (20.center);
		\draw [qWire] (28.center) to (23.center);
		\draw [qWire] (22.center) to (31);
		\draw [qWire] (29.center) to (24.center);
		\draw [qWire] (27.center) to (30.center);
	\end{pgfonlayer}
\end{tikzpicture}
\eeq

where to see that these are orthogonal projectors we use Eq.~\eqref{eq:AALeak} to note that
\[%
\InputIfFileExists{Diagrams/orthogProj3.tikz}{}{\input{./figures/Diagrams/orthogProj3.tikz}} =%
\InputIfFileExists{Diagrams/orthogProj4.tikz}{}{\input{./figures/Diagrams/orthogProj4.tikz}}\]
and then we can use Eq.~\eqref{eq:leakonstate} to rewrite this as:
\[%
\InputIfFileExists{Diagrams/orthogProj2.tikz}{}{\input{./figures/Diagrams/orthogProj2.tikz}} = %
\InputIfFileExists{Diagrams/orthogProj5.tikz}{}{\input{./figures/Diagrams/orthogProj5.tikz}}\]
and finally, use the definition of the projectors  (Eq.~\eqref{eq:deforthogproj}) to conclude that
\[%
\InputIfFileExists{Diagrams/orthogProj1.tikz}{}{\input{./figures/Diagrams/orthogProj1.tikz}} = %
\InputIfFileExists{Diagrams/orthogProj6.tikz}{}{\input{./figures/Diagrams/orthogProj6.tikz}}\]
that is, the processes are indeed orthogonal projectors.  It is easy to see, that as $\Pi_A+\Pi_B+\Pi_X = \mathds{1}_{A\otimes A}$ that $\pi_a^\chi+\pi_b^\chi+\pi_x^\chi=\mathds{1}_A$, that is, they define a decomposition $A=a\oplus b\oplus x$.  Next let us examine $\pi_x^\chi$, if we insert the definition of $\Pi_X$ then we find that:
\beq
\begin{tikzpicture}
	\begin{pgfonlayer}{nodelayer}
		\node [style=none] (15) at (0, 0) {$=$};
		\node [style=small box] (16) at (-1.25, 0) {$\pi^\chi_x$};
		\node [style=none] (17) at (-1.25, 1.25) {};
		\node [style=none] (18) at (-1.25, -1.25) {};
		\node [style=none] (20) at (1.25, 0.5) {};
		\node [style=none] (21) at (1.25, -0.5) {};
		\node [style=none] (22) at (1.75, -0.5) {};
		\node [style=none] (23) at (1.75, 0.5) {};
		\node [style=none] (24) at (2.75, 0.5) {};
		\node [style=none] (25) at (3.25, 0.5) {};
		\node [style=none] (26) at (3.25, -0.5) {};
		\node [style=none] (27) at (2.75, -0.5) {};
		\node [style=none] (28) at (1.75, 1) {};
		\node [style=none] (29) at (2.75, 1.75) {};
		\node [style=none] (30) at (2.75, -1.75) {};
		\node [style=point] (31) at (1.75, -1.25) {$\chi$};
		\node [style=upground] (32) at (1.75, 1.25) {};
		\node [style=none] (33) at (2.25, 0) {$\Pi_X$};
		\node [style=none] (34) at (4.25, 0) {$=\sum_{ij}$};
		\node [style=none] (35) at (5.5, 0.5) {};
		\node [style=none] (36) at (5.5, -0.5) {};
		\node [style=none] (37) at (6, -0.5) {};
		\node [style=none] (38) at (6, 0.5) {};
		\node [style=none] (39) at (7.5, 0.5) {};
		\node [style=none] (40) at (6.5, 0.5) {};
		\node [style=none] (41) at (6.5, -0.5) {};
		\node [style=none] (42) at (7.5, -0.5) {};
		\node [style=none] (43) at (6, 1) {};
		\node [style=none] (44) at (7.5, 1.75) {};
		\node [style=none] (45) at (7.5, -1.75) {};
		\node [style=point] (46) at (6, -1.25) {$\chi$};
		\node [style=upground] (47) at (6, 1.25) {};
		\node [style=none] (48) at (6, 0) {$\Pi_i$};
		\node [style=none] (49) at (7, 0.5) {};
		\node [style=none] (50) at (7, -0.5) {};
		\node [style=none] (51) at (7.5, -0.5) {};
		\node [style=none] (52) at (7.5, 0.5) {};
		\node [style=none] (53) at (8, 0.5) {};
		\node [style=none] (54) at (8, -0.5) {};
		\node [style=none] (55) at (7.5, 0) {$\Pi_j$};
	\end{pgfonlayer}
	\begin{pgfonlayer}{edgelayer}
		\draw [qWire] (17.center) to (16);
		\draw [qWire] (16) to (18.center);
		\draw (20.center) to (25.center);
		\draw (25.center) to (26.center);
		\draw (26.center) to (21.center);
		\draw (21.center) to (20.center);
		\draw [qWire] (28.center) to (23.center);
		\draw [qWire] (22.center) to (31);
		\draw [qWire] (29.center) to (24.center);
		\draw [qWire] (27.center) to (30.center);
		\draw (35.center) to (40.center);
		\draw (40.center) to (41.center);
		\draw (41.center) to (36.center);
		\draw (36.center) to (35.center);
		\draw [qWire] (43.center) to (38.center);
		\draw [qWire] (37.center) to (46);
		\draw [qWire] (44.center) to (39.center);
		\draw [qWire] (42.center) to (45.center);
		\draw (49.center) to (53.center);
		\draw (53.center) to (54.center);
		\draw (54.center) to (50.center);
		\draw (50.center) to (49.center);
	\end{pgfonlayer}
\end{tikzpicture}
\eeq
recalling that the sum over $i$ and $j$ avoids the case where $i=j=1$ we therefore have:
\beq
\begin{tikzpicture}
	\begin{pgfonlayer}{nodelayer}
		\node [style=small box] (16) at (-2, 0) {$\pi^\chi_x$};
		\node [style=none] (17) at (-2, 1.25) {};
		\node [style=none] (18) at (-2, -1.25) {};
		\node [style=none] (34) at (0.5, 0) {$=\sum_{j\neq 1}$};
		\node [style=none] (39) at (2.5, 0.5) {};
		\node [style=none] (42) at (2.5, -0.5) {};
		\node [style=none] (44) at (2.5, 1.75) {};
		\node [style=none] (45) at (2.5, -1.75) {};
		\node [style=none] (49) at (2, 0.5) {};
		\node [style=none] (50) at (2, -0.5) {};
		\node [style=none] (51) at (2.5, -0.5) {};
		\node [style=none] (52) at (2.5, 0.5) {};
		\node [style=none] (53) at (3, 0.5) {};
		\node [style=none] (54) at (3, -0.5) {};
		\node [style=none] (55) at (2.5, 0) {$\Pi_j$};
	\end{pgfonlayer}
	\begin{pgfonlayer}{edgelayer}
		\draw [qWire] (17.center) to (16);
		\draw [qWire] (16) to (18.center);
		\draw [qWire] (44.center) to (39.center);
		\draw [qWire] (42.center) to (45.center);
		\draw (49.center) to (53.center);
		\draw (53.center) to (54.center);
		\draw (54.center) to (50.center);
		\draw (50.center) to (49.center);
	\end{pgfonlayer}
\end{tikzpicture}
\eeq
and therefore that:
\beq
\begin{tikzpicture}
	\begin{pgfonlayer}{nodelayer}
		\node [style=small box] (16) at (2, 0) {$\pi^\chi_a$};
		\node [style=none] (17) at (2, 1.5) {};
		\node [style=none] (18) at (2, -1.5) {};
		\node [style=none] (34) at (0.5, 0) {$=$};
		\node [style=none] (39) at (-1.25, 0.5) {};
		\node [style=none] (42) at (-1.25, -0.5) {};
		\node [style=none] (44) at (-1.25, 1.5) {};
		\node [style=none] (45) at (-1.25, -1.5) {};
		\node [style=none] (49) at (-1.75, 0.5) {};
		\node [style=none] (50) at (-1.75, -0.5) {};
		\node [style=none] (51) at (-1.25, -0.5) {};
		\node [style=none] (52) at (-1.25, 0.5) {};
		\node [style=none] (53) at (-0.75, 0.5) {};
		\node [style=none] (54) at (-0.75, -0.5) {};
		\node [style=none] (55) at (-1.25, 0) {$\Pi_1$};
		\node [style=small box] (56) at (5, 0) {$\pi^\chi_b$};
		\node [style=none] (57) at (5, 1.5) {};
		\node [style=none] (58) at (5, -1.5) {};
		\node [style=none] (59) at (3.5, 0) {$+$};
	\end{pgfonlayer}
	\begin{pgfonlayer}{edgelayer}
		\draw [qWire] (17.center) to (16);
		\draw [qWire] (16) to (18.center);
		\draw [qWire] (44.center) to (39.center);
		\draw [qWire] (42.center) to (45.center);
		\draw (49.center) to (53.center);
		\draw (53.center) to (54.center);
		\draw (54.center) to (50.center);
		\draw (50.center) to (49.center);
		\draw [qWire] (57.center) to (56);
		\draw [qWire] (56) to (58.center);
	\end{pgfonlayer}
\end{tikzpicture}
\eeq
that is, we have a decomposition $A_1 = a\oplus b$. However, by assumption $A_1$ is simple, therefore it must be the case that either $A_1=a$ and $b=0$ or $A_1=b$ and $a=0$. On the level of projectors this means either, that $\pi^\chi_a=\Pi_1$ and $\pi^\chi_b=0$, or, that $\pi^\chi_b=\Pi_1$ and $\pi^\chi_a=0$. Which we can lift to the definition of the projectors on the joint system as saying that, for each $\chi \in A_1$ that either 
\beq%
\InputIfFileExists{Diagrams2/projector1.tikz}{}{\input{./figures/Diagrams2/projector1.tikz}}\ \ = \ \ 0 \qquad \text{or} \qquad %
\InputIfFileExists{Diagrams2/projector2.tikz}{}{\input{./figures/Diagrams2/projector2.tikz}}\ \ = \ \ 0 \eeq
but not both. By symmetry we can similarly argue that, for each $\chi'\in A_1$ that either
\beq\begin{tikzpicture}
	\begin{pgfonlayer}{nodelayer}
		\node [style=none] (0) at (2, 0.5) {};
		\node [style=none] (1) at (2, -0.5) {};
		\node [style=none] (2) at (1.5, -0.5) {};
		\node [style=none] (3) at (0.5, -0.5) {};
		\node [style=none] (4) at (0, -0.5) {};
		\node [style=none] (5) at (0, 0.5) {};
		\node [style=none] (6) at (0.5, 0.5) {};
		\node [style=none] (7) at (1.5, 0.5) {};
		\node [style=none] (8) at (1.5, -0.5) {};
		\node [style=point] (9) at (1.5, -1.25) {$\chi'$};
		\node [style=none] (10) at (1.5, 1.25) {};
		\node [style=none] (11) at (0.5, 1.25) {};
		\node [style=none] (12) at (0.5, -1.25) {};
		\node [style=none] (13) at (1, 0) {$\Pi_B$};
	\end{pgfonlayer}
	\begin{pgfonlayer}{edgelayer}
		\draw [qWire] (8.center) to (9);
		\draw (0.center) to (5.center);
		\draw (5.center) to (4.center);
		\draw (4.center) to (1.center);
		\draw (1.center) to (0.center);
		\draw [qWire] (10.center) to (7.center);
		\draw [qWire] (11.center) to (6.center);
		\draw [qWire] (3.center) to (12.center);
	\end{pgfonlayer}
\end{tikzpicture}
\ \ = \ \ 0 \qquad \text{or} \qquad \begin{tikzpicture}
	\begin{pgfonlayer}{nodelayer}
		\node [style=none] (0) at (2, 0.5) {};
		\node [style=none] (1) at (2, -0.5) {};
		\node [style=none] (2) at (1.5, -0.5) {};
		\node [style=none] (3) at (0.5, -0.5) {};
		\node [style=none] (4) at (0, -0.5) {};
		\node [style=none] (5) at (0, 0.5) {};
		\node [style=none] (6) at (0.5, 0.5) {};
		\node [style=none] (7) at (1.5, 0.5) {};
		\node [style=none] (8) at (1.5, -0.5) {};
		\node [style=point] (9) at (1.5, -1.25) {$\chi'$};
		\node [style=none] (10) at (1.5, 1.25) {};
		\node [style=none] (11) at (0.5, 1.25) {};
		\node [style=none] (12) at (0.5, -1.25) {};
		\node [style=none] (13) at (1, 0) {$\Pi_C$};
	\end{pgfonlayer}
	\begin{pgfonlayer}{edgelayer}
		\draw [qWire] (8.center) to (9);
		\draw (0.center) to (5.center);
		\draw (5.center) to (4.center);
		\draw (4.center) to (1.center);
		\draw (1.center) to (0.center);
		\draw [qWire] (10.center) to (7.center);
		\draw [qWire] (11.center) to (6.center);
		\draw [qWire] (3.center) to (12.center);
	\end{pgfonlayer}
\end{tikzpicture}
\ \ = \ \ 0 \eeq
but not both.
Next, without loss of generality, consider some $\chi$ such that:
\[\begin{tikzpicture}
	\begin{pgfonlayer}{nodelayer}
		\node [style=none] (0) at (2, 0.5) {};
		\node [style=none] (1) at (2, -0.5) {};
		\node [style=none] (2) at (1.5, -0.5) {};
		\node [style=none] (3) at (0.5, -0.5) {};
		\node [style=none] (4) at (0, -0.5) {};
		\node [style=none] (5) at (0, 0.5) {};
		\node [style=none] (6) at (0.5, 0.5) {};
		\node [style=none] (7) at (1.5, 0.5) {};
		\node [style=none] (8) at (1.5, -0.5) {};
		\node [style=point] (9) at (1.5, -1.25) {$\chi'$};
		\node [style=none] (10) at (1.5, 1.25) {};
		\node [style=none] (11) at (0.5, 1.25) {};
		\node [style=none] (12) at (0.5, -1.25) {};
		\node [style=none] (13) at (1, 0) {$\Pi_B$};
	\end{pgfonlayer}
	\begin{pgfonlayer}{edgelayer}
		\draw [qWire] (8.center) to (9);
		\draw (0.center) to (5.center);
		\draw (5.center) to (4.center);
		\draw (4.center) to (1.center);
		\draw (1.center) to (0.center);
		\draw [qWire] (10.center) to (7.center);
		\draw [qWire] (11.center) to (6.center);
		\draw [qWire] (3.center) to (12.center);
	\end{pgfonlayer}
\end{tikzpicture}
\ \ = \ \ 0 \]
This means that
\begin{align}\forall \chi'\in A_1 \quad %
\InputIfFileExists{Diagrams2/projector4.tikz}{}{\input{./figures/Diagrams2/projector4.tikz}}\ \ = \ \ 0\quad \implies\quad \forall \chi\in A_1 \quad \begin{tikzpicture}
	\begin{pgfonlayer}{nodelayer}
		\node [style=none] (0) at (-0.25, 0.5000001) {};
		\node [style=none] (1) at (-0.25, -0.5000001) {};
		\node [style=none] (2) at (1.75, -0.5000001) {};
		\node [style=none] (3) at (0.25, -0.5000001) {};
		\node [style=none] (4) at (2.25, -0.5000001) {};
		\node [style=none] (5) at (2.25, 0.5000001) {};
		\node [style=none] (6) at (1.75, 0.5000001) {};
		\node [style=none] (7) at (0.25, 0.5000001) {};
		\node [style=none] (8) at (1.75, -0.5000001) {};
		\node [style=point] (9) at (1.75, -1.25) {$\chi'$};
		\node [style=none] (10) at (0.25, 1.25) {};
		\node [style=none] (11) at (1.75, 1.25) {};
		\node [style=none] (12) at (1, -0) {$\Pi_C$};
		\node [style=point] (13) at (0.25, -1.25) {$\chi$};
	\end{pgfonlayer}
	\begin{pgfonlayer}{edgelayer}
		\draw [qWire](8.center) to (9);
		\draw (0.center) to (5.center);
		\draw (5.center) to (4.center);
		\draw (4.center) to (1.center);
		\draw (1.center) to (0.center);
		\draw [qWire](10.center) to (7.center);
		\draw [qWire](11.center) to (6.center);
		\draw [qWire](3.center) to (13);
	\end{pgfonlayer}
\end{tikzpicture}
\ \ \neq \ \ 0 \quad\\ \implies \quad \forall \chi'\in A_1\quad \begin{tikzpicture}
	\begin{pgfonlayer}{nodelayer}
		\node [style=none] (0) at (2, 0.5) {};
		\node [style=none] (1) at (2, -0.5) {};
		\node [style=none] (2) at (1.5, -0.5) {};
		\node [style=none] (3) at (0.5, -0.5) {};
		\node [style=none] (4) at (0, -0.5) {};
		\node [style=none] (5) at (0, 0.5) {};
		\node [style=none] (6) at (0.5, 0.5) {};
		\node [style=none] (7) at (1.5, 0.5) {};
		\node [style=none] (8) at (1.5, -0.5) {};
		\node [style=point] (9) at (1.5, -1.25) {$\chi'$};
		\node [style=none] (10) at (1.5, 1.25) {};
		\node [style=none] (11) at (0.5, 1.25) {};
		\node [style=none] (12) at (0.5, -1.25) {};
		\node [style=none] (13) at (1, 0) {$\Pi_C$};
	\end{pgfonlayer}
	\begin{pgfonlayer}{edgelayer}
		\draw [qWire] (8.center) to (9);
		\draw (0.center) to (5.center);
		\draw (5.center) to (4.center);
		\draw (4.center) to (1.center);
		\draw (1.center) to (0.center);
		\draw [qWire] (10.center) to (7.center);
		\draw [qWire] (11.center) to (6.center);
		\draw [qWire] (3.center) to (12.center);
	\end{pgfonlayer}
\end{tikzpicture}
\ \ \neq \ \ 0 \quad \implies \forall \chi'\in A_1\quad \begin{tikzpicture}
	\begin{pgfonlayer}{nodelayer}
		\node [style=none] (0) at (2, 0.5) {};
		\node [style=none] (1) at (2, -0.5) {};
		\node [style=none] (2) at (1.5, -0.5) {};
		\node [style=none] (3) at (0.5, -0.5) {};
		\node [style=none] (4) at (0, -0.5) {};
		\node [style=none] (5) at (0, 0.5) {};
		\node [style=none] (6) at (0.5, 0.5) {};
		\node [style=none] (7) at (1.5, 0.5) {};
		\node [style=none] (8) at (1.5, -0.5) {};
		\node [style=point] (9) at (1.5, -1.25) {$\chi'$};
		\node [style=none] (10) at (1.5, 1.25) {};
		\node [style=none] (11) at (0.5, 1.25) {};
		\node [style=none] (12) at (0.5, -1.25) {};
		\node [style=none] (13) at (1, 0) {$\Pi_B$};
	\end{pgfonlayer}
	\begin{pgfonlayer}{edgelayer}
		\draw [qWire] (8.center) to (9);
		\draw (0.center) to (5.center);
		\draw (5.center) to (4.center);
		\draw (4.center) to (1.center);
		\draw (1.center) to (0.center);
		\draw [qWire] (10.center) to (7.center);
		\draw [qWire] (11.center) to (6.center);
		\draw [qWire] (3.center) to (12.center);
	\end{pgfonlayer}
\end{tikzpicture}
\ \ = \ \ 0
\end{align}
Moreover, it is clear that this last statement holds also for $\chi'\in \bigoplus_{j>1} A_j$ hence is true for arbitrary states, hence, local tomography, implies that
\[%
\InputIfFileExists{Diagrams2/projector5.tikz}{}{\input{./figures/Diagrams2/projector5.tikz}}\ \ = \ \ 0 \]
We therefore have $A_1\otimes A_1=B\oplus C = C$ and so $A_1\otimes A_1$ is not decomposable. This contradicts our initial assumption. Hence, if $A_i$ and $A_j$ are simple then so is $A_i\otimes A_j$.

The second main step in our proof is to note that $\mathsf{Dim}[A\otimes A]=\mathsf{Dim}[A]^2$ which follows from local tomography. We will moreover show that $\mathsf{Rank}(A\otimes A) = \mathsf{Rank}(A)^2$.  It is clear that $\mathsf{Rank}(A\otimes B)\geq \mathsf{Rank}(A)\mathsf{Rank}(B)$ as given a set of perfectly distinguishable states for $A$,  $\{s_i^A\}_{i=1}^{\mathsf{Rank}(A)}$ and a set and perfectly distinguishable states for $B$, $\{s_j^B\}_{j=1}^{\mathsf{Rank}(B)}$ then the set $\{s_i^A\otimes s_j^B\}$ defines a set perfectly distinguishable states for $A\otimes B$. 
We then know that (using tomographic locality):
\[%
\InputIfFileExists{Diagrams/discardComposition.tikz}{}{\input{./figures/Diagrams/discardComposition.tikz}}\]
and as the discarding map is an internal effect we have a spectral decomposition of this effect given by the composite of the spectral decomposition of the two individual discarding maps. Hence, $\mathsf{Rank}(A\otimes B)\leq \mathsf{Rank}(A)\mathsf{Rank}(B)$, the conjunction of these two inequalities then gives the required result.

 Finally, we will show that the above results imply that the only EJAs that we can consider are C* algebras. Again consider the decomposition of $A$ into simple components, $A=\bigoplus_i A_i$ and consider the self composite of this $A\otimes A = \bigoplus_{ij}A_i\otimes A_j$. Our first result shows that $A_i\otimes A_j$ must be simple, and our second result puts constraints on the dimension and rank of these simple EJA. In particular, let us consider the summand $A_1\otimes A_1$. If we let $A_1 = \mathds{H}_n$ then we find that $\mathds{H}_n\otimes \mathds{H}_n$ must be a simple EJA with $\mathsf{Rank}=n^2$ and $\mathsf{Dim}=n^2(2n-1)^2$, it is straightforward to check that this does not exist for any integer $n>1$. herefore, the quaternionic case is ruled out. Next we turn to the real case, $\mathds{R}_n\otimes \mathds{R}_n$ and see that this requires a simple EJA with $\mathsf{Rank}=n^2$ and $\mathsf{Dim}=\frac{n^2(n+1)^2}{4}$ it is again straightforward to check that this does not exist for any integer $n>1$.
Considering the octonionic case we have for $\mathds{O}_3\otimes \mathds{O}_3$ that $\mathsf{Rank}=3^2$ and $\mathsf{Dim}=3^6$ it is easy to check that there is no such simple EJA. Finally we consider the spin factors $\mathbf{Spin}_K\otimes \mathbf{Spin}_K$ which require that $\mathsf{Rank}=4$ and $\mathsf{Dim}=K^2$. We find a solution here only for the case that $K=4$, however, this is the situation when $\mathbf{Spin}_4=\mathds{C}_2$ i.e.\ the cone is the Bloch ball. Hence, the only summands which can compose in the correct way are $\mathds{C}_n$. We can then check that the standard quantum tensor product i.e.\ $\mathds{C}_n\otimes\mathds{C}_m:=\mathds{C}_{nm}$ is the only choice consistent with our constraints as it has $\mathsf{Rank}=nm$ and  $\mathsf{Dim}=n^2m^2$ as required. Therefore we have shown that the only EJAs which satisfy our postulates are those of the form $\bigoplus_i \mathds{C}_{n_i}$, i.e., those that are C* algebras.
\endproof

We have therefore demonstrated that the state spaces of our systems correspond to the state spaces of C*-Algebras;  moreover 
processes in our theory from a system $A$ to $B$ correspond to the set of \emph{all} CP maps between the associated C*-Algebras. We know that processes must be CP maps (they are linear and map bipartite states to bipartite states). All that needs to be proved is that every CP map corresponds to a process in the theory.  This is exactly what we showed in Lem.~\ref{lem:NRH}.

This completes the reconstruction as we have demonstrated that our postulates are equivalent to systems being finite-dimensional C*-algebras and processes being completely positive trace-preserving maps between them: this is precisely the process-theoretic description of quantum theory (Ex.~\ref{Ex:CStar}) we were aiming for.  Note that this does not mean that every C*-algebra is necessarily in the theory, for example, classical theory satisfies all of our postulates, as does the theory made up of classical theory together with tensors and direct sums of qubits. This, however, \emph{should} be the target of a reconstruction. A reconstruction should tell us what can exist not what does exist. Indeed, this is precisely what we have in traditional presentations of quantum theory, we are told that to every physical system there is an associated Hilbert space, not that to every Hilbert space there is an associated physical system.

\section{Distinguishing quantum and classical systems}

 In the previous section, we reconstructed the full process-theoretic description of quantum theory. However, it is still of interest to ask the question: 
how can we single out the fully quantum and classical subtheories? We now introduce two novel ways to do this, and, moreover, demonstrate that they are equivalent for any process theory with cups \& caps.

\begin{proposition}The purity of cups (and/or caps) restricts us to quantum systems.
\end{proposition}
\proof
In Ex.~\ref{Ex:DaggerCompact} we defined the cup for C*-algebraic systems. It is simple to see that this has the following dilation:
\[%
\InputIfFileExists{Diagrams2/CStarCupDilation.tikz}{}{\input{./figures/Diagrams2/CStarCupDilation.tikz}}\]
where $\{k\}$ is some set of perfectly distinguishable states. Considering the definition of purity for states (Ex.~\ref{Ex:PureStates}), we find that for the cup to be pure, any dilation must separate. This is the case only if $k$ takes a single value, hence it is pure if and only if the C*-algebra has a single  summand, that is, when it is a fully quantum system.
\endproof

In other words, it is really the existence of correlations mediated by \emph{pure} states (and/or effects) that is the distinguishing feature of quantum theory. Any other apparently odd feature of quantum theory should, in principle, be able to be traced back to this.

\begin{proposition}The triviality of leaks restricts to quantum systems.\end{proposition}
\proof
For quantum theory all leaks are trivial, and for any other C*-algebra, we can construct a non-trivial leak by leaking the `which branch' information. That is,
\[%
\InputIfFileExists{Diagrams2/CStarLeak.tikz}{}{\input{./figures/Diagrams2/CStarLeak.tikz}}\]
is trivial only when $k$ takes a single value and the system is fully quantum.
\endproof

This can be seen as a generalised no-broadcasting theorem, or, that information gain always causes disturbance of a quantum system. This does not seem to be a particularly surprising feature of nature. At least, as soon as one takes the view that measurements are physical and should be described by an interaction between systems. Then, even going back to Newton, it is not surprising that a measurement should have an impact on the system being measured.

The fact that we have these two ways to characterise fully quantum systems is also not surprising, for it can generally be shown that:
\begin{proposition}
For any process theory with cups \& caps, purity of the cup (or cap) is equivalent to the triviality of all leaks.
\end{proposition}
\proof
Firstly note that there is a one-to-one correspondence between the leaks for a system and dilations of the cup. That is, any leak, $l$, will define a dilation of the cup, $L$ as:
\[%
\InputIfFileExists{Diagrams/Quantum1.tikz}{}{\input{./figures/Diagrams/Quantum1.tikz}}\]
Moreover, any dilation, $S$, of the cup will define a leak, $s$ as:
\[%
\InputIfFileExists{Diagrams/Quantum2.tikz}{}{\input{./figures/Diagrams/Quantum2.tikz}}\]
Then it is clear, that if the leak is trivial then the corresponding dilation factorises and vice versa:
\[%
\InputIfFileExists{Diagrams/Quantum3.tikz}{}{\input{./figures/Diagrams/Quantum3.tikz}}\]
Triviality of all leaks is therefore equivalent to factorisation of all dilations of the cup. This is precisely what it means for the cup to be pure (see Ex.~\ref{Ex:PureStates}). The proof for the cap proceeds similarly.
\endproof

Similarly to how we have singled out the quantum systems, one could ask how the classical systems can be distinguished. Intuitively, these can be seen as those that are maximally leaking, and which have the minimally pure cups \& caps. All of the other systems therefore have intermediate strength of leaks and intermediate purity of cups \& caps. To formalise this notion, we would have to introduce a measure of purity (for example $\mathsf{tr}(\rho^2)$) and a measure of strength of leaks (for example the quality of a leak \cite{selby2017leaks}). It is interesting to note that fully quantum systems are much easier to single out from this perspective ---as they do not require us to introduce such measures--- and so, the fully quantum theory appears to be, in some sense, the more fundamental, or natural, subtheory.

\section{Related work}

\subsection{Purification}\label{Sec:StandardPurification}

In \cite{selby2017leaks} we showed that leaks must be taken into account when defining purity for general processes in general process theories. In particular, in the case of classical theory, this refined definition is essential for purity of the identity process. Given this notion of purity, we can define the symmetric purification postulate, which holds for both quantum and classical theory.

The essential component of the standard purification postulate, for distinguishing quantum and classical theory, is the purity of the cup. Indeed, we can show that the conjunction of symmetric purification, and the existence of a pure cup, implies the standard purification postulate.

\begin{proposition} Any theory with symmetric purifications and  pure cups satisfies the standard purification postulate.
\end{proposition}
\proof
Firstly we note that given a symmetric purification $F$ of any process $f$, we can bend the `discarded' input into an output using a cup:
\[%
\InputIfFileExists{Diagrams/DerivingPurification1.tikz}{}{\input{./figures/Diagrams/DerivingPurification1.tikz}}\]
so $\mathcal{F}$ provides a standard purification for $f$.
\endproof

An alternative way of viewing this is that the pure cups \& caps available for fully quantum systems, allow us to arbitrarily interchange uncertainty about the past, and uncertainty about the future. In contrast, in classical theory past uncertainty and future uncertainty play distinct roles.

One may wonder whether we can obtain the standard `essential uniqueness' property in a similar way. This turns out to be the case in theories where processes are reversible if and only  if they are pure and causal. For such theories we have:
\begin{proposition} Any theory with pure cups, essentially unique symmetric purifications, in which pure causal processes are reversible has essentially unique standard purifications.
\end{proposition}
\proof
Note that any purification can be considered as a symmetric purification where the input system is trivial:
\[%
\InputIfFileExists{Diagrams/DerivingPurification2.tikz}{}{\input{./figures/Diagrams/DerivingPurification2.tikz}}\]
Therefore, if we have two such purifications $F$ and $G$ then the essential uniqueness condition of symmetric purification implies that:
\[%
\InputIfFileExists{Diagrams/DerivingPurification3.tikz}{}{\input{./figures/Diagrams/DerivingPurification3.tikz}}\]
where $R$ is a pure causal process. By assumption, this means that $R$ is reversible and hence this is precisely the standard essential uniqueness condition for purification.
\endproof
Whilst this assumption that pure causal processes are reversible may seem natural ---as it holds in the cases of quantum and classical theory where the pure causal processes are respectively unitaries and permutations--- it does not hold for general theories. For instance, take the process theory to be a restricted form of quantum theory, which has the same states and measurements as quantum theory, but a restricted set of dynamics. Specifically, restricting to a single unitary which is a rotation about $Z$ of $\sqrt{2}\pi$. In this theory we can generate other rotations, but they will all be by an angle of $n\sqrt{2}\pi$, and so, we can never reverse a transformation. That is, we have unitaries that are pure and causal but are not reversible.

If we want to obtain the standard essential uniqueness property precisely we could have assumed that the  vise $R$ was reversible. However, this was stronger than necessary for the reconstruction, so we did not make that assumption there.

\subsection{GPTs}

The most recent reconstructions of quantum theory have been in the generalised probabilistic theory (GPT) framework. These typically describe systems as finite-dimensional, regular, closed cones, along with an intersecting hyperplane specifying the normalised states. Transformations are then described as linear maps and, in particular, effects are linear functionals on the state space.  In Prop.~\ref{lem:ConvexCones} of our reconstruction we obtain the essence of this structure from our classical interface. In fact, our classical interface can be interpreted as representing the probabilistic layer of the GPT framework  in a process-theoretic manner.

This classical interface however, does not provide all of the structure of the GPT framework. Specifically, we do not obtain that the state space is closed.  This is typically assumed in the study of GPTs. Closure is ---from an operational perspective--- a very natural assumption, as, up to any finite error, a state space should be indistinguishable from its closure. However, it is  not clear yet how this property should be understood from the process theoretic perspective.

Despite the lack of closure, we are still able to use some tools commonly found in the GPT literature---in particular the Koecher-Vinberg theorem. To do this, we focused on the states of the theory. However, our postulates apply equally well to arbitrary processes, and so, many of our results can immediately be generalised. We have  omitted such generalisations in an attempt to keep the mathematics of the paper to a minimum, and present only the results that directly pertain to the reconstruction.

\subsection{CPTs}

The formalism of Cateogrical Probabilistic Theories (CPTs) \cite{gogioso2017categorical} is another route to subsuming the GPT framework under a categorical or process-theoretic framework. CPTs are closely related to the conjunction of postulates 1 and 2, that is, a process theory with a classical interface. Specifically a CPT is defined (Def.~1 of \cite{gogioso2017categorical}) as a process theory subject to three mathematical requirements. The first of these is part of the definition of a classical interface, whilst the second two can be derived from the classical interface.

The classical interface is a mathematically strictly stronger assumption, as the CPT structure can be derived from it, but there are elements of the classical interface that cannot be derived from the CPT structure. For instance, the existence of \emph{all} classically-controlled processes is not a feature of a generic CPT. However, the classical interface has a clear physical interpretation, whilst the requirements of a CPT (in particular requirement 2) do not. One direction for future work is to further explore the connections between these two formalisms. For example, to determine exactly what must be added to the CPT formalism to derive the existence of a classical interface.

\section{Future work}

Whilst the postulates of this reconstruction are  diagrammatic the proof at times utilises some standard linear algebraic techniques. Moreover, whilst the postulates are defined at the level of processes we often only use them in the context of states. These are, to some extent, against the spirit of process theories, as such, it would be interesting in future work to try to make the proof of the reconstruction process-theoretic along with postulates. Indeed, as discussed earlier, it seems plausible that, by using the full strength of the postulates, there may be a much simpler and more direct way to go about the reconstruction. Another more recent categorical reconstruction of quantum theory \cite{tullReconstruction,tullSuperposition} may help providing a route to this.

As with all reconstructions, we assume the existence of a classical interface for the theory representing  how we interact with the world, for example, deciding the experiments to perform, and  obtaining classical data as outputs.
However, at some level we expect this  classical interface to be an emergent feature of quantum theory, and so it would be interesting to see if we can move beyond this  probabilistic approach. Can we instead express everything in diagrammatic terms,
and find some replacement for the classical interface? For example, can we capture the copiability and deletability of classical data via `spiders' \cite{CPav, CPaqPav, CKbook}?

A second issue with the classical interface (as it is defined here) is that it immediately limits the scope of the reconstruction to finite dimensional quantum theory. It would be interesting to see whether our axioms could reconstruct quantum theory even if we removed this limitation, that is, if we removed the `finite' part of finite tomography. If not, it would be interesting to explore these alternative theories and how close they are to quantum theory. One potential approach to this would be to follow the works of \cite{gogioso2016infinite,gogiosotowards} or of \cite{abramsky1999nuclear,coecke2016pictures,abramsky2012h}, which demonstrate how to extend various diagrammatic features ---such as cups \& caps--- to the infinite dimensional setting.

There are many recent results in quantum foundations and in quantum information \cite{sikora2017simple,lee2015generalised,lee2016higher,lee2017no,chiribella2015entanglement,chiribella2015operational,chiribella2014distinguishability,TowardsThermo,Microcanonical,HOP} which use the standard notion of purification in the derivation but where the result is also valid in classical theory. It therefore seems plausible that the same results could be obtained using our notion of symmetric purification, and so, apply to a wider range of theories. Another recent research direction in quantum foundations is in formulating quantum theory in a causally neutral \cite{leifer2013towards,oreshkov2011quantum} or time symmetric \cite{oreshkov2015operational,aharonov2010time} way or with indefinite causal order \cite{HardyCausaloid,araujo2016purification,kissinger2017picturing,brukner2015bounding,chiribella2012perfect}. This notion of purification may be more applicable in such situations as it does not distinguish between input and output systems.

We presented the cups \& caps of postulate~\ref{def:compactStructure} as a relaxation of a  basic process-theoretic constraints on compositionality by freely allowing inputs to be connected to inputs and outputs to be connected to outputs.  One could equally ask what  structure we obtain if we relax the other constraints on compositionality. In particular, relaxing the constraint that only pairs of inputs, and outputs are connected means that we need some sort of notion of `multi-wires'. This naturally gives the structure of `spiders' described in \cite{CKbook}.  Moreover, allowing for different system types to be connected leads to many other typical quantum processes. It is the subject of ongoing work to investigate how much of the structure of quantum theory can be described in such  terms, and, how to reconcile such a perspective with causality and the classical interface.

\section*{Acknowledgements}
The authors would like to thank Markus M\"uller for helpful comments on an earlier draft of the paper which have significantly improved the clarity of the presentation. The authors would also like to thank Jon Barrett, Matty Hoban, David Jennings, Ciar\'{a}n Lee,
Kenji Nakahira, Miguel Navascu\'es, Rob Spekkens, and Sean Tull for useful comments and suggestions.  The authors would also like to thank the referees whose careful reading and insightful comments greatly improved the manuscript twice.

CMS acknowledges the support by the Pacific Institute for the Mathematical Sciences, by a Faculty of Science Grand Challenge Award at the University of Calgary, EPSRC doctoral training grant and by Oxford-Google DeepMind Graduate Scholarship. JHS acknowledges the support of EPSRC via the Controlled Quantum Dynamics Centre for Doctoral Training and the Perimeter Institute for Theoretical Physics. Research at Perimeter Institute is supported by the Government of Canada through the Department of Innovation, Science and Economic Development Canada and by the Province of Ontario through the Ministry of Research, Innovation and Science. This
research was also supported in part by the Foundation for
Polish Science through IRAP project co-financed by EU
within Smart Growth Operational Programme (contract
no. 2018/MAB/5). This project/publication  was made possible through the support of a grant  from the John Templeton Foundation. The opinions expressed in this publication are those of the author(s) and do not necessarily reflect the views of the John Templeton Foundation.

\bibliographystyle{plainnat}
\bibliography{bibliography}

\appendix
\section{Consequences of a classical interface}\label{App:ClassInterface}
\subsection{Proof of Proposition \ref{lem:UniqueClassicalControl}}\label{proof:unqiueCC}
\proof
Let us assume that we have two different processes $F^0$ and $F^1$ which classically control the same set of processes $\{f_i\}$, that is:
\[\forall i \quad %
\InputIfFileExists{Diagrams/UniqueCC1.tikz}{}{\input{./figures/Diagrams/UniqueCC1.tikz}}\]
Now, let us characterise these two processes $F^\alpha$ where $\alpha \in \{0,1\}$ using tomography (Def.\ref{def:Tomography}). That is, consider composing $F^\alpha$ with arbitrary tester-vises, $\tau$:
\[%
\InputIfFileExists{Diagrams/UniqueCC2.tikz}{}{\input{./figures/Diagrams/UniqueCC2.tikz}}\]
if, for all $\tau$, this is independent of $\alpha$ then tomography tells us that $F^0=F^1$. Let us now show that this is indeed the case. Firstly let us define a pair of vises $\tau^\alpha$:
\[%
\InputIfFileExists{Diagrams/UniqueCC3.tikz}{}{\input{./figures/Diagrams/UniqueCC3.tikz}}\quad =:\quad %
\InputIfFileExists{Diagrams/UniqueCC4.tikz}{}{\input{./figures/Diagrams/UniqueCC4.tikz}} \]
and note that these have only classical inputs and outputs so are themselves classical processes. We can therefore decompose the classical identity and use distributivity of the classical sum over classical diagrams:
\[%
\InputIfFileExists{Diagrams/UniqueCC5.tikz}{}{\input{./figures/Diagrams/UniqueCC5.tikz}}\quad =\quad %
\InputIfFileExists{Diagrams/UniqueCC6.tikz}{}{\input{./figures/Diagrams/UniqueCC6.tikz}} \]
Then using the definition of the $\tau^\alpha$ and the definition of the $F^\alpha$ as classically controlled processes for the set $\{f_i\}$ we obtain:
\[=%
\InputIfFileExists{Diagrams/UniqueCC7.tikz}{}{\input{./figures/Diagrams/UniqueCC7.tikz}}\quad =\quad %
\InputIfFileExists{Diagrams/UniqueCC8.tikz}{}{\input{./figures/Diagrams/UniqueCC8.tikz}} \]
Then simply note that this is independent of $\alpha$ for all $\tau$ to complete the proof.
\endproof

\subsection{Proof of Proposition \ref{lem:ConvexCones}}\label{proof:diagSums}
\proof
 Recall that we propose a sum of any finite set of processes to be defined as:
\beq 
\InputIfFileExists{Sums/Sums1.tikz}{}{\input{./figures/Sums/Sums1.tikz}}
\eeq
where $F$ is the  unique (Prop.~\ref{lem:UniqueClassicalControl}) classically controlled process satisfying:
\beq\label{eq:controlS}
\InputIfFileExists{Sums/Sums2.tikz}{}{\input{./figures/Sums/Sums2.tikz}}.
\eeq
 and
\[%
\begin{tikzpicture}
	\begin{pgfonlayer}{nodelayer}
		\node [style=downground] (0) at (0, -0.5) {};
		\node [style=none] (1) at (0, 0.5) {};
		\node [style={right label}] (2) at (0, 0.25) {};
	\end{pgfonlayer}
	\begin{pgfonlayer}{edgelayer}
		\draw [cWire] (1.center) to (0);
	\end{pgfonlayer}
\end{tikzpicture}}=\sum_i %
\begin{tikzpicture}
	\begin{pgfonlayer}{nodelayer}
		\node [style=point] (0) at (0, -0) {$i$};
		\node [style=none] (1) at (0, 1) {};
	\end{pgfonlayer}
	\begin{pgfonlayer}{edgelayer}
		\draw [style=cWire] (1.center) to (0);
	\end{pgfonlayer}
\end{tikzpicture}}\]
First, as a sanity check, note that this is equivalent to the standard sum for classical processes. This follows immediately from the definition of $F$ and by decomposing the identity process:
\[%
\InputIfFileExists{Diagrams/CSumCheck1.tikz}{}{\input{./figures/Diagrams/CSumCheck1.tikz}}\ = \ \sum_i %
\InputIfFileExists{Diagrams/CSumCheck2.tikz}{}{\input{./figures/Diagrams/CSumCheck2.tikz}}\ = \ \sum_i %
\InputIfFileExists{Diagrams/CSumCheck3.tikz}{}{\input{./figures/Diagrams/CSumCheck3.tikz}} \ = \ \sum_i\ %
\InputIfFileExists{Diagrams/CSumCheck4.tikz}{}{\input{./figures/Diagrams/CSumCheck4.tikz}}\]
The sum that we are defining is therefore an extension of the classical sum to the full process theory. In particular, this means that, as scalars are classical (see the discussion after Def.~\ref{def:FullSubTheory}), that the sum of scalars is the standard sum of non-negative real numbers that we would expect.
Next,given this proposed definition, we need to check that this is indeed  a valid summation, i.e.\ that it satisfies the conditions of Def.~\ref{def:DiagrammaticSums}. The key step is to show that:
\beq\label{eq:Distributivity}
\InputIfFileExists{Sums/Dist1.tikz}{}{\input{./figures/Sums/Dist1.tikz}}
\eeq
holds for any diagram $\chi$. To check this, first note that the LHS of this is defined through Eq.~\eqref{eq:DefiningSums} as:
\[%
\InputIfFileExists{Sums/Dist2.tikz}{}{\input{./figures/Sums/Dist2.tikz}}\]
On the other hand, to understand the RHS let us first define:
\[%
\InputIfFileExists{Sums/Dist3.tikz}{}{\input{./figures/Sums/Dist3.tikz}}\]
then, using Eq.~\eqref{eq:DefiningSums} again, the RHS is equal to:
\[%
\InputIfFileExists{Sums/Dist4.tikz}{}{\input{./figures/Sums/Dist4.tikz}}\]
where we are using classical control to define $G$ such that it satisfies:
\[%
\InputIfFileExists{Sums/Dist5.tikz}{}{\input{./figures/Sums/Dist5.tikz}}\]
It is then simple to see that:
\[\forall i \quad %
\InputIfFileExists{Sums/Dist6.tikz}{}{\input{./figures/Sums/Dist6.tikz}}\]
which, using local tomography of classical theory, implies that:
\[%
\InputIfFileExists{Sums/Dist8.tikz}{}{\input{./figures/Sums/Dist8.tikz}}\]
We therefore find that:
\[%
\InputIfFileExists{Sums/Dist7L.tikz}{}{\input{./figures/Sums/Dist7L.tikz}}\ \ =%
\InputIfFileExists{Sums/Dist7.tikz}{}{\input{./figures/Sums/Dist7.tikz}}\ \ =\ \ %
\InputIfFileExists{Sums/Dist7R.tikz}{}{\input{./figures/Sums/Dist7R.tikz}} \]
and so, Eq.~\ref{eq:SumsDistribute} is satisfied for all
diagrams $\chi$. Hence, the sums are free to move around diagrams. The other constraints on the sum (e.g.\ commutativity etc.) are immediately inherited from the equivalent property of the classical sum.
\endproof

\subsection{Proof of Proposition \ref{prop:CausalTheory}}\label{proof:CST}
\proof
First, note that if we have local causal-compatible tomographic tests then we have:
\beq%
\InputIfFileExists{Diagrams/LocalCausalTomography.tikz}{}{\input{./figures/Diagrams/LocalCausalTomography.tikz}}\label{eq:causcompatproof}\eeq
Next, we want to characterise the set of processes in the full theory that are causal-compatible, that is, the set of processes which will not lead to any non-causal classical processes. By assumption the tomographic-tests are causal-compatible, hence, a minimal requirement for any other process $f$ to be causal-compatible is that it must satisfy:
\beq%
\InputIfFileExists{Diagrams/LocalCausalTomography1.tikz}{}{\input{./figures/Diagrams/LocalCausalTomography1.tikz}}\label{eq:CausCompatGeneral}\eeq
We will now see that this is a necessary and sufficient constraint, which, can be moreover expressed as the usual causality condition by suitably defining discarding maps. To see this consider the special case of eq.~\eqref{eq:CausCompatGeneral} in which $B$ is trivial, that is:
\[%
\InputIfFileExists{Diagrams/LocalCausalTomography2.tikz}{}{\input{./figures/Diagrams/LocalCausalTomography2.tikz}}\]
We therefore find, for all causal-compatible effects, $f$, that:
\[%
\InputIfFileExists{Diagrams/LocalCausalTomography3.tikz}{}{\input{./figures/Diagrams/LocalCausalTomography3.tikz}}\]
and hence, by tomography that:
\[%
\InputIfFileExists{Diagrams/LocalCausalTomography4.tikz}{}{\input{./figures/Diagrams/LocalCausalTomography4.tikz}}\]
We can therefore take this to be a unique discarding map for system $A$:
\beq%
\InputIfFileExists{Diagrams/LocalCausalTomography5.tikz}{}{\input{./figures/Diagrams/LocalCausalTomography5.tikz}}\label{eq:defCausGeneral}\eeq
Given this definition, then our causal-compatibility condition (eq.~\eqref{eq:CausCompatGeneral}) can be reexpressed as a causality constraint. To begin we substitute eq.~\ref{eq:defCausGeneral} into eqs.~\eqref{eq:CausCompatGeneral} and \eqref{eq:causcompatproof} to obtain
\[%
\InputIfFileExists{Diagrams/LocalCausalTomography6.tikz}{}{\input{./figures/Diagrams/LocalCausalTomography6.tikz}} \]
respectively.
Then, by tomography we find that $f$ must satisfy the usual causality condition:
\[%
\InputIfFileExists{Diagrams/LocalCausalTomography7.tikz}{}{\input{./figures/Diagrams/LocalCausalTomography7.tikz}}.\]
This constraint is clearly necessary, however, how do we see that it is moreover sufficient? This is simple, just note that the causal-compatibility condition merely requires that any process with a classical input and no outputs is the classical discarding, as we have now demonstrated that there is a unique effect for every system then this is just a special case.
\endproof

\subsection{Proof of Proposition \ref{lem:LinearityAndConvexCones}}\label{proof:GPTStruct}
\proof
Firstly  recall, as noted underneath Def.~\ref{def:FullSubTheory}, that the scalars in the theory are classical and hence are non-negative real numbers. It is then clear that the state space of a given system $A$ is a convex cone $C_A$, as it is closed under linear combinations with non-negative real coefficients, i.e.:
\[%
\InputIfFileExists{Sums/linearCombination.tikz}{}{\input{./figures/Sums/linearCombination.tikz}}\]
is a valid state.

Allowing the coefficients of linear combinations to be negative, the cone extends naturally to a real ordered vector space, spanned and ordered by the cone itself. By construction, the cone is then full dimensional (i.e., it spans the vector space), and it is simple to show that it is pointed (i.e.\ the zero-vector is in the cone, and it is the unique vector for which $v$ and $-v$ are in the cone). Moreover the cone is finite-dimensional: this immediately follows from finite tomography, as it implies that a state is characterised by a finite number of real values.

Thanks to distributivity of sums (Eq.~\ref{eq:SumsDistribute}), any process $f:A\rightarrow B$ induces a positive linear map between the vector space spanned by the states of system $A$ and the vector space spanned by the states of system $B$. This is defined as:
\[%
\InputIfFileExists{Sums/linearMap1.tikz}{}{\input{./figures/Sums/linearMap1.tikz}}\]
 where linearity follows immediately as a special case of Eq.~\ref{eq:SumsDistribute} by:
 \[%
\InputIfFileExists{Sums/linearMap2.tikz}{}{\input{./figures/Sums/linearMap2.tikz}}\]
As it maps the convex cone of states of $A$ to the convex cone of states of $B$, the induced linear map is positive. Moreover, the map is \emph{completely positive}, as it maps bipartite states to bipartite states when applied locally to a subsystem.

An immediate corollary is that effects are linear functionals on the cone of states. Therefore, $%
\begin{tikzpicture}[scale=0.5]
	\begin{pgfonlayer}{nodelayer}
		\node [style=none] (0) at (0, 0.25) {};
		\node [style=none] (1) at (0, -0.5) {};
		\node [style=upground] (2) at (0, 0.5) {};
	\end{pgfonlayer}
	\begin{pgfonlayer}{edgelayer}
		\draw (0.center) to (1.center);
	\end{pgfonlayer}
\end{tikzpicture}}\circ s=1$ defines a hyperplane. This hyperplane intersects the cone, but not the origin: first consider

\[%
\begin{tikzpicture}
	\begin{pgfonlayer}{nodelayer}
		\node [style=point] (0) at (0, -0.5) {$s$};
		\node [style=none] (1) at (0, 0.5) {};
		\node [style=upground] (2) at (0, 0.75) {};
	\end{pgfonlayer}
	\begin{pgfonlayer}{edgelayer}
		\draw [qWire] (1.center) to (0);
	\end{pgfonlayer}
\end{tikzpicture}}= 0\ \ \iff\ \ %
\InputIfFileExists{Diagrams/TraceZero2.tikz}{}{\input{./figures/Diagrams/TraceZero2.tikz}}= 0\ \ \iff\ \ %
\InputIfFileExists{Diagrams/TraceZero3.tikz}{}{\input{./figures/Diagrams/TraceZero3.tikz}}= 0\ \ \iff\ \ %
\begin{tikzpicture}
	\begin{pgfonlayer}{nodelayer}
		\node [style=point] (0) at (0, -0.5) {$s$};
		\node [style=none] (1) at (0, 0.5) {};
	\end{pgfonlayer}
	\begin{pgfonlayer}{edgelayer}
		\draw [qWire] (1.center) to (0);
	\end{pgfonlayer}
\end{tikzpicture}}= 0 \]

Therefore, for any state $s\neq 0$, there is some scalar $r_s$ such that $%
\begin{tikzpicture}[scale=0.5]
	\begin{pgfonlayer}{nodelayer}
		\node [style=none] (0) at (0, 0.25) {};
		\node [style=none] (1) at (0, -0.5) {};
		\node [style=upground] (2) at (0, 0.5) {};
	\end{pgfonlayer}
	\begin{pgfonlayer}{edgelayer}
		\draw (0.center) to (1.center);
	\end{pgfonlayer}
\end{tikzpicture}}\circ(r_s s)=1$; hence the hyperplane intersects the cone. This is equivalent to the statement that $%
\begin{tikzpicture}[scale=0.5]
	\begin{pgfonlayer}{nodelayer}
		\node [style=none] (0) at (0, 0.25) {};
		\node [style=none] (1) at (0, -0.5) {};
		\node [style=upground] (2) at (0, 0.5) {};
	\end{pgfonlayer}
	\begin{pgfonlayer}{edgelayer}
		\draw (0.center) to (1.center);
	\end{pgfonlayer}
\end{tikzpicture}}$ is in the interior of the dual to the state cone.
\endproof

\section{Classification of leaks and pure processes}\label{app:CStar}
We now classify the leaks and pure processes for quantum theory.  First however, recall the representation of a finite dimensional C*-algebra $\mathcal{A}$ from Ex.~\ref{Ex:CStar},
\[\mathcal{A}=\bigoplus_i A_i\quad\text{where}\quad A_i=\mathcal{B}(\mathcal{H}_i)\]
and recall that processes in the processes theory are completely positive maps between these C*-algebras. We now introduce a few processes and basic results that will be used in this appendix.

\ben
\item Given a C*-algebra $\mathcal{A}=\bigoplus_iA_i$ there is a decomposition of the identity into orthogonal projectors
\[%
\InputIfFileExists{CStarDiagrams/decompositionIdentity.tikz}{}{\input{./figures/CStarDiagrams/decompositionIdentity.tikz}}\]
\item such that each of these projectors \emph{splits} through an irreducible C*-algebra, that is
\[%
\InputIfFileExists{CStarDiagrams/splittingIdempotent.tikz}{}{\input{./figures/CStarDiagrams/splittingIdempotent.tikz}}\quad\text{where}\quad%
\InputIfFileExists{CStarDiagrams/splitingIdempotent2.tikz}{}{\input{./figures/CStarDiagrams/splitingIdempotent2.tikz}}\]
 note that these triangles are not states and effects as they have both inputs and outputs, the use of the triangle is simply to distinguish them from generic processes,  linear algebraically, they are the coordinate projections (mapping $\mathcal{A}\to A_i$) and inclusion maps (mapping $A_i\to \mathcal{A}$),
\item this decomposition of the identity provides us with a \emph{matrix representation} of processes with input $\mathcal{A}=\bigoplus_{i=1}^n A_i$ and output $\mathcal{B}=\bigoplus_{j=1}^m B_j$ as follows
\[%
\InputIfFileExists{CStarDiagrams/matrixRepresentation.tikz}{}{\input{./figures/CStarDiagrams/matrixRepresentation.tikz}}\quad \sim \quad \left(f_{ij}\right)_{i=1,..,n}^{j=1,...,m}\]
\item where each $f_{ij}$ defines a map between the irreducible C*-algebras $A_i$ and $B_j$ as
\[%
\InputIfFileExists{CStarDiagrams/quantumProcesses.tikz}{}{\input{./figures/CStarDiagrams/quantumProcesses.tikz}}\]
\item Note the discarding map for the C*-algebra and for the irreducible components are related by
\[%
\InputIfFileExists{CStarDiagrams/decompTrace.tikz}{}{\input{./figures/CStarDiagrams/decompTrace.tikz}}\]
\item Finally,  for this process theory, it can be shown that
\[%
\InputIfFileExists{CStarDiagrams/traceZero.tikz}{}{\input{./figures/CStarDiagrams/traceZero.tikz}}.\]
This, in fact, applies more generally than just to quantum theory as it can be derived from local tomography, that any effect can appear in some decomposition of the discarding map, and that scalars are non-negative real numbers (utilising the fact that if the sum of a set of non-negative numbers is zero then every element must be zero).
\een

Recalling the definition of pure processes (Def.~\ref{def:pureeq}), we see that to understand what the pure processes are we must understand what the leaks are for these systems. That is, what leaks do we have for a C*-algebra?
For fully quantum systems all leaks are trivial, that is, any leak separates:
\[%
\InputIfFileExists{CStarDiagrams/quantumLeak.tikz}{}{\input{./figures/CStarDiagrams/quantumLeak.tikz}}\]
however, in classical theory, and more general C*-algebras the leaks are more interesting as we will now demonstrate.

\begin{proposition}\label{Prop:LeakClassification}
Given a C*-algebra, $\mathcal{A} = \bigoplus_i A_i$ any leak can be written as:
\[%
\InputIfFileExists{CStarDiagrams/CStarLeaks.tikz}{}{\input{./figures/CStarDiagrams/CStarLeaks.tikz}}\]
 where the $\rho_i$ are normalised quantum states. Note that the connecting system on the right hand diagram is a classical system, so the `leaked information' is always essentially classical.
\end{proposition}
\proof
Note first that any leak for $\mathcal{A}$ defines a leak for each of the $A_i$ by pre- and post- composing with the relevant projector, hence, as the $A_i$ are fully quantum systems these leaks must be constant:
\beq\label{proof:leak1}
\InputIfFileExists{CStarDiagrams/leaks1.tikz}{}{\input{./figures/CStarDiagrams/leaks1.tikz}}
\eeq
We therefore have
\[%
\InputIfFileExists{CStarDiagrams/leaks2.tikz}{}{\input{./figures/CStarDiagrams/leaks2.tikz}}\]
However, by decomposing the identity as a sum of the projectors, using the defining equation of a leak, and  that the $\rho_i$ are normalised, we find
\[%
\InputIfFileExists{CStarDiagrams/leaks3.tikz}{}{\input{./figures/CStarDiagrams/leaks3.tikz}}\]
and hence if $i\neq j$
\[%
\InputIfFileExists{CStarDiagrams/leaks4.tikz}{}{\input{./figures/CStarDiagrams/leaks4.tikz}}\ \ = \ \ 0 \qquad \text{and so}\qquad %
\InputIfFileExists{CStarDiagrams/leaks7.tikz}{}{\input{./figures/CStarDiagrams/leaks7.tikz}}\ \ = \ \ 0\]
Combining this with Eq.~\eqref{proof:leak1} provides us with the result. Classical control then allows us to write this in the form
\[%
\InputIfFileExists{CStarDiagrams/leaks5.tikz}{}{\input{./figures/CStarDiagrams/leaks5.tikz}} \quad \text{where}\qquad %
\InputIfFileExists{CStarDiagrams/leaks6.tikz}{}{\input{./figures/CStarDiagrams/leaks6.tikz}}\]
which completes the proof.
\endproof

We are now in a position to understand what the pure processes are for quantum theory.

\begin{proposition}\label{Prop:PureProcesses}
Pure processes in quantum theory are processes whose matrix representation has i) pure processes as matrix elements, and ii) at most a single non-zero process in each row and column.
\end{proposition}
\proof
Consider a pure process $f:\mathcal{A}\to\mathcal{B}$ where $\mathcal{A}=\bigoplus_iA_i$ and $\mathcal{B}=\bigoplus_jB_j$, and its matrix representation $f=\sum_{ij}f_{ij}$.
Now given the leak of system $\mathcal{B}$, which leaks the `which branch' information to a classical system
\[%
\InputIfFileExists{CStarDiagrams/whichBranchLeak.tikz}{}{\input{./figures/CStarDiagrams/whichBranchLeak.tikz}}\]
 we find that
\[%
\InputIfFileExists{CStarDiagrams/pure2.tikz}{}{\input{./figures/CStarDiagrams/pure2.tikz}}\ \ := \ \ \sum_{ij}f_{ij}l_{ij}\]
 The first equality follows from realising that in classical theory composing the cap with the broadcasting map in this way is just the discarding map, i.e.:
\[\begin{tikzpicture}
	\begin{pgfonlayer}{nodelayer}
		\node [style={white dot}] (0) at (0, -0) {};
		\node [style=none] (1) at (0, -0.9999999) {};
		\node [style=none] (2) at (-0.7499999, 0.9999999) {};
		\node [style=none] (3) at (0.7499999, 0.9999999) {};
		\node [style=none] (4) at (2, -0) {$=$};
		\node [style=upground] (5) at (4, 0.25) {};
		\node [style=none] (6) at (4, -0) {};
		\node [style=none] (7) at (4, -0.9999999) {};
	\end{pgfonlayer}
	\begin{pgfonlayer}{edgelayer}
		\draw [style=cWire] (1.center) to (0);
		\draw [style=cWire, bend left, looseness=1.00] (0) to (2.center);
		\draw [style=cWire, bend left=90, looseness=2.00] (2.center) to (3.center);
		\draw [style=cWire, bend left, looseness=1.00] (3.center) to (0);
		\draw[style=cWire] (6.center) to (7.center);
	\end{pgfonlayer}
\end{tikzpicture}\]
The second from the fact that the same `which branch' information is leaked each time, i.e.:
\[\begin{tikzpicture}
	\begin{pgfonlayer}{nodelayer}
		\node [style={white dot}] (0) at (0, 0.75) {};
		\node [style=none] (1) at (0.75, 1.75) {};
		\node [style=none] (2) at (2, -0) {$=$};
		\node [style=none] (3) at (-0.75, 1.75) {};
		\node [style=none] (4) at (-1.75, 1.75) {};
		\node [style=leak] (5) at (-1.75, -0.4999999) {};
		\node [style=none] (6) at (-1.75, -1.75) {};
		\node [style=leak] (7) at (3.75, 0.75) {};
		\node [style=none] (8) at (3.75, 1.75) {};
		\node [style=none] (9) at (3.75, -0) {};
		\node [style=leak] (10) at (3.75, -0.75) {};
		\node [style=none] (11) at (3.75, -0) {};
		\node [style=none] (12) at (3.75, -1.75) {};
		\node [style=none] (13) at (4.75, 1.75) {};
		\node [style=none] (14) at (5.75, 1.75) {};
	\end{pgfonlayer}
	\begin{pgfonlayer}{edgelayer}
		\draw [style=cWire, in=23, out=-90, looseness=1.00] (1.center) to (0);
		\draw [style=cWire, in=157, out=-90, looseness=1.00] (3.center) to (0);
		\draw [style=qWire] (4.center) to (5);
		\draw [style=qWire] (5) to (6.center);
		\draw [style=qWire] (8.center) to (7);
		\draw [style=qWire] (7) to (9.center);
		\draw [style=qWire] (11.center) to (10);
		\draw [style=qWire] (10) to (12.center);
		\draw [style=cWire, in=-98, out=15, looseness=0.75] (5) to (0);
		\draw [style=cWire, in=-90, out=15, looseness=0.75] (7) to (13.center);
		\draw [style=cWire, in=-90, out=30, looseness=0.75] (10) to (14.center);
	\end{pgfonlayer}
\end{tikzpicture}\]
The third comes from the definition of purity (Def.~\ref{def:pureeq}) and the above characterisation of leaks (Prop. \ref{Prop:LeakClassification}), and the final equality is obtained by decomposing the two leaks into the branches.

We therefore find that $f_{ij}l_{ij}=f_{ij}$ and so either $f_{ij}=0$ or $l_{ij}=1$. However, as we know that $l$ is causal, this means that $\sum_j l_{ij}=1$ for all $i$, and so for each $i$ there is only a single value of $j$ such that $l_{ij}=1$. Therefore, for that $i$, all other values of $j$ must result in $f_{ij}=0$. That is, the matrix representation of $f$ has at most a single non-zero element in each row. We can make an equivalent argument starting with the leak at the bottom pushing it through to the top. In this case we find that there is at most one non-zero element in each column.

Now note that every dilation $F$ of $f$ must be given by some leak, which means that
\[%
\InputIfFileExists{CStarDiagrams/pure3.tikz}{}{\input{./figures/CStarDiagrams/pure3.tikz}}\]
therefore
\[%
\InputIfFileExists{CStarDiagrams/pure4.tikz}{}{\input{./figures/CStarDiagrams/pure4.tikz}}\]
and so every dilation of each of the $f_{ij}$ must separate. In other words, the $f_{ij}$ are pure quantum processes.

To summarise, from the `commutativity' conditions, we find that a pure process $f$ maps each  summand of $\mathcal{A}$ to at most one  summand of $\mathcal{B}$ and vice versa. Furthermore, from the `all dilations are leaks' condition we find that these maps between  summands are themselves pure.
\endproof

\section{Symmetric purification in quantum theory}\label{App:ProofSymPurif}

Given the characterisation of the pure processes in App.~\ref{app:CStar}, we can show that quantum theory satisfies the symmetric purification postulate.

\begin{theorem} \label{app:symmetricPurification}Completely positive maps between C*-algebras have symmetric purifications.\end{theorem}
\proof
First let us consider the quantum case. We know (from Stinespring's dilation theorem \cite{Stinespring}) that any process $f:A\to B$ can be purified to a process $\mathcal{F}:A\to B\otimes C$ where $C=A\otimes B$, and so it is simple to see that it can also be purified in a symmetric way $F:A\otimes B \to B \otimes A$:
\[%
\InputIfFileExists{CStarDiagrams/symPurifProof1.tikz}{}{\input{./figures/CStarDiagrams/symPurifProof1.tikz}}\]
where we have used the fact that the caps are pure for fully quantum systems, and that the composite of pure processes is pure. It therefore just remains to check that any two such purifications are related in the correct way. Consider two such  symmetric purifications $F$ and $G$, then we can use the cup ---as it is pure for fully quantum systems--- to define two  standard purifications (see Sec.~\ref{Sec:StandardPurification}) as:
\[%
\InputIfFileExists{CStarDiagrams/symPurifProof2.tikz}{}{\input{./figures/CStarDiagrams/symPurifProof2.tikz}}\]
these must be related by a reversible, and hence causal, transformation $r$:
\[%
\InputIfFileExists{CStarDiagrams/symPurifProof3.tikz}{}{\input{./figures/CStarDiagrams/symPurifProof3.tikz}}\]
Using the caps for a second time therefore means that:
\beq\label{eq:quantumSymmetricPurification}
\InputIfFileExists{CStarDiagrams/symPurifProof4.tikz}{}{\input{./figures/CStarDiagrams/symPurifProof4.tikz}}
\eeq
where the two processes forming the  vise $R$ (see Eq.~\ref{eq:circuitfragment}) are given by:
\[\begin{tikzpicture}
	\begin{pgfonlayer}{nodelayer}
		\node [style=none] (0) at (-15, 0.5) {};
		\node [style=none] (1) at (-14, 1.25) {};
		\node [style=none] (2) at (-14.5, -1.25) {};
		\node [style=none] (3) at (-12, -0) {$:=$};
		\node [style=none] (4) at (-15, 1.25) {};
		\node [style=none] (5) at (-14, 0.5) {};
		\node [style=none] (6) at (-14.5, -0.5) {};
		\node [style=none] (7) at (-15.5, 0.5) {};
		\node [style=none] (8) at (-13.5, -0.5) {};
		\node [style=none] (9) at (-14.5, -0) {$x_R$};
		\node [style=none] (10) at (-14.5, -0.5) {};
		\node [style=none] (11) at (-13.5, 0.5) {};
		\node [style=none] (12) at (-15.5, -0.5) {};
		\node [style=none] (13) at (-10, 1.25) {};
		\node [style=none] (14) at (-7.5, 0.5) {};
		\node [style=none] (15) at (-7.5, -0.5) {};
		\node [style=none] (16) at (-9.25, -1.25) {};
		\node [style=none] (17) at (-8, 1.25) {};
		\node [style=none] (18) at (-9.25, -0.5) {};
		\node [style=none] (19) at (-7.5, 0.5) {};
		\node [style=none] (20) at (-9, 1.25) {};
		\node [style=none] (21) at (-10.5, -0.5) {};
		\node [style=none] (22) at (-10, 0.5) {};
		\node [style=none] (23) at (-8, 0.5) {};
		\node [style=none] (24) at (-10.5, 0.5) {};
		\node [style=none] (25) at (-9, 0.5) {};
		\node [style=none] (26) at (-13.75, 0.5) {};
		\node [style=none] (27) at (-13.75, 1.25) {};
	\end{pgfonlayer}
	\begin{pgfonlayer}{edgelayer}
		\draw [qWire] (4.center) to (0.center);
		\draw [qWire] (1.center) to (5.center);
		\draw (12.center) to (8.center);
		\draw (8.center) to (11.center);
		\draw (11.center) to (7.center);
		\draw (7.center) to (12.center);
		\draw [qWire] (6.center) to (10.center);
		\draw [qWire] (10.center) to (2.center);
		\draw [qWire] (13.center) to (22.center);
		\draw [qWire, bend right=90, looseness=1.25] (22.center) to (25.center);
		\draw [qWire] (25.center) to (20.center);
		\draw [qWire] (17.center) to (23.center);
		\draw [qWire, in=-90, out=90, looseness=1.00] (18.center) to (23.center);
		\draw [style=thick gray dashed edge] (14.center) to (15.center);
		\draw [style=thick gray dashed edge] (15.center) to (21.center);
		\draw [style=thick gray dashed edge] (21.center) to (24.center);
		\draw [style=thick gray dashed edge] (24.center) to (19.center);
		\draw [qWire] (18.center) to (16.center);
		\draw [qWire] (27.center) to (26.center);
	\end{pgfonlayer}
\end{tikzpicture}
\quad\text{and}\quad
\begin{tikzpicture}
	\begin{pgfonlayer}{nodelayer}
		\node [style=none] (0) at (0.25, -2) {};
		\node [style=none] (1) at (0.25, -0.75) {};
		\node [style=none] (2) at (1.25, -1.25) {};
		\node [style=none] (3) at (1.25, -2) {};
		\node [style=none] (4) at (1.25, -1.25) {};
		\node [style=none] (5) at (1.25, -0.75) {};
		\node [style=none] (6) at (-4, 0.5) {};
		\node [style=none] (7) at (-3.25, -0.5) {};
		\node [style=none] (8) at (-3.5, -1.25) {};
		\node [style=none] (9) at (-1.5, -0) {$:=$};
		\node [style=none] (10) at (-4, 1.25) {};
		\node [style=none] (11) at (-3.25, -1.25) {};
		\node [style=none] (12) at (-3.5, -0.5000001) {};
		\node [style=none] (13) at (-5, 0.5000001) {};
		\node [style=none] (14) at (-3, -0.5000001) {};
		\node [style=none] (15) at (-4, -0) {$y_R$};
		\node [style=none] (16) at (-3.5, -0.5000001) {};
		\node [style=none] (17) at (-3, 0.5000001) {};
		\node [style=none] (18) at (-5, -0.5000001) {};
		\node [style=none] (19) at (-4.5, -1.25) {};
		\node [style=none] (20) at (-4.5, -0.5000001) {};
		\node [style=none] (21) at (0.75, -0.25) {$r$};
		\node [style=none] (22) at (0, -0.75) {};
		\node [style=none] (23) at (0, 0.25) {};
		\node [style=none] (24) at (1.5, 0.25) {};
		\node [style=none] (25) at (1.5, -0.75) {};
		\node [style=none] (26) at (0.25, 0.25) {};
		\node [style=none] (27) at (1.5, 1.25) {};
		\node [style=none] (28) at (1.25, 0.25) {};
		\node [style=none] (29) at (1.25, 0.25) {};
		\node [style=none] (30) at (2.25, 0.25) {};
		\node [style=none] (31) at (2.25, -2) {};
		\node [style=none] (32) at (-0.25, -1.25) {};
		\node [style=none] (33) at (2.75, -1.25) {};
		\node [style=none] (34) at (2.75, -1.25) {};
		\node [style=none] (35) at (2.75, 1.25) {};
		\node [style=none] (36) at (-0.25, 1.25) {};
		\node [style=none] (37) at (1.5, 2) {};
	\end{pgfonlayer}
	\begin{pgfonlayer}{edgelayer}
		\draw [qWire] (1.center) to (0.center);
		\draw [qWire] (5.center) to (4.center);
		\draw [qWire] (3.center) to (2.center);
		\draw [qWire] (10.center) to (6.center);
		\draw [qWire] (7.center) to (11.center);
		\draw [qWire] (20.center) to (19.center);
		\draw (18.center) to (14.center);
		\draw (14.center) to (17.center);
		\draw (17.center) to (13.center);
		\draw (13.center) to (18.center);
		\draw [qWire] (12.center) to (16.center);
		\draw [qWire] (16.center) to (8.center);
		\draw (23.center) to (24.center);
		\draw (24.center) to (25.center);
		\draw (25.center) to (22.center);
		\draw (22.center) to (23.center);
		\draw [qWire, in=90, out=-90, looseness=1.00] (27.center) to (26.center);
		\draw [qWire] (28.center) to (29.center);
		\draw [qWire, bend left=90, looseness=2.00] (28.center) to (30.center);
		\draw [qWire] (30.center) to (31.center);
		\draw [style=thick gray dashed edge] (32.center) to (33.center);
		\draw [style=thick gray dashed edge] (36.center) to (32.center);
		\draw [style=thick gray dashed edge] (36.center) to (35.center);
		\draw [style=thick gray dashed edge] (35.center) to (34.center);
		\draw [qWire] (37.center) to (27.center);
	\end{pgfonlayer}
\end{tikzpicture}
\]
It is simple to check that causality of $r$ implies that $R$ satisfies Eq.~\ref{eq:reversiblyConnectedConstraint} as required.

Next we turn to the general C*-algebraic case. Consider a process $f:\mathcal{A}\to\mathcal{B}$ where $\mathcal{A}=\bigoplus_i A_i$ and $\mathcal{B}=\bigoplus_j B_j$ with $\{A_i\}$ and $\{B_j\}$ irreducible C*-algebras, i.e.\ fully quantum systems. It is simple to check that this does have a symmetric purification by noting that we can define a dilation of a process $f$ by symmetrically purifying the quantum maps of matrix representation
\[%
\InputIfFileExists{CStarDiagrams/newCStarProof.tikz}{}{\input{./figures/CStarDiagrams/newCStarProof.tikz}}\]
It is then simple to check that this dilation is moreover pure and hence a symmetric purification of $f$. Like for the quantum case, the more interesting part to check is Eq.~\eqref{eq:ReversiblyConnectedDefinition}.

Given some process $f$ with two symmetric purifications $F$ and $G$, i.e.
\[%
\InputIfFileExists{CStarDiagrams/symPurifProof6.tikz}{}{\input{./figures/CStarDiagrams/symPurifProof6.tikz}}\]
let us define:
\[%
\InputIfFileExists{CStarDiagrams/symProofCStar1.tikz}{}{\input{./figures/CStarDiagrams/symProofCStar1.tikz}}\quad\qquad \text{and}\quad\qquad %
\InputIfFileExists{CStarDiagrams/symProofCStar4.tikz}{}{\input{./figures/CStarDiagrams/symProofCStar4.tikz}}
\]
Noting that these are purifications of the same fully quantum process, that is
\[%
\InputIfFileExists{CStarDiagrams/symPurifProof5.tikz}{}{\input{./figures/CStarDiagrams/symPurifProof5.tikz}}\]
and so using the above result for fully quantum systems we obtain:
\[%
\InputIfFileExists{CStarDiagrams/symProofCStar2.tikz}{}{\input{./figures/CStarDiagrams/symProofCStar2.tikz}}\]
and therefore:
\[%
\InputIfFileExists{CStarDiagrams/symProofCStar3.tikz}{}{\input{./figures/CStarDiagrams/symProofCStar3.tikz}}\]
where in the last step we have used classical control to construct $R$ and defined the forwards and backwards leaks as:
\[%
\InputIfFileExists{CStarDiagrams/CStarLeak1.tikz}{}{\input{./figures/CStarDiagrams/CStarLeak1.tikz}}\qquad\text{and}\qquad%
\InputIfFileExists{Diagrams/CStarLeak2v2.tikz}{}{\input{./figures/Diagrams/CStarLeak2v2.tikz}}
\qquad\text{respectively.}\]
That $R$ satisfies Eq.~\ref{eq:reversiblyConnectedConstraint} follows directly from the quantum case.
\endproof

\end{document}